\documentclass[a4paper,11pt]{article}
\pdfoutput=1 
\usepackage{jheppub}

\graphicspath{{figures}}
\usepackage{color}
\usepackage{physics}
\usepackage{graphicx}
\usepackage{csvsimple}
\usepackage{caption}
\usepackage{subcaption}

\newcommand\cCompletelyConfinedPZero{1.738(29)}
\newcommand\dCompletelyConfinedPZero{0.022(31)}
\newcommand\cConPpointTwo{1.729(17)}
\newcommand\dConPpointTwo{0.086(20)}
\newcommand\cMixPpointTwo{1.731(21)}
\newcommand\dMixPpointTwo{-0.076(24)}
\newcommand\cConPlusMixPpointTwo{1.733(19)}
\newcommand\dConPlusMixPpointTwo{0.021(22)}
\newcommand\cConPpointTwoFive{1.724(23)}
\newcommand\dConPpointTwoFive{0.127(28)}
\newcommand\cMixPpointTwoFive{1.708(20)}
\newcommand\dMixPpointTwoFive{-0.109(23)}
\newcommand\cConPlusMixPpointTwoFive{1.718(18)}
\newcommand\dConPlusMixPpointTwoFive{-0.006(21)}
\newcommand\cTZeroTwoFive{2.0887(30)}
\newcommand\dTZeroTwoFive{0.0011(31)}

\title{Linear confinement in the partially-deconfined phase}

\author[a]{Vaibhav Gautam,}
\author[a]{Masanori Hanada,}
\author[b]{Jack Holden,}
\author[c]{Enrico Rinaldi}

\preprint{DMUS-MP-22/14, RIKEN-iTHEMS-Report-22}

\affiliation[a]{Department of Mathematics, University of Surrey,
Guildford, Surrey, GU2 7XH, UK}
\affiliation[b]{STAG Research Centre, University of Southampton,
Southampton, SO17 1BJ, UK}
\affiliation[c]{Physics  Department,  University  of  Michigan,
Ann  Arbor,  MI  48109, United States}
\affiliation[c]{Theoretical  Quantum  Physics  Laboratory, Cluster for Pioneering Research (CPR), RIKEN,
Wako, Saitama 351-0198, Japan}
\affiliation[c]{Interdisciplinary Theoretical and Mathematical Sciences (iTHEMS) Program, RIKEN,
Wako, Saitama 351-0198, Japan}
\affiliation[c]{Center for Quantum Computing (RQC), RIKEN,
Wako, Saitama 351-0198, Japan}

\abstract{
We consider the partially-deconfined saddle of large-$N$ pure Yang-Mills theory lying between confined and deconfined phases, in which the color degrees of freedom split into confined and deconfined sectors. Based on the microscopic mechanism of deconfinement, we argue that a flux tube is formed in the confined sector and a linear confinement potential is generated. 
The string tension should not depend on the size of the confined sector. 
We provide evidence for the case of the finite-temperature strong-coupling lattice gauge theory. In particular, we make analytic predictions assuming linear confinement in the confined sector, and then confirm these by numerical simulations. We discuss some implications of the conjecture to QCD and holography. 
}

\begin{document} 
\maketitle
\flushbottom

\section{Introduction and conjecture}
\hspace{0.51cm}
For SU($N$) gauge theories at large $N$, the $N$-dependence of the energy $E$ and entropy $S$ provides us with a simple characterization of confined and deconfined phases. Typically, $E\sim S\sim N^2$ in the deconfined phase because $N^2$ gluons contribute. 
On the other hand, in the confined phase, individual colors are not visible and hence $E\sim S\sim N^0$ is expected up to the zero-point energy. Such a characterization has been very useful in the context of holography~\cite{Witten:1998zw,Sundborg:1999ue,Aharony:2003sx}. 
Furthermore, there is good reason to expect that this characterization is related to other characterizations of confinement and deconfinement, such as the disappearance of the linear confinement potential in the deconfined phase. 
The $O(N^2)$ entropy and energy come from the condensation of long strings~\cite{Polyakov:1978vu,Susskind:1979up} with length of order $N^2$ and such long strings naturally explain the disappearance of the confinement potential. 
See Sec.~\ref{sec:string_condensation} for details. 

\begin{figure}[htbp]
  \begin{center}
  \includegraphics[width=0.5\textwidth]{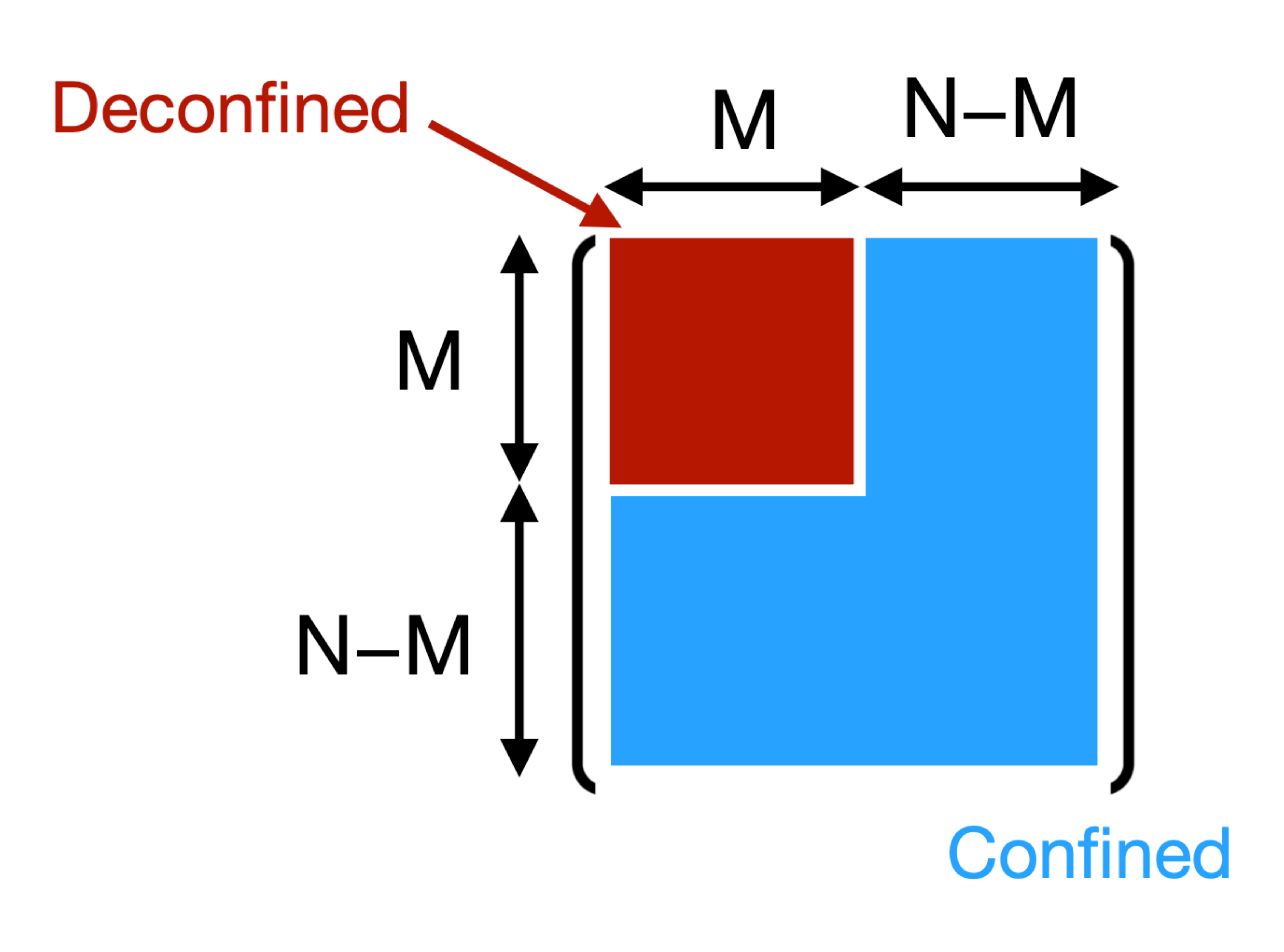}
  \end{center}
  \caption{In the partially-deconfined phase (equivalently, partially-confined phase), color degrees of freedom split into the confined and deconfined sectors. In this paper, we use $M$ to denote the size of the deconfined sector. 
  }\label{fig:Partial-deconfinement}
\end{figure}

Between confined and deconfined phases of SU($N$) gauge theory there exists the partially-deconfined phase (equivalently, partially-confined phase), in which a SU($M$) subgroup deconfines; see Fig.~\ref{fig:Partial-deconfinement}. The size of the deconfined sector $M$ can change from 0 (complete confinement) to $N$ (complete deconfinement)~\cite{Hanada:2016pwv,Berenstein:2018lrm,Hanada:2018zxn,Hanada:2019czd}. 
The underlying mechanism identified in Ref.~\cite{Hanada:2020uvt} does not depend on the details of the theories.  
Let us consider large-$N$ QCD with $N_f$ flavors in the fundamental representation. We take $\frac{N_f}{N}$ sufficiently large and quark masses sufficiently small such that the confinement/deconfinement transition is not of first order, thus resembling the real-world SU(3) QCD~\cite{Aoki:2006we}. In such a case, the partially-deconfined phase is thermodynamically stable; see Ref.~\cite{Hanada:2019kue} for an explicit analysis at finite volume and weak coupling. This opens up the interesting possibility that the so-called QCD crossover region is the partially-deconfined phase. 

In the past, partial deconfinement was studied based on the $N$-dependence of the free energy and entropy. We expect that another characterization of confinement --- that we cannot separate quarks without forming a color singlet --- is valid for the confined sector in the partially-deconfined phase as well, but there has so far been no direct confirmation.
In this paper, we make solid progress regarding this point. 

As a concrete example, we consider pure Yang-Mills theory, which exhibits a first-order confinement/deconfinement transition. There is a partially-deconfined saddle separating two minima of the free energy (completely-confined phase and completely-deconfined phase); see Fig.~\ref{fig:M,E,P-vs-T}. Although this saddle is thermodynamically unstable, it is connected to the stable saddle in QCD as $\frac{N_f}{N}$ or quark mass are varied\footnote{
Pure Yang-Mills can be regarded as the heavy-quark-mass limit or $\frac{N_f}{N}\to 0$ limit.
} and we expect qualitative similarity between stable and unstable saddles. 
In the string-condensation picture, only the SU($M$) chromo-electric strings are condensed. If we take a probe quark and antiquark from the deconfined sector, then they can interact with condensed strings and should not have the confinement potential. On the other hand, if we take probes from the confined sector, we expect the linear confinement potential, because there are no condensed strings in this sector. 
\begin{figure}[htbp]
  \begin{center}
  \includegraphics[width=0.32\textwidth]{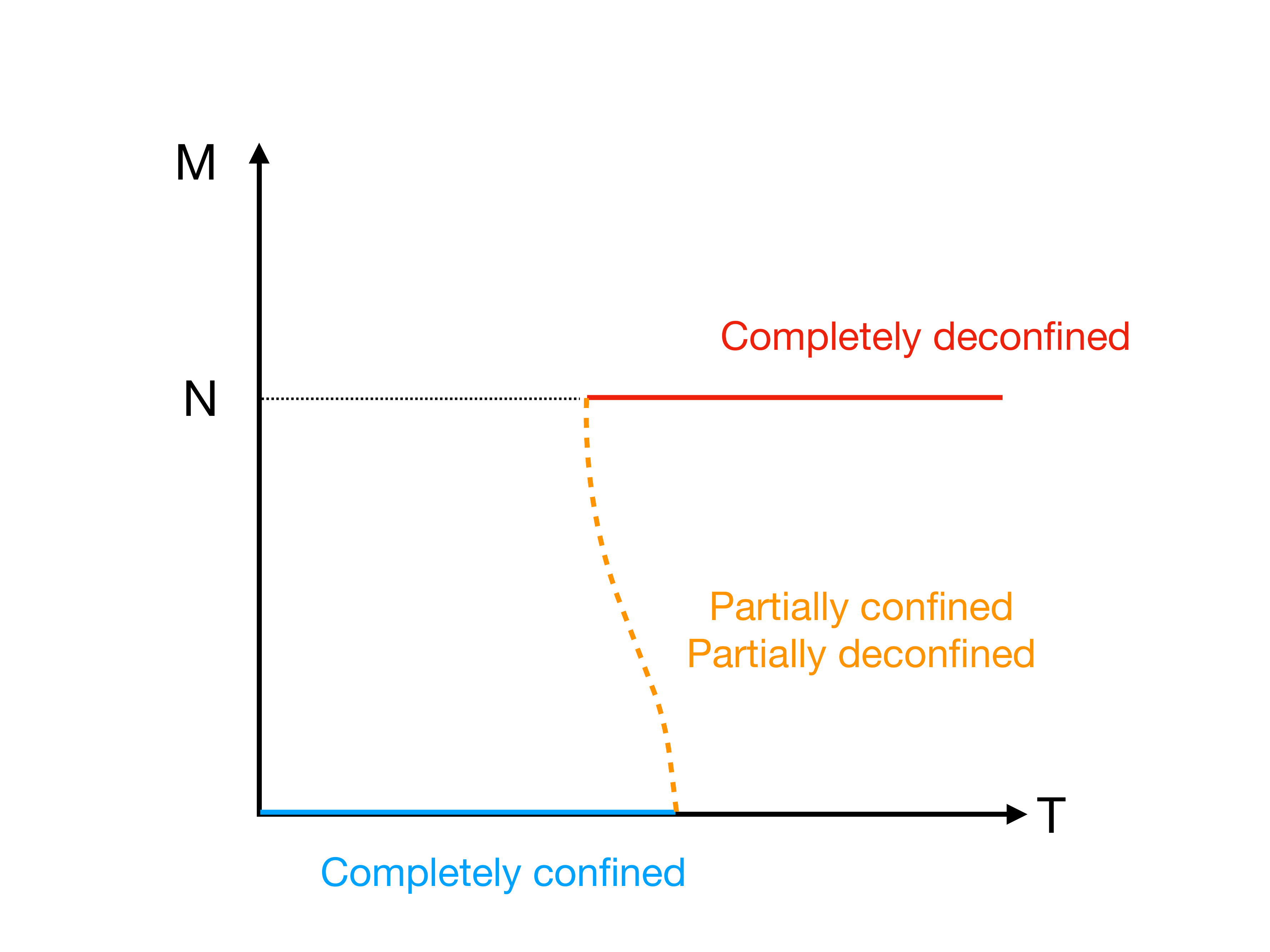}
  \includegraphics[width=0.32\textwidth]{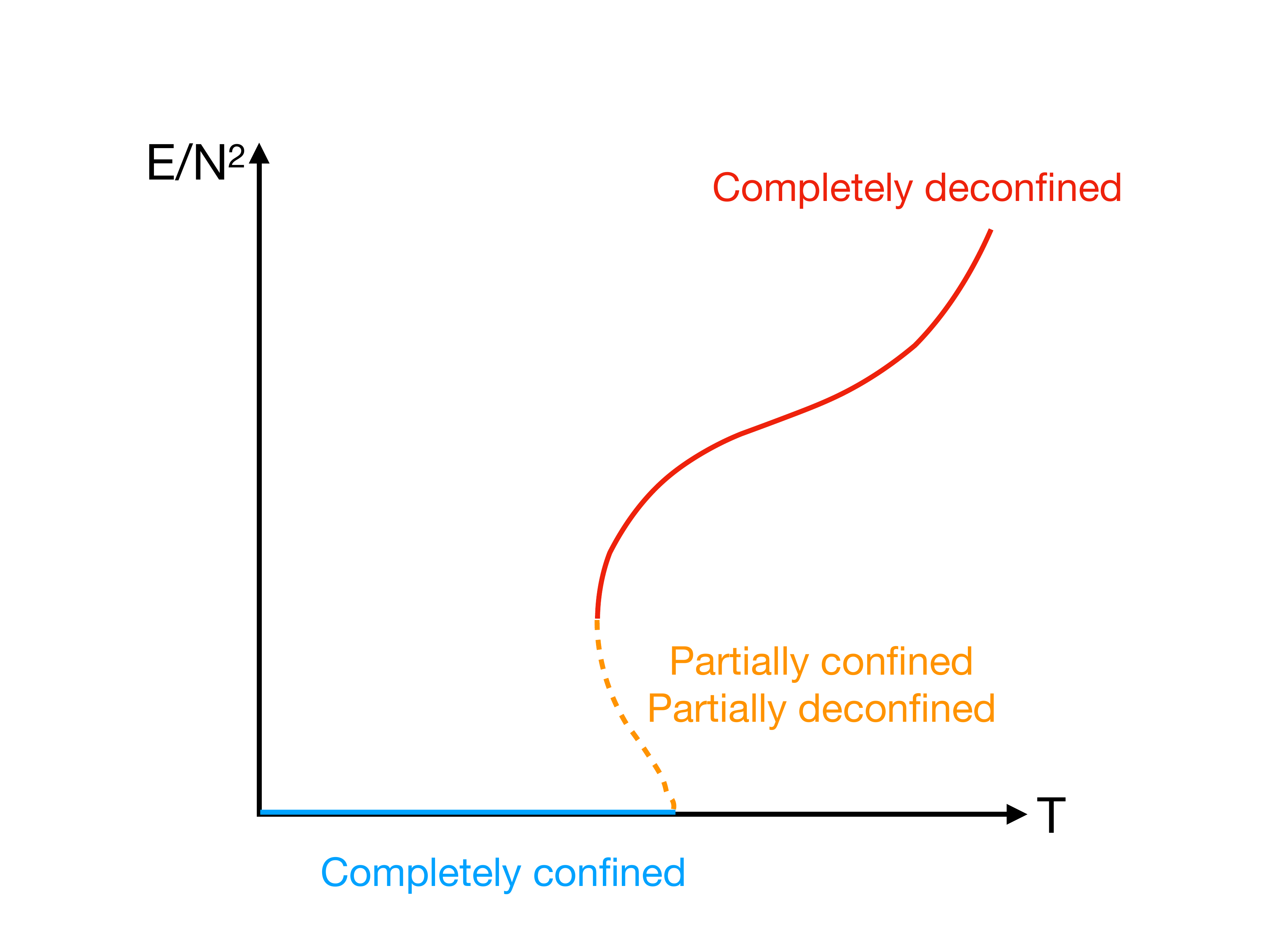}
  \includegraphics[width=0.32\textwidth]{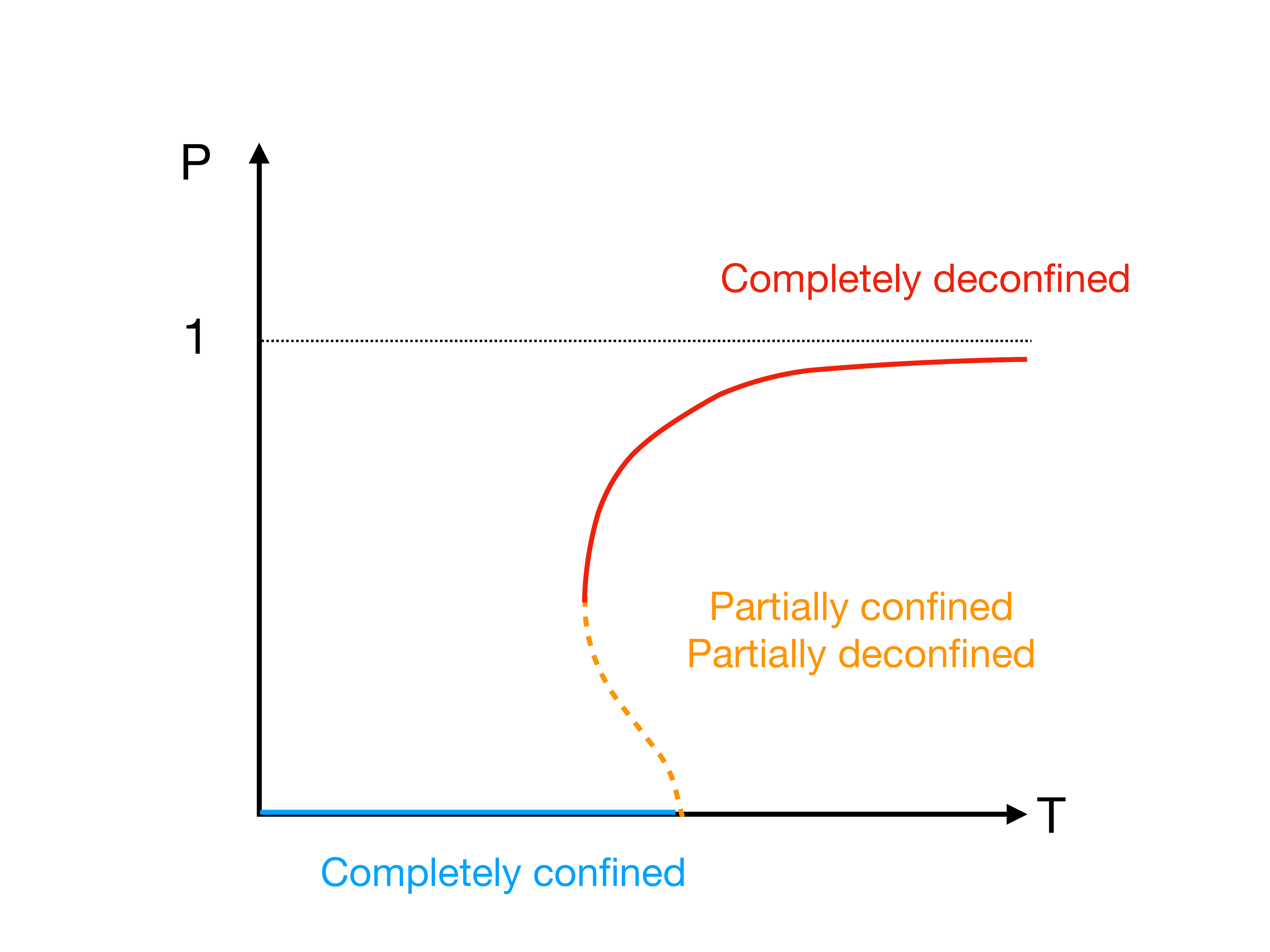}
  \end{center}
  \caption{Sketches of the temperature dependence of $M$, $\frac{E}{N^2}$ and $P$ for the pure Yang-Mills theory.
  Blue, orange and red lines are completely-confined phase, partially-deconfined saddle (equivalently, partially-confined saddle) and completely-deconfined phase. The partially-deconfined saddle is the maximum of free energy at fixed temperature.  
  See Ref.~\cite{Hanada:2018zxn} for other examples including a thermodynamically stable partially-deconfined phase.
  }\label{fig:M,E,P-vs-T}
\end{figure}

We will test this statement by using a gauge fixing that separates confined and deconfined sectors in a controlled manner. 
Specifically, we will consider the strong-coupling lattice gauge theory at finite temperature and study the two-point function of the Polyakov loop.
There are several reasons we consider this theory. First of all, partial deconfinement is a generic property of gauge theories with confinement/deconfinement transition including this model. Secondly, we can use the Eguchi-Kawai equivalence (large-$N$ volume reduction) which makes our numerical simulation tractable. Furthermore, 
for this particular setup, we can make a few analytic predictions assuming the formation of the flux tube in the confined sector, and we can confirm the prediction numerically. 
Because such analytic predictions are not available in the continuum limit, we can perform a stronger test by focusing on the strong-coupling lattice gauge theory.
The Polyakov loop ${\rm Tr}{\cal P}(\vec{x})$ is the trace of the holonomy ${\cal P}(\vec{x})$ along the temporal circle defined by 
\begin{align}
{\cal P}(\vec{x}) =\left[{\rm P}\exp\left(
i\oint dt A_t(t,\vec{x}) 
\right)\right]
\end{align}
where ${\rm P}$ stands for the path ordering and ${\vec{x}}$ is the spatial point. 
The two-point function decays exponentially in the completely-confined phase as 
\begin{align}
\left\langle
{\rm Tr}{\cal P}(\vec{x}) 
\cdot
{\rm Tr}{\cal P}(\vec{y}) 
\right\rangle
\sim
e^{-\beta\sigma|\vec{x}-\vec{y}|}, 
\end{align}
where $\beta=T^{-1}$ is the inverse temperature and $\sigma$ is the string tension. 
We will perform a gauge fixing such that colors split into the confined and deconfined sectors, respectively, and define the Polyakov loops in those sectors,  ${\rm Tr}{\cal P}_{{\rm con}}(\vec{x})$ and ${\rm Tr}{\cal P}_{{\rm dec}}(\vec{x})$. Then, we show that ${\rm Tr}{\cal P}_{{\rm con}}(\vec{x})$ exhibits the exponential decay   
\begin{align}
\left\langle
{\rm Tr}{\cal P}_{{\rm con}}(\vec{x})
\cdot
{\rm Tr}{\cal P}_{{\rm con}}(\vec{y}) 
\right\rangle
\sim
e^{-\beta\sigma|\vec{x}-\vec{y}|} 
\label{conjecture:con-con}
\end{align}
with the same string tension $\sigma$.
We also show that 
\begin{align}
\left\langle
{\rm Tr}{\cal P}_{{\rm con}}(\vec{x})
\cdot
{\rm Tr}{\cal P}_{{\rm dec}}(\vec{y}) 
\right\rangle
\sim
e^{-\beta\sigma|\vec{x}-\vec{y}|}.   
\label{conjecture:con-dec}
\end{align}
We interpret the relations \eqref{conjecture:con-con} and \eqref{conjecture:con-dec} as consequences of the formation of flux tubes and linear confinement in the confined sector. For the strong-coupling lattice gauge theory, we confirm these relations quantitatively, including the concrete value of the string tension $\sigma$ and the overall constant factor. 

This paper is organized as follows. In Sec.~\ref{sec:Theoretical_background}, we use the operator formalism to explain partial deconfinement. In Sec.~\ref{sec:define-lattice-model}, we define the lattice action and set the simulation strategy. In Sec.~\ref{sec:determining-M}, we determine the size of the deconfined sector $M$ at each simulation point by using the relation between $M$ and the phase distribution of the Polyakov loop~\cite{Hanada:2020uvt}. In Sec.~\ref{sec:flux-tubes}, we provide evidence of the linear confinement potential in the confined sector of the partially-deconfined phase. Theoretical considerations are provided in Sec.~\ref{sec:flux-tube-theory}. Then, numerical evidence is given in Sec.~\ref{sec:flux-tube-simulation}. We define the Wilson loop equivalent to the two-point function of the Polyakov loop, and study it by using the Eguchi-Kawai equivalence in the large-$N$ limit.

\section{Theoretical background}\label{sec:Theoretical_background}
\hspace{0.51cm}

\subsubsection*{Hamiltonian formulation}
\hspace{0.51cm}
The meaning of partial deconfinement is clearer if we look directly at quantum states by working in the operator formalism.  In this section, we describe the partially-deconfined phase by using the Hamiltonian formulation by Kogut and Susskind~\cite{Kogut:1974ag}. 
We consider the (3+1)-d Yang-Mills theory with U($N$) gauge group. The adaptation of this theory to the path-integral formalism, as actually used in our simulations, is described in Sec.~\ref{sec:define-lattice-model}. The purpose of this section is to explain aspects of partial deconfinement using the string condensation picture~\cite{Polyakov:1978vu,Susskind:1979up}.

The Kogut-Susskind Hamiltonian consists of the electric and magnetic terms, 
\begin{eqnarray}
\hat{H}
=
\hat{H}_{\rm E}
+
\hat{H}_{\rm B}.  
\end{eqnarray}
The electric part of the Hamiltonian becomes
\begin{eqnarray}
\hat{H}_{\rm E}
=
\frac{1}{2}
\sum_{\vec{n}}\sum_{\mu=1}^3\sum_{\alpha=1}^{N^2}\left(\hat{E}_{\mu,\vec{n}}^{\alpha}\right)^2. 
\end{eqnarray}
The magnetic term (plaquette) is dropped in the strong coupling limit~\cite{Susskind:1979up}. 
The commutation relations are 
\begin{eqnarray}
\left[
\hat{U},
\hat{U}
\right]
=
\left[
\hat{U},
\hat{U}^\dagger
\right]
=
\left[
\hat{U}^\dagger,
\hat{U}^\dagger
\right]
=
0.   
\end{eqnarray}
\begin{eqnarray}
\left[
\hat{E}_{\mu,\vec{n}}^\alpha,
\hat{U}_{\nu,\vec{n}'}
\right]
=
\delta_{\mu\nu}\delta_{\vec{n}\vec{n}'}\tau_\alpha\hat{U}_{\nu,\vec{n}'}, 
\qquad
\left[
\hat{E}_{\mu,\vec{n}}^\alpha,
\hat{U}^\dagger_{\nu,\vec{n}'}
\right]
=
-\delta_{\mu\nu}\delta_{\vec{n}\vec{n}'}\hat{U}_{\nu,\vec{n}'}^\dagger\tau_\alpha,   
\end{eqnarray}
\begin{eqnarray}
\left[
\hat{E}_{\mu,\vec{n}}^\alpha,
\hat{E}_{\nu,\vec{n}'}^\beta
\right]
=
-if^{\alpha\beta\gamma}\delta_{\mu\nu}\delta_{\vec{n}\vec{n}'}\hat{E}^\gamma_{\nu,\vec{n}'},  
\end{eqnarray}
where $f^{\alpha\beta\gamma}$ is the structure constant of U($N$), and 
\begin{eqnarray}
\sum_{\alpha=1}^{N^2}\tau^{\alpha}_{pq}\tau^{\alpha}_{rs}
=
\frac{\delta_{ps}\delta_{qr}}{N}, 
\qquad
\sum_{\alpha=1}^{N^2}\left(\tau^{\alpha}\tau^\alpha\right)_{pq}=\delta_{pq}. 
\end{eqnarray}

The ground state of this string-coupling lattice gauge theory Hamiltonian $|{\rm g.s.}\rangle$ satisfies $\hat{E}_{\mu,\vec{n}}^\alpha|{\rm g.s.}\rangle = 0$ for any $\alpha,\mu$ and $\vec{n}$. Hence, let us use the notation $|E=0\rangle$ to denote the ground state.

The operator $\hat{U}_{\mu,\vec{n}}$ is interpreted as the coordinate of the group manifold U($N$) for the link variable on the site $\vec{n}$ in the $\mu$-direction. The operators $\hat{U}$ and $\hat{E}$ are defined on the extended Hilbert space $\mathcal{H}_{\rm ext}$ that contains gauge non-singlet modes. 
A convenient basis of $\mathcal{H}_{\rm ext}$ is the coordinate representation, 
\begin{eqnarray}
{\cal H}_{\rm ext}
=
\otimes_{\mu,\vec{n}}{\cal H}_{\mu,\vec{n}}
\sim
\otimes_{\mu,\vec{n}}
\left(
\oplus_{g\in{\rm U}(N)}
|g\rangle_{\mu,\vec{n}}
\right), 
\end{eqnarray}
 where
 \begin{eqnarray}
\hat{U}_{\mu,\vec{n}}
|g\rangle_{\mu,\vec{n}}
=
g|g\rangle_{\mu,\vec{n}}
\qquad 
g\in{\rm U}(N). 
\end{eqnarray}
More precisely, we will consider only the Hilbert space of square-integrable wave functions on U($N$): ${\cal H}_{\mu,\vec{n}}=L^2({\rm U}(N))$, where $L^2({\rm U}(N))$ is the set of square-integrable functions from ${\rm U}(N)$ to ${\mathbb C}$. The ground state is the constant mode, 
\begin{align}
|E=0\rangle
=
\otimes_{\mu,\vec{n}}
|E=0\rangle_{\mu,\vec{n}},
\qquad
|E=0\rangle_{\mu,\vec{n}}
=
\frac{1}{\sqrt{{\rm vol U}(N)}}\int_{{\rm U}(N)} dg
|g\rangle_{\mu,\vec{n}}. 
\end{align}
Namely, the wave function $\langle g|E=0\rangle$ is constant. 

Let $G=\prod_{\vec{n}}[{\rm U}(N)]_{\vec{n}}$ be the group of all local gauge transformations. 
Gauge transformation by $\hat{\Omega}=\otimes_{\vec{n}}\hat{\Omega}_{\vec{n}}$ is defined by 
\begin{align}
\hat{\Omega}
\left(
\otimes_{\mu,\vec{n}}
|g\rangle_{\mu,\vec{n}}
\right)
=
\otimes_{\mu,\vec{n}}
\left(
\hat{\Omega}_{\vec{n}}|g\rangle_{\mu,\vec{n}}
\right)
=
\otimes_{\mu,\vec{n}}
|\Omega_{\vec{n}}g\Omega_{\vec{n}}^{-1}\rangle_{\mu,\vec{n}}. 
\end{align}Note that the ground state is gauge-invariant:
\begin{align}
\hat{\Omega}|E=0\rangle= |E=0\rangle. 
\end{align}
We can define the projection operator to the gauge-invariant Hilbert space $\mathcal{H}_{\rm inv}$ as
\begin{align}
\hat{\pi}
=
\frac{1}{{\rm Vol}G}\int_Gd\Omega \hat{\Omega}, 
\label{def-gauge-projector}
\end{align}
where the integral is taken by using the Haar measure. By using this projection operator, canonical partition function at temperature $T$ can be written in two ways as
\begin{align}
Z(T)
=
{\rm Tr}_{\mathcal{H}_{\rm inv}}\left(e^{-\hat{H}/T}\right)
=
{\rm Tr}_{\mathcal{H}_{\rm ext}}\left(\hat{\pi}e^{-\hat{H}/T}\right). 
\label{canonical_partition_function}
\end{align}
The latter expression is directly related to the path-integral formalism with gauge field $A_t$, as shown in Appendix~\ref{appendix:gauge_fixing}. The U($N$)-element $\Omega$ corresponds to the Polyakov loop in the path-integral formalism~\cite{Hanada:2020uvt}. The insertion of $\hat{\pi}$ means we should identify the states connected by gauge transformation, as we identify field configurations connected by gauge transformation in the classical theory. In other words, all states on the gauge orbit are physically equivalent. This makes the meaning of `gauge fixing' in the extended Hilbert space clear: it selects a point from the gauge orbit, just as in the path-integral formalism or even in the classical theory. In the SU($M$)-deconfined phase, we can take a gauge such that the upper-left $M\times M$ block is deconfined, as in Fig.~\ref{fig:Partial-deconfinement}.

\subsubsection*{Strings and Interactions}
\hspace{0.51cm}

A closed string is created by the Wilson loop $\hat{W}^{\rm (closed)}_C={\rm Tr}(\hat{U}_{\mu,\hat{n}}\hat{U}_{\nu,\hat{n}+\hat{\mu}}\cdots)$, where the product of $\hat{U}$'s in the trace is taken along the closed path $C$. 
An open string is created by the open Wilson line along an open path $C'$, denoted by $\hat{W}^{\rm (open)}_{C'}$, which is a product of $\hat{U}$'s without trace. The closed string is gauge invariant. 

Suppose that there is a closed loop without self-intersection (i.e. no link appears twice). Then, the state $|W^{\rm (closed)}_C\rangle\equiv\hat{W}^{\rm (closed)}_C|0\rangle$ is an energy eigenstate, and the energy is $\frac{L}{2}$, where $L$ is the length of the loop (i.e., the number of links consisting the loop). This can be seen as follows. Firstly, note that
\begin{align}
\sum_\alpha\left(\hat{E}_{\mu,\vec{n}}^{\alpha}\right)^2\hat{W}_C^{\rm (closed)}|E=0\rangle
=
\sum_\alpha\left[
\hat{E}_{\mu,\vec{n}}^{\alpha},\left[\hat{E}_{\mu,\vec{n}}^{\alpha},\hat{W}_C^{\rm (closed)}\right]
\right]|E=0\rangle. 
\end{align}
$\hat{E}_{\mu,\vec{n}}^{\alpha}$ commutes with $\hat{W}^{\rm (closed)}_C$ unless the latter contains $\hat{U}_{\mu,\vec{n}}$ or $\hat{U}^\dagger_{\mu,\vec{n}}$. When $\hat{U}_{\mu,\vec{n}}$ is contained, 
\begin{align}
\sum_\alpha\left[
\hat{E}_{\mu,\vec{n}}^{\alpha},\left[\hat{E}_{\mu,\vec{n}}^{\alpha},\hat{W}_C^{\rm (closed)}\right]
\right]|E=0\rangle
&=
\sum_\alpha{\rm Tr}\left(\left[
\hat{E}_{\mu,\vec{n}}^{\alpha},\left[\hat{E}_{\mu,\vec{n}}^{\alpha},\hat{U}_{\mu,\hat{n}}\right]\right]\hat{U}_{\nu,\hat{n}+\hat{\mu}}\cdots\right)|E=0\rangle
\nonumber\\
&=
\sum_\alpha{\rm Tr}\left(\tau^\alpha\tau^\alpha\hat{U}_{\mu,\hat{n}}\hat{U}_{\nu,\hat{n}+\hat{\mu}}\cdots\right)|E=0\rangle
\nonumber\\
&=
{\rm Tr}\left(\hat{U}_{\mu,\hat{n}}\hat{U}_{\nu,\hat{n}+\hat{\mu}}\cdots\right)|E=0\rangle
\nonumber\\
&=
\hat{W}_C^{\rm (closed)}|E=0\rangle. 
\end{align}
The same holds when $\hat{U}^\dagger_{\mu,\hat{n}}$ is contained. 
Therefore, $\hat{H}_{\rm E}|W^{\rm (closed)}_C\rangle=\frac{L}{2}|W^{\rm (closed)}_C\rangle$. The same holds for any multi-string states, including closed or open strings, as long as there is no intersection. 

Next, suppose a link  $\hat{U}_{\mu,\vec{n}}$ appears twice in the loop, while other links appear only once. Let us write such a loop as
${\rm Tr}(\hat{V}_1\hat{U}_{\mu,\vec{n}}\hat{V}_2\hat{U}_{\mu,\vec{n}})$. Then, 
\begin{align}
\lefteqn{
\sum_\alpha\left[
\hat{E}_{\mu,\vec{n}}^{\alpha},\left[\hat{E}_{\mu,\vec{n}}^{\alpha},{\rm Tr}(\hat{V}_1\hat{U}_{\mu,\vec{n}}\hat{V}_2\hat{U}_{\mu,\vec{n}})\right]
\right]|E=0\rangle
}
\nonumber\\
&=
2{\rm Tr}(\hat{V}_1\hat{U}_{\mu,\vec{n}}\hat{V}_2\hat{U}_{\mu,\vec{n}})|E=0\rangle
+
2\sum_\alpha{\rm Tr}(\hat{V}_1[\hat{E}_{\mu,\vec{n}}^{\alpha},\hat{U}_{\mu,\vec{n}}]\hat{V}_2[\hat{E}_{\mu,\vec{n}}^{\alpha},\hat{U}_{\mu,\vec{n}}])|E=0\rangle
\nonumber\\
&=
2{\rm Tr}(\hat{V}_1\hat{U}_{\mu,\vec{n}}\hat{V}_2\hat{U}_{\mu,\vec{n}})|E=0\rangle
+
2\sum_\alpha
{\rm Tr}(\hat{V}_1\tau^\alpha\hat{U}_{\mu,\vec{n}}\hat{V}_2\tau^\alpha\hat{U}_{\mu,\vec{n}})|E=0\rangle
\nonumber\\
&=
2{\rm Tr}(\hat{V}_1\hat{U}_{\mu,\vec{n}}\hat{V}_2\hat{U}_{\mu,\vec{n}})|E=0\rangle
+
\frac{2}{N}
{\rm Tr}(\hat{V}_1\hat{U}_{\mu,\vec{n}}){\rm Tr}(\hat{V}_2\hat{U}_{\mu,\vec{n}})|E=0\rangle. 
\end{align}
Therefore, 
\begin{align}
&
\hat{H}_E\left({\rm Tr}(\hat{V}_1\hat{U}_{\mu,\vec{n}}\hat{V}_2\hat{U}_{\mu,\vec{n}})|E=0\rangle\right)
\nonumber\\
\qquad
&=
\frac{L}{2}{\rm Tr}(\hat{V}_1\hat{U}_{\mu,\vec{n}}\hat{V}_2\hat{U}_{\mu,\vec{n}})|E=0\rangle
+
\frac{1}{N}
{\rm Tr}(\hat{V}_1\hat{U}_{\mu,\vec{n}}){\rm Tr}(\hat{V}_2\hat{U}_{\mu,\vec{n}})|E=0\rangle. 
\end{align}
The second term can be understood as the splitting of a string into two strings. 
In the same manner, we can show that two strings can be joined to form one string at an intersection, 
\begin{align}
&
\hat{H}_E\left({\rm Tr}(\hat{V}_1\hat{U}_{\mu,\vec{n}}){\rm Tr}(\hat{V}_2\hat{U}_{\mu,\vec{n}})|E=0\rangle\right)
\nonumber\\
\qquad
&=
\frac{L}{2}{\rm Tr}(\hat{V}_1\hat{U}_{\mu,\vec{n}}){\rm Tr}(\hat{V}_2\hat{U}_{\mu,\vec{n}})|E=0\rangle
+
\frac{1}{N}{\rm Tr}(\hat{V}_1\hat{U}_{\mu,\vec{n}}\hat{V}_2\hat{U}_{\mu,\vec{n}})|E=0\rangle.  
\end{align}
In general, such splitting and joining can take place at any intersection. 
\subsubsection*{Confinement}
\hspace{0.51cm}
Let us consider low-energy gauge-invariant states consisting of a small number of closed strings with total length $L\sim N^0$. Then, there is at most order $N^0$ number of intersections. The interaction (splitting or joining) at each intersection is suppressed by $\frac{1}{N}$, and hence the interaction is negligible at large $N$. The energy of the system is simply $\frac{L}{2}$. Such states are in the confined phase. 

If we introduce a probe quark-antiquark pair connected by the open Wilson line, the energy increases linearly as $E_{q\bar{q}}=\frac{L}{2}$. This leads to the linear confinement potential.

\subsubsection*{Deconfinement and string condensation}\label{sec:string_condensation}
\hspace{0.51cm}
In the deconfined phase~\cite{Polyakov:1978vu,Susskind:1979up}, long strings with length of order $N^2$ condense. There are many intersections, and thus the $1/N$-suppressed interactions accompanying each intersection pile up and become non-negligible as a whole. Intuitively, if we introduce a short open string, it interacts with a condensed long string and forms a long open string, which allows us to separate quark and antiquark without making the string longer.

\subsubsection*{Partial deconfinement}
\hspace{0.51cm}
Let $\hat{U}_{\rm dec}$ be the SU($M$)-subsector. The SU($M$)-deconfined states can be constructed by acting with long traces of $\hat{U}_{\rm dec}$'s on $|E=0\rangle$~\cite{Hanada:2019czd}.\footnote{See Refs.~~\cite{Hanada:2021ipb,Hanada:2021swb,Gautam:2022akq} for more general characterization.}  By using the Wilson loops restricted to the SU($M$)-subsector 
\begin{align}
\hat{W}_{{\rm dec},C}={\rm Tr}(\hat{U}_{{\rm dec};\mu,\vec{n}}\hat{U}_{{\rm dec};\nu,\vec{n}+\hat{\mu}}\cdots),
\end{align}
we can construct multi-string states 
\begin{align}
\hat{W}_{{\rm dec},C}\hat{W}_{{\rm dec},C'}\cdots|E=0\rangle,
\end{align}
and then we can take a linear combination of such states.  Such states are SU($M$)-invariant, but not SU($N$)-invariant. If we want to get an SU($N$)-invariant state, we can act with the projector $\hat{\pi}$. 

The reason that such a phase is favored is understood from \eqref{canonical_partition_function}~\cite{Hanada:2020uvt}. We write the projection operator $\hat{\pi}$ explicitly as 
\begin{align}
Z(T)
=
\frac{1}{{\rm Vol}(G)}\int_Gd\Omega 
{\rm Tr}_{\mathcal{H}_{\rm ext}}\left(\hat{\Omega}e^{-\hat{H}/T}\right). 
\label{partition_function_2}
\end{align}
From this expression, we can see that each energy eigenstate\footnote{
Here, $E$ is the energy and not the electric field. For the ground state, we used the notation $|E=0\rangle$ by meaning that the electric field $E$ is zero. Coincidentally, the energy of the ground state is also zero. 
} $|E\rangle\in\mathcal{H}_{\rm ext}$ has a contribution 
\begin{align}
\frac{1}{{\rm Vol}(G)}\int_Gd\Omega e^{-E/T}
\langle E|\hat{\Omega}|E\rangle. 
\end{align}
If $|E\rangle$ is an SU($M$)-deconfined state, we have $\hat{\Omega}|E\rangle=|E\rangle$ for $\Omega\in{\rm U}(N-M)$, such that $\Omega$ does not act on the SU($M$)-subsector. Hence, there is an enhancement factor ${\rm Vol}({\rm U}(N-M))\sim e^{(N-M)^2}$. This factor makes smaller $M$ more favorable. At each fixed energy, this factor and other entropy factors compete and a certain value of $M$ between $0$ and $N$ is determined as the most favorable value that maximizes the entropy. This is essentially the same mechanism as the Bose-Einstein condensation~\cite{Hanada:2020uvt}. 

Let $\hat{U}_{\rm con}$ be the other $N^2-M^2$ elements. Operators that consist of a number $L \sim O(N^0)$ of $\hat{U}_{\rm con}$'s increase the energy by $\frac{L}{2}$. They cannot be joined to the condensed strings. The energy eigenstate remains an energy eigenstate. Naturally, we expect that the color flux in the confined sector forms a flux tube and exhibits the linear potential $E_{q\bar{q}}=\frac{L}{2}$ while the deconfined sector does not. We will discuss this later, together with numerical results.

\section{Methods} \label{sec:define-lattice-model}
\hspace{0.51cm}
We will perform our simulations on a lattice-discretized model of U($N$) Yang-Mills theory. Our simulations will involve constraining specific quantities so that we can remain in the partially-deconfined phase, and also so that we can separate the confined and deconfined sectors.
Specifically, we use the Eguchi-Kawai model, which is equivalent to U($N$) Yang-Mills theory in the large-$N$ limit. The introduction of the constraints amounts to the study of the microcanonical ensemble rather than canonical ensemble, plus gauge fixing. Simulations of lattice gauge theories in the microcanonical ensemble (where the energy is fixed, instead of the temperature)~\cite{Langfeld:2012ah} play a very important role in the numerical study of phase transitions~\cite{Aarts:2023vsf}.
\subsection{The lattice regularization}

\subsubsection{Path-integral formulation}
\hspace{0.51cm}

To set up our lattice model, we will begin with the path integral formulation. We consider Yang-Mills theory on a $d$-dimensional spatial lattice with continuous time $t$. We will focus on $d=3$.
Let $\vec{n}$ be the spatial points labeled by $d$ integers, and $U_{\mu,\vec{n}}(t)$ be the U($N$) link variable on the link connecting $\vec{n}$ and $\vec{n}+\hat{\mu}$. Here, $\hat{\mu}$ is the unit vector along the $\mu$-direction. 
The gauge transformation is defined by 
\begin{align}
U_{\mu,\vec{n}}(t)
\to
\Lambda_{\vec{n}}^{-1}(t) U_{\mu,\vec{n}}(t)\Lambda_{\vec{n}+\hat{\mu}}(t). 
\end{align}  
Here $\Lambda_{\vec{n}}(t)$ and $\Lambda_{\vec{n}+\hat{\mu}}(t)$ are $N\times N$ unitary matrices that describe the local gauge transformation at points $\vec{n}$ and $\vec{n}+\hat{\mu}$, respectively. 
We introduce a gauge field $A_{\vec{n}}(t)$ that transforms as 
\begin{align}
A_{\vec{n}}(t)
\to
\Lambda_{\vec{n}}^{-1}(t) A_{\vec{n}}(t)\Lambda_{\vec{n}}(t)
+
i\Lambda_{\vec{n}}^{-1}(t)\partial_t\Lambda_{\vec{n}}(t). 
\end{align}  
The covariant derivative $D_tU_{\mu,\vec{n}}$, which transform as $D_tU_{\mu,\vec{n}}\to\Lambda_{\vec{n}}^{-1}(D_tU_{\mu,\vec{n}})\Lambda_{\vec{n}+\hat{\mu}}$, is defined by 
\begin{align}
D_tU_{\mu,\vec{n}}
=
\partial_t U_{\mu,\vec{n}}
-
iA_{\vec{n}}U_{\mu,\vec{n}}
+
iU_{\mu,\vec{n}}A_{\vec{n}+\hat{\mu}}. 
\end{align}
We will work in the strong coupling limit, by which we mean that the action contains only the electric term and not the magnetic term.
Hence the Euclidean action at temperature $T=\beta^{-1}$ is\footnote{
We use the same symbol $S$ for the action and entropy, assuming the risk of confusion is low.
} 
\begin{align}
S
=
\frac{1}{2g^2}\sum_{\vec{n}}\int_0^\beta dt
{\rm Tr}\left(
(D_tU_{\mu,\vec{n}}(t))(D_tU_{\mu,\vec{n}}(t))^\dagger
\right). 
\end{align}

The operator $\hat{\Omega}$ in \eqref{partition_function_2} corresponds to the Polyakov loop in the path-integral formalism~\cite{Hanada:2020uvt} (see Appendix~\ref{appendix:gauge_fixing}). In the large-$N$ limit, the distribution of the phases of the Polyakov loop is related to the symmetry of the typical quantum states dominating the partition function. 
The size of the deconfined sector $M$ can be determined from the distribution of the phases~\cite{Hanada:2020uvt}, as explained in Sec.~\ref{sec:determining-M}.

\subsubsection{Eguchi-Kawai reduction} \label{sec:EK}
\hspace{0.51cm}
In the large-$N$ limit, some features of the strong-coupling lattice gauge theory do not depend on the lattice size. Therefore, we can use the single-site model, which is called the Eguchi-Kawai model, to learn about the infinite-volume theory. 
We use conventions close to those in Ref.~\cite{Hanada:2014noa}.

In the Eguchi-Kawai model, we have only one spatial point, so we drop $\vec{n}$ from the expressions for the strong-coupling lattice gauge theory. 
We will employ the gauge field $A$ and unitary link variables $U_\mu$, both of which are function of (Euclidean) time $t$.
The gauge transformation is defined by 
\begin{align}
U_\mu
\to
\Lambda^{-1} U_\mu\Lambda
\end{align}  
and
\begin{align}
A
\to
\Lambda^{-1} A\Lambda
+
i\Lambda^{-1}\partial_t\Lambda. 
\end{align}  
The covariant derivative $D_tU_\mu$, which transform as $D_tU_\mu\to \Lambda^{-1}(D_tU_\mu)\Lambda$, is defined by 
\begin{align}
D_tU_\mu
=
\partial_t U_\mu
-
i[A,U_\mu]. 
\end{align}
The Euclidean action is 
\begin{align}
S
=
\frac{1}{2g^2}\int_0^\beta dt{\rm Tr}(D_tU_\mu)^2. 
\end{align}

This action is invariant under the global (i.e. $t$-independent) U(1)$^d$ center symmetry generated by 
\begin{align}
U_\mu
\to
e^{i\theta_\mu}U_\mu. 
\end{align}
As long as this symmetry is unbroken, the Eguchi-Kawai model and infinite-volume lattice are equivalent, in the sense that various properly-normalized quantities agree. 
This is the so-called Eguchi-Kawai equivalence \cite{Eguchi:1982nm}.\footnote{
In the original work by Eguchi and Kawai, all dimensions including time are reduced to a point. In most references, `Eguchi-Kawai model' and `Eguchi-Kawai equivalence' are used for the original setup. 
} 

In the strong coupling limit, the center symmetry is not spontaneously broken. 
To confirm this in our simulations as a sanity check, we calculated the Wilson loop wrapped on the spatial direction, 
\begin{align}
\frac{1}{Nn_td}\sum_{\mu,d}|{\rm Tr} U_{\mu,t}|. 
\end{align}
If the center symmetry is not broken, this quantity should be zero up to $1/N$-suppressed terms. Our simulations affirmed this.

\subsubsection{Gauge fixing} \label{sec:gauge-fixing-static-daigonal}
\hspace{0.51cm}
To make the separation to confined and deconfined phases easier, we take the static diagonal gauge
(used in Ref.~\cite{Watanabe:2020ufk} for the same purpose), 
\begin{align}
A=\frac{1}{\beta}\cdot{\rm diag}(\alpha_1,\cdots,\alpha_N),
\qquad
-\pi<\alpha_i\le\pi. 
\end{align} 
Associated with this gauge fixing, we add the Faddeev-Popov term  
\begin{align}
S_{\rm F.P.}
= 
-
\sum_{i<j}2\log\left|\sin\left(\frac{\alpha_i-\alpha_j}{2}\right)\right|
\label{eq:Faddeev-Popov}
\end{align}
to the action. 
This fixes SU($N$) down to S$_N$. In Sec.~\ref{sec:constrained_simulation}, we explain how the confined and deconfined phases can be separated by fixing the residual S$_N$ symmetry appropriately. 
\subsubsection{Lattice regularization for the time dimension}
\hspace{0.51cm}
Finally, we place this theory on the lattice in the time direction, too, with the below action:
\begin{align}
S
=
\frac{N}{2a}
\sum_{\mu=1}^{d}\sum_{t=1}^{n_t}
{\rm Tr}\left(
\textbf{1}_N
-
U_{\mu,t}VU_{\mu,t+1}^\dagger V^\dagger 
\right)
+
{\rm h.c.}
+
S_{\rm F.P.}, 
\label{EK-action-lattice}
\end{align}
where $V={\rm diag}(e^{i\alpha_1/n_t},\cdots,e^{i\alpha_N/n_t})$. 
Here $a$ is the lattice spacing, and $\beta=an_t$ is the inverse temperature, $\beta=T^{-1}$. 
This is the lattice action we use in our simulations.
We will focus on $d=3$.
(We will make one more alteration by adding terms that constrain the Polyakov loop; see Sec.~\ref{sec:constrained_simulation}.)
\subsection{Polyakov loop and constrained simulations}\label{sec:constrained_simulation}
\hspace{0.51cm}
The Polyakov loop under the gauge fixing described above is $P=\frac{1}{N}{\rm Tr}{\cal P}$, where

\begin{align}
{\cal P}
=
{\rm diag}(e^{i\alpha_1},\cdots,e^{i\alpha_N}).
\end{align}
This is the quantity we will measure in the standard (unconstrained) simulation. In addition, we will run two kinds of constrained simulation that make use of this definition of the Polyakov loop.

\subsubsection{The microcanonical ensemble and constrained simulation of the first kind}
\hspace{0.51cm}
Via the Euclidean path integral, we can study the thermodynamic properties of a theory in the canonical ensemble, in which temperature $T$ is the controlling parameter. If the confinement/deconfinement phase transition is of first order, however, it is more convenient to study the microcanonical ensemble, in which the energy $E$ is the controlling parameter~\cite{Mason:2022aka}.

In canonical thermodynamics, temperature is fixed and free energy is minimized. In microcanonical thermodynamics, energy is fixed and entropy is maximized. Usually, the partially-deconfined phase is unstable in the canonical ensemble, but can be stable in the microcanonical ensemble.
If the spatial volume is large and the transition is of first order, the partially-deconfined phase will be unstable even in the microcanonical ensemble. Some part of space is occupied by the completely-deconfined phase while the rest is filled by the completely-confined phase, and the partially-deconfined phase is realized only at the interface of these two phases. When the spatial volume is small, such a spatial splitting can be avoided. 
This is sometime exemplified by gauge theories compactified on a sphere. For matrix models, including the Eguchi-Kawai model, spatial splitting cannot take place because `space' is just a single point. The large-$N$ volume independence connects the partially-deconfined phase in the Eguchi-Kawai, which is microcanonically stable, to the partially-deconfined phase of large-volume theory which is not stable even in microcanonical thermodynamics.
The Polyakov loop $P$ increases monotonically with $E$ (Fig.~\ref{fig:M,E,P-vs-T}). Hence, by fixing $P$ we also fix $E$ and can access the information of the microcanonical ensemble. 

For the first constrained simulation, we add the following term to fix the Polyakov loop:
\begin{align}
\Delta S
=
\left\{
\begin{array}{cc}
\frac{g_{\rm P}}{2}\left(
|P|-(P_{\rm fix}+\delta)
\right)^2 & (|P| > P_{\rm fix}+\delta)\\
0 & (P_{\rm fix}-\delta \le |P| \le P_{\rm fix}+\delta)\\
\frac{g_{\rm P}}{2}\left(
|P|-(P_{\rm fix}-\delta)
\right)^2& (|P| < P_{\rm fix}- \delta)
\end{array}
\right.
\label{eq:1st-kind-constraint}
\end{align}

We take $g_{\rm P}$ sufficiently large, so that the value of $|P|$ is fixed to a small window $P_{\rm fix}-\delta \le |P| \le P_{\rm fix}+\delta$. The purpose of this constraint is to probe the partially-confined saddle, which is otherwise unstable in the canonical ensemble. Essentially, we use the density-of-state method by fixing $P$, and hence $E$. The size of the deconfined sector $SU(M)$ will depend on our choice of $|P|$. The precise relation is explained in Sec.~\ref{sec:determining-M}.
Note that the constraint term \eqref{eq:1st-kind-constraint} does not require a specific gauge choice (although we took the static diagonal gauge) and hence we can extract information from the microcanonical thermodynamics in a gauge-invariant manner.

\subsubsection{Fixing of residual gauge symmetry and constrained simulation of the second kind} \label{sec:constrained-second-kind}
\hspace{0.51cm}
For the second kind of constrained simulation, we want to take this SU($M$)-partially-deconfined phase and separate the confined and the deconfined degrees of freedom. 
As we will explain in Sec.~\ref{sec:determining-M}, this can be achieved by, firstly, taking the static diagonal gauge, and then fixing the remaining S$_N$ gauge redundancy down to S$_M\times$S$_{N-M}$ by reordering the eigenvalues such that the $N-M$ confined eigenvalues $\alpha_{M+1},\cdots,\alpha_N$ constitute a uniform distribution. With this gauge fixing, we separate $N^2$ color degrees of freedom into the deconfined sector ($M\times M$ upper-left block) and the confined sector, as depicted in Fig.~\ref{fig:Partial-deconfinement}~\cite{Watanabe:2020ufk}.

The most obvious approach to fixing S$_N$ in this way is, after having performed the constrained simulation of first kind, to sort the resulting $\alpha$'s appropriately. In this paper, we take another approach which is technically simpler. 
We define
\begin{align}
P_{\rm dec}=\frac{1}{M}\sum_{j=1}^Me^{i\alpha_j}
\end{align}
and
\begin{align}
P_{\rm con}=\frac{1}{N-M}\sum_{j=M+1}^Ne^{i\alpha_j}. 
\end{align}

We want to fix $P_{\rm dec}$ and $P_{\rm con}$ to be $P_{\rm fix}$ and $0$, respectively. 
Hence we will add 
\begin{align}
\Delta S_{\rm dec}
=
\left\{
\begin{array}{cc}
\frac{g_{\rm P}}{2}\left(
|P_{\rm dec}|-(P_{\rm fix}+\delta)
\right)^2 & (|P_{\rm dec}| > P_{\rm fix}+\delta)\\
0 & (P_{\rm fix}-\delta \le |P_{\rm dec}| \le P_{\rm fix}+\delta)\\
\frac{g_{\rm P}}{2}\left(
|P_{\rm dec}|-(P_{\rm fix}-\delta)
\right)^2& (|P_{\rm dec}| < P_{\rm fix}- \delta)
\end{array}
\right.
\end{align}
and
\begin{align}
\Delta S_{\rm con}
=
\left\{
\begin{array}{cc}
\frac{g_{\rm P}}{2}\left(
|P_{\rm con}|-\delta
\right)^2 & (|P_{\rm con}| > \delta)\\
0 & (|P_{\rm con}| \le \delta),
\end{array}
\right.
\end{align}
taking $g_{\rm P}$ sufficiently large. 
Fixing $P=\frac{M}{N}P_{\rm dec}+\frac{N-M}{N}P_{\rm con}$ ensures we are on the partially-confined saddle, while fixing $P_{\rm con}$ to zero and $P_{\rm dec}$ to nonzero enforces the gauge fixing.
Note that $P_{\rm con}=0$ does not necessarily guarantee the uniform phase distribution in the confined sector unless the relationship between $M$ and $P_{\rm fix}$ is set correctly following the procedures explained in Sec.~\ref{sec:determining-M}.
With the correct choice of $M$ and $P_{\rm fix}$, this constraint is equivalent to the constraint of the first kind plus permutations of phases. Some explicit checks are provided in Appendix~\ref{appendix:constraint-sanity-check}.
\section{The size of the deconfined sector in the partially-deconfined phase}\label{sec:determining-M}
\hspace{0.51cm}

In this section, we explain how we can separate confined and deconfined sectors. The starting point is the relationship between the operator formalism and the path-integral formalism discussed in Appendix~\ref{appendix:gauge_fixing}.

The canonical partition function is written as \eqref{partition_function_2}. Let us write the expression again:
\begin{align}
Z(T)
=
\frac{1}{{\rm Vol}(G)}\int_Gd\Omega 
{\rm Tr}_{\mathcal{H}_{\rm ext}}\left(\hat{\Omega}e^{-\hat{H}/T}\right). 
\label{partition_function_2_again}
\end{align}
For the microcanonical ensemble, we can obtain a similar expression for the density of states by inserting the projection operator $\hat{\pi}=\frac{1}{{\rm Vol}(G)}\int_Gd\Omega\hat{\Omega}$. 
As explained in Appendix~\ref{appendix:gauge_fixing}, this $\Omega$ corresponds to the Polyakov loop in the path-integral formalism~\cite{Hanada:2020uvt}. 

\begin{figure}[htbp]
  \begin{center}
  \includegraphics[width=1.0\textwidth]{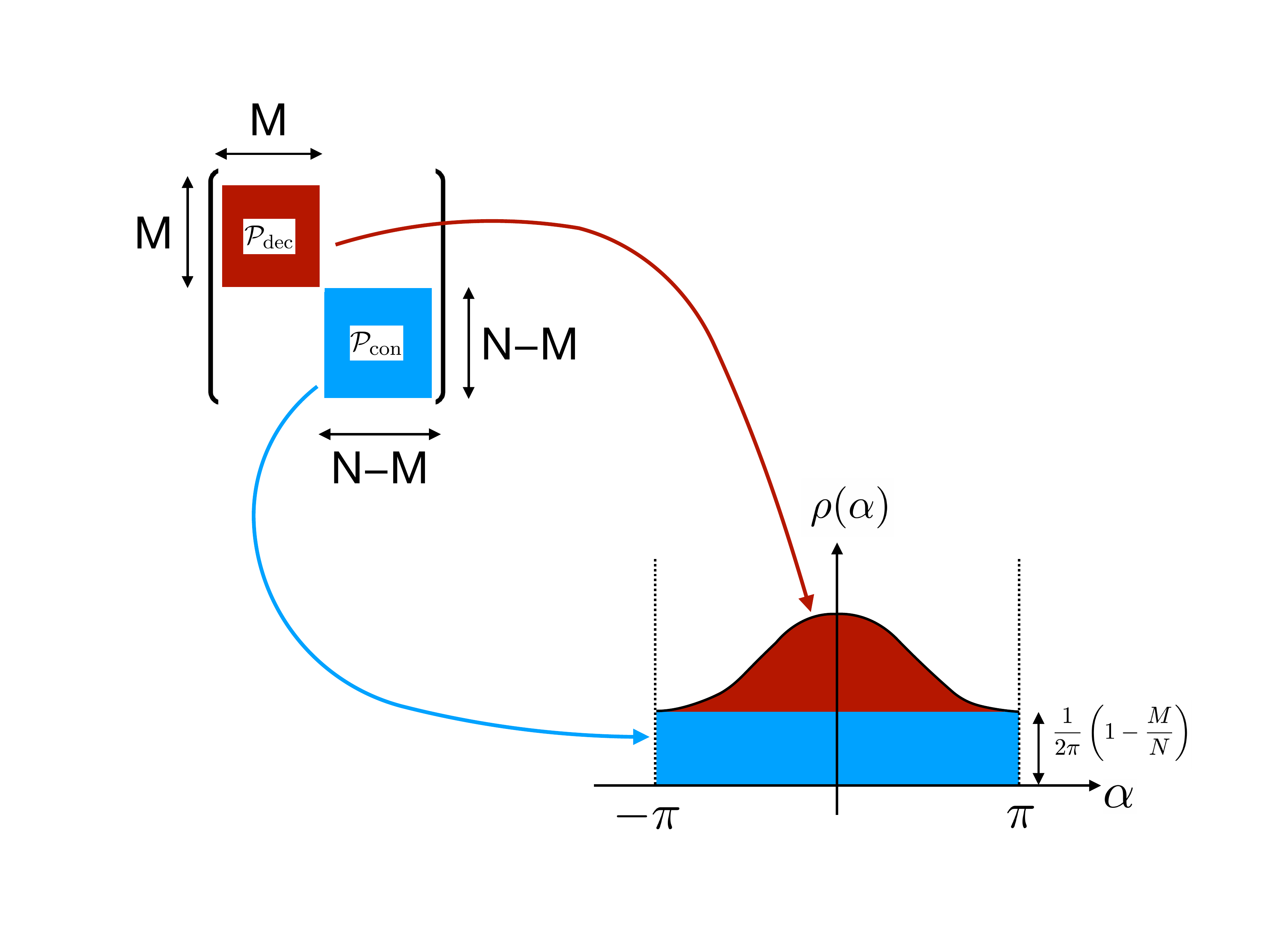}
  \end{center}
  \caption{Sketches of the distribution of the phases of Polyakov loop in the partially-deconfined phase. Constant offset comes from the confined Polyakov loop $\mathcal{P}_{\rm con}$ while the non-uniform part comes from the deconfined Polyakov loop $\mathcal{P}_{\rm dec}$.
  }\label{fig:phase-distribution}
\end{figure}

An energy eigenstate $\ket{E}\in\mathcal{H}_{\rm ext}$ contributes to the partition function as
\begin{align}
\frac{e^{-E/T}}{{\rm Vol}(G)}\int_Gd\Omega 
\bra{E}\hat{\Omega}\ket{E}. 
\end{align}
The SU($M$)-partially-deconfined states are characterized by invariance under $\mathrm{SU}(N-M)\subset\mathrm{SU}(N)$~\cite{Hanada:2020uvt}. 
Namely, 
\begin{align}
\hat{\Omega}\ket{E}=\ket{E}\, , 
\qquad
\Omega\in\mathrm{SU}(N-M)\subset\mathrm{SU}(N)
\end{align}
for SU($M$)-deconfined states. Specifically, by choosing the SU($N-M$) to correspond to the lower-right $(N-M)\times (N-M)$-block, we can choose the SU($M$)-deconfined sector to be the upper-left $M\times M$-block as in Fig.~\ref{fig:Partial-deconfinement}. This choice of embedding of SU($N-M$) into SU($N$) fixes SU($N$) down to $\mathrm{SU}(M)\times\mathrm{SU}(N-M)$. $\Omega$ takes the following form:
\begin{align}
    \Omega=
    \left(
    \begin{array}{cc}
    \mathcal{P}_{\rm dec} & 0\\
    0 & \mathcal{P}_{\rm con}
    \end{array}
    \right)\, .
    \label{eq:gauge-fixing-Polyakov}
\end{align}
Here, $\mathcal{P}_{\rm con}$ can be any element of SU($N-M$), and the generic phase distribution in this part is uniform in the limit of $N-M\to\infty$. The phases of $\mathcal{P}_{\rm dec}$ and $\mathcal{P}_{\rm con}$ are $\alpha_1,\cdots,\alpha_M$ and $\alpha_{M+1},\cdots,\alpha_N$, respectively. From these, we can determine the distribution of the phases $\rho_{\rm dec}(\alpha)$ and $\rho_{\rm con}(\alpha)$. The latter is constant,
\begin{align}
    \rho_{\rm con}(\alpha)=\frac{1}{2\pi}\, , 
\end{align}
while the former is not and its smallest value is zero. 
For the model under consideration, we can fix center symmetry in such a way that
\begin{align}
    \rho_{\rm dec}(\pm\pi)=0\, . 
    \label{phase:dec_is_GWW}
\end{align}
The full distribution is 
\begin{align}
    \rho(\alpha)
    &=
    \left(1-\frac{M}{N}\right)\cdot\rho_{\rm con}(\alpha)
    +
    \frac{M}{N}\cdot\rho_{\rm dec}(\alpha)
    \nonumber\\
    &=
    \frac{1}{2\pi}\cdot\left(1-\frac{M}{N}\right)
    +
    \frac{M}{N}\cdot\rho_{\rm dec}(\alpha)\, . 
    \label{phase:full=con+dec}
\end{align}
See Fig.~\ref{fig:phase-distribution}. 
Constrained simulation of the second kind enforces this separation by constraining $P_{\rm dec}$ to be the appropriate value for each $M$ and setting $P_{\rm con}$ to be zero. In the large-$N$ limit, this is equivalent to the constrained simulation of the first kind plus sorting of the phases, because the distribution becomes continuous and sample-by-sample fluctuation is suppressed.  

For our numerical analysis via the Euclidean path integral, we prefer the Polyakov loop to have the form \eqref{eq:gauge-fixing-Polyakov} at any Euclidean time $t$ so that the SU($M$)-deconfined sector is always in the upper-left $M\times M$-block. The static diagonal gauge is suitable for this purpose: because the gauge field $A_t$ is constant, the Polyakov loop does not depend on Euclidean time $t$. (If $A_t$ is not static, the Polyakov loop can depend on $t$, although the phases do not.) By appropriately fixing the residual S$_N$ symmetry, we can have the same embedding visualized in Fig.~\ref{fig:Partial-deconfinement} at any $t$.

The model under consideration has a first-order confinement/deconfinement transition around $T_c=\frac{1}{2\log(2d-1)}$ \cite{Hanada:2014noa}. 
It is easy to see the two-state signal when $N$ is not too large. When $N$ is large, we can see one of two phases depending on the initial configuration for the simulation.  

In order to determine the size of deconfined sector for each $N$, let us take $P$ to be real and positive (i.e., $P=|P|$) configuration-by-configuration, by using the U(1) center symmetry. 
 For each fixed value of $P$, we collect many samples to determine the distribution of the phases $\alpha$, denoted by $\rho(\alpha)$. 
Because of \eqref{phase:dec_is_GWW} and \eqref{phase:full=con+dec}, the minimum value of $\rho(\alpha)$ (which should be at $\alpha=\pm\pi$) is related to the size of the deconfined sector $M$ by~\cite{Hanada:2020uvt} 
\begin{align*}
\rho(\pm\pi)=\frac{1}{2\pi}\left(1-\frac{M}{N}\right).
\end{align*}
Note that this relation is precise in the large-$N$ limit. 
After determining $M$ for each $P$ and $N$, we can perform the constrained simulation of the second kind with $P_{\rm dec}=\frac{N}{N-M}P(M)$. 

The constrained-simulation methods in this paper are essentially the same as the one used in Ref.~\cite{Watanabe:2020ufk}. 
The only difference is that the models studied in Ref.~\cite{Watanabe:2020ufk} had the simpler relation $P=\frac{M}{2N}$, which does not hold in the model under consideration in this paper. Therefore, we must determine the relationship between $P$ and $M$ by numerically determining $\rho(\pm\pi)$ in the constrained simulation of the first kind.

In general, $\rho(\alpha)$ can be written as
\begin{align}
\label{eq:rho-alpha}
\rho(\alpha)
=
\frac{1}{2\pi}
+
\sum_{k=1}^\infty
\tilde{\rho}_k\cos(k\alpha). 
\end{align}
We determine the coefficients $\tilde{\rho}_k$ based on the simulation data, by using a Bayesian inference procedure with the likelihood of the data $\alpha$ given by Eq.~\eqref{eq:rho-alpha}. 

Each of the $n_{\rm config}$ configurations contains $N$ phases $\alpha_1,\cdots,\alpha_N$. 
Following the usual assumption of the self-averaging nature, i.e., the phase distribution does not depend on samples for sufficiently large $N$, we suppose that each $\alpha_i$ is obtained with a probability $\rho(\alpha_i)$, regardless of the values of the other phases.
Then, for a given model distribution $\rho(\alpha)$ specified by a fixed set of coefficients $\{\tilde{\rho}_k\}$, the probability that $\{\alpha\}=(\alpha_1,\cdots,\alpha_N)$ is obtained is simply the product of the individual phases probabilities:
\begin{align}
\prod_{i=1}^N
\rho(\alpha_i). 
\end{align}

Taking into account all configurations (assumed to be independent), we obtain the likelihood function for the data $\{\alpha\}$, given the parameters $\{\tilde{\rho}_k\}$:
\begin{align}
\label{eq:rho-likelihood}
L(\{\alpha^{(1)}\},\cdots,\{\alpha^{(n_{\rm config})}\}|\{\tilde{\rho}_k\})
=
\prod_{n=1}^{n_{\rm config}}
\prod_{i=1}^{N}
\rho(\alpha_i^{(n)}). 
\end{align}

We use Bayes rule to compute the posterior distribution over the parameters $\{\tilde{\rho}_k\}$ given a set of observations corresponding to the data $\{\alpha^{(1)}\},\cdots,\{\alpha^{(n_{\rm config})}\}$.
In practice, we need to truncate the Fourier expansion in Eq.~\eqref{eq:rho-alpha} at some order $\Lambda$, setting $\tilde{\rho}_k=0$ for $k>\Lambda$, and we choose a uniform prior distribution centered around zero and with bounds $\pm 0.1$.
At fixed order $\Lambda$ we compute the posterior over exactly $\Lambda$ parameters by using a dynamical Nested Sampling algorithm~\cite{skilling2004nested,Buchner2021nestedsampling} implemented in the python library \texttt{ultranest}~\cite{Buchner2021ultranest}.

Ultimately, we do not care about the values of the model parameters $\{\tilde{\rho}_k\}$, but we want to use their posterior distribution to sample all models $\rho(\alpha)$ that are compatible with the data.
With these samples we have direct access to the expectation value of $\rho(\pi)=\frac{1}{2\pi}\left(1-\frac{M}{N}\right)$, hence to the expectation value of $M$ and its corresponding error bar, given by the $16$\% and $84$\% quantiles of the posterior predictive distribution.

We tried a few different values of the Fourier expansion order $\Lambda$ and found that results are consistent within error bars for $\Lambda \in \{2,3,4,5\}$. For our analysis we choose to consider $\Lambda=3$ out of simplicity.

The values of $M$ for $T=0.29$ obtained this way are summarized in Table~\ref{table:N-L-P-M}.
For each fixed $N$ and $P$, there is a weak dependence on the lattice size $n_t$.

\begin{table}
 \centering
  \begin{tabular}{|c||c|c|c|}
        \hline
        $P$ & $N$ & $n_t$ & $M$\\
        \hline
        \hline
        0.2 & 16 & 16 & $7.44^{+0.12}_{-0.11}$\\
            &     & 24 & $7.62^{+0.15}_{-0.16}$\\
            &     & 32 & $7.56^{+0.11}_{-0.12}$\\
            & 24 & 16 & $10.99^{+0.18}_{-0.18}$\\
            &     & 24 & $11.13^{+0.19}_{-0.19}$\\
            &     & 32 & $11.12^{+0.22}_{-0.22}$\\
            & 32 & 16 & $14.51^{+0.21}_{-0.19}$\\
            &     & 24 & $14.77^{+0.22}_{-0.22}$\\
            &     & 32 & $14.91^{+0.21}_{-0.23}$\\
            & 64 & 16 & $28.70^{+0.31}_{-0.31}$\\
            &     & 24 & $29.04^{+0.31}_{-0.34}$\\
            &     & 32 & $29.06^{+0.29}_{-0.29}$\\
            & 96 & 16 & $42.78^{+0.40}_{-0.38}$\\
            &     & 24 & $43.19^{+0.48}_{-0.49}$\\
        \hline
        0.25 & 16 & 16 & $10.06^{+0.09}_{-0.10}$\\
            &     & 24 & $9.99^{+0.13}_{-0.13}$\\
            &     & 32 & $10.18^{+0.14}_{-0.15}$\\
            & 24 & 16 & $14.85^{+0.17}_{-0.16}$\\
            &     & 24 & $14.92^{+0.17}_{-0.17}$\\
            &     & 32 & $15.18^{+0.11}_{-0.12}$\\
            & 32 & 16 & $19.52^{+0.18}_{-0.15}$\\
            &     & 24 & $19.81^{+0.22}_{-0.23}$\\
            &     & 32 & $19.92^{+0.19}_{-0.18}$\\
            & 64 & 16 & $38.19^{+0.30}_{-0.28}$\\
            &     & 24 & $38.77^{+0.37}_{-0.36}$\\
            &     & 32 & $39.11^{+0.33}_{-0.34}$\\
            & 96 & 16 & $56.50^{+0.35}_{-0.34}$\\
            &     & 24 & $57.43^{+0.42}_{-0.40}$\\
        \hline
  \end{tabular}
  \caption{\label{table:N-L-P-M}
  $P$ vs $M$ at $T=0.29$. The central value of $M$ is the median of the posterior predictive distribution. The lower and upper bound are the $16$\% and $84$\% quantiles, respectively.
  }
\end{table}

\begin{table}
 \centering
  \begin{tabular}{|c||c|c|c|c|}
        \hline
        $P$ & $N$ & $M$ at $T=0.29$ & $M$ at $T=0.30$ & $M$ at $T=0.31$\\
        \hline
        \hline
        0.2 & 16 & $7.62^{+0.15}_{-0.16}$ & $7.75^{+0.15}_{-0.15}$ & $7.83^{+0.16}_{-0.16}$\\
            & 24 & $11.13^{+0.19}_{-0.19}$ & $11.31^{+0.19}_{-0.20}$ & $11.68^{+0.19}_{-0.20}$\\
            & 32 & $14.77^{+0.22}_{-0.22}$ & $15.03^{+0.20}_{-0.20}$ & $15.23^{+0.22}_{-0.22}$\\
        \hline
        0.25 & 16 & $9.99^{+0.13}_{-0.13}$ & $10.36^{+0.15}_{-0.14}$ & $10.59^{+0.14}_{-0.14}$\\
            & 24 & $14.92^{+0.17}_{-0.17}$ & $15.61^{+0.17}_{-0.16}$ & $15.95^{+0.16}_{-0.17}$\\
            & 32 & $19.81^{+0.22}_{-0.23}$ & $20.50^{+0.18}_{-0.16}$ & $21.07^{+0.20}_{-0.19}$\\
        \hline
  \end{tabular}
  \caption{\label{table:N-L-P-M-T-dependence}
  $P$ vs $M$ at three different temperatures for $n_t=24$. The central value of $M$ is the median of the posterior predictive distribution. The lower and upper bound are the $16$\% and $84$\% quantiles, respectively.
  }
\end{table}

In this paper, we are trying to study the partially-deconfined saddle.
The temperature of the saddle changes slightly with $P$, while we varied $P$ at fixed $T$. 
We are implicitly assuming that a slight difference of temperature does not have a significant effect if $P$ is fixed.
The validity of this assumption requires that the saddle is insensitive with respect to the temperature. 
To test this, we looked at the value of $M$ for different values of temperature at fixed $P$.
See Table~\ref{table:N-L-P-M-T-dependence} for the relation between $P$ and $M$ for $T=0.29, 0.30$ and $T=0.31$.
The dependence on temperature is rather mild (mostly compatible within error bars).
Hence, we assume other quantities such as the string tension are not sensitive to the small change of temperature.
We leave an explicit confirmation of this assumption as a future work.

\section{Flux tube in partially-deconfined phase}\label{sec:flux-tubes}
\hspace{0.51cm}

\subsection{Theoretical expectations}\label{sec:flux-tube-theory}
\begin{figure}[htbp]

  \begin{center}
   \includegraphics[width=100mm]{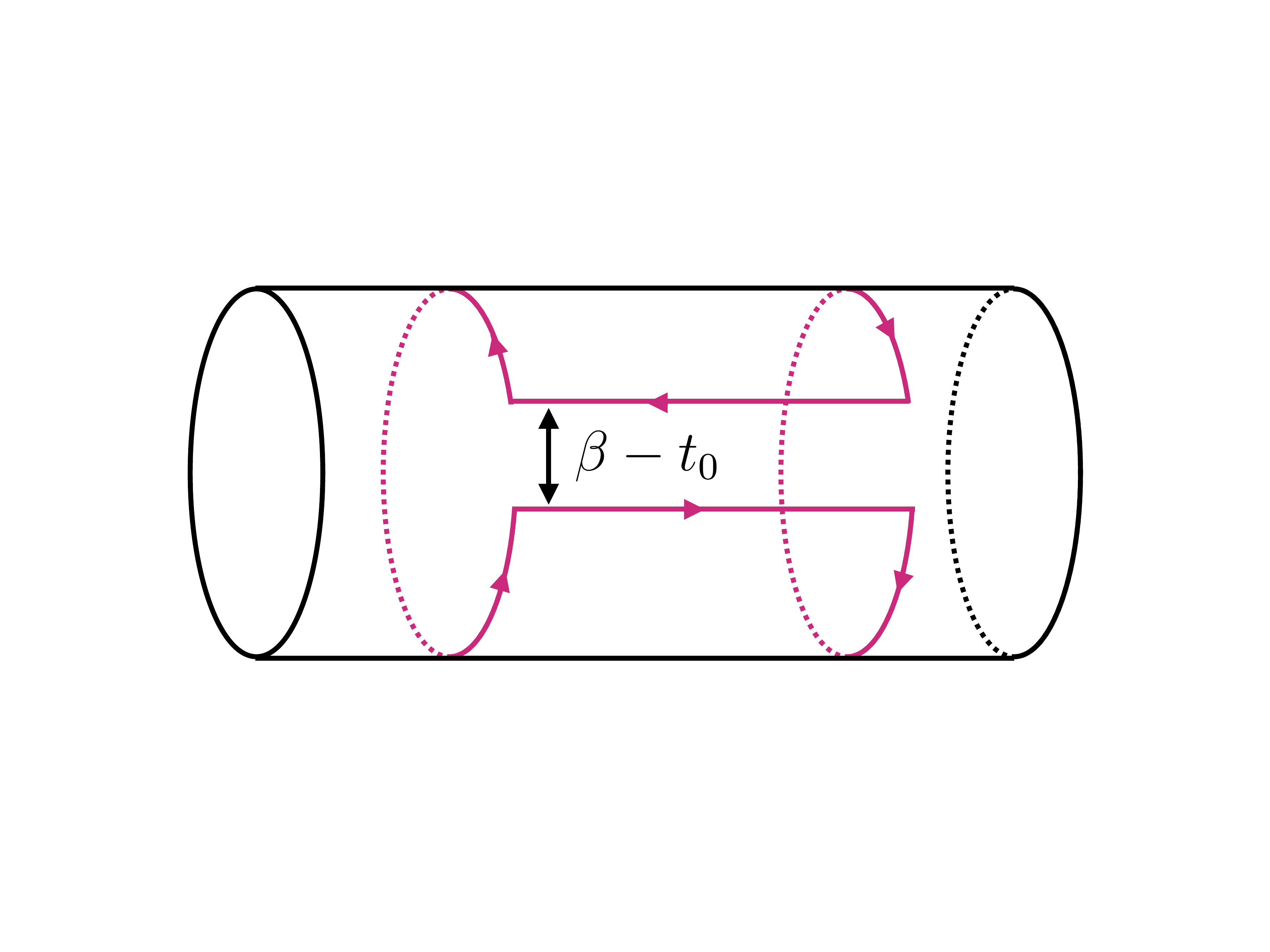}
  \end{center}
  \caption{
The temporal Wilson loop $W_{\mu}(L,t_0)$ considered in this work. 
A spatial Wilson line with length $L$ is created, goes through Euclidean time evolution around the temporal circle, and is then annihilated. 
By exchanging the role of time and space, we can interpret this also as the propagation of an open string along the compactified space over Euclidean time $L$. We will focus on the case of $t_0=\beta$.
  }\label{fig:Wilson_loop_2}
\end{figure}

\begin{figure}[htbp]
  \begin{center}
   \includegraphics[width=100mm]{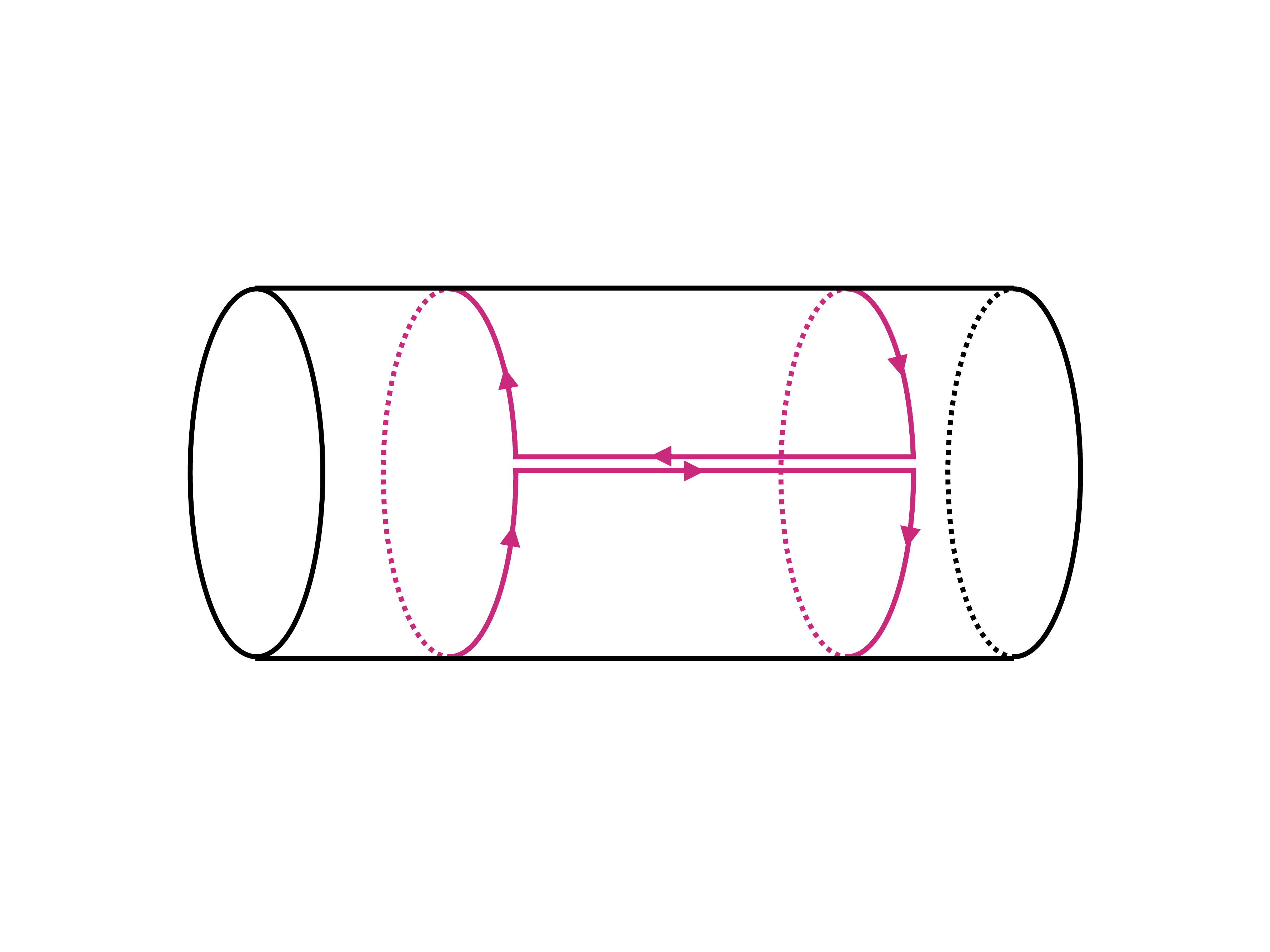}
  \end{center}
  \caption{
Temporal Wilson loop $W_{\mu}(L,\beta)$. 
Such a Wilson loop is equivalent to the two-point function of Polyakov loops. 
  }\label{fig:Wilson_loop}
\end{figure}

In this section, we demonstrate \eqref{conjecture:con-con} and \eqref{conjecture:con-dec} using Eguchi-Kawai equivalence. The Eguchi-Kawai model is easier for numerical purposes, but the lack of the spatial dimensions forces us to rewrite the two-point functions on the left-hand side of \eqref{conjecture:con-con} and \eqref{conjecture:con-dec} to slightly different forms. 
We define the temporal Wilson loop with temporal extent $0<t_0\le\beta$ and spatial extent $L$ in the lattice unit, as shown in See Fig.~\ref{fig:Wilson_loop_2}. When $t_0=\beta$ (Fig.~\ref{fig:Wilson_loop}), it is equivalent to the two-point function of the Polyakov loop, as we will see shortly.
Thanks to Eguchi-Kawai equivalence, we can calculate this Wilson loop by using the Eguchi-Kawai model.

Let us elucidate the physical meaning of this loop. 
Let $F^\dagger$ be a creation operator for two heavy probe particles $\phi$ and $\chi$ connected by a Wilson line~\cite{Nair:2005iw}, 
\begin{align}
\hat{F}_\mu^\dagger(\vec{x},L)=\hat{\phi}^\dagger_a(\vec{x})\hat{U}_{\mu,ab}(\vec{x},\vec{x}+L\hat{\mu})\hat{\chi}_b^\dagger(\vec{x}+L\hat{\mu}).
\label{Wilson_line_creation}  
\end{align}
Then $W_\mu(L,t_0)$ is 
\begin{align}
W_\mu(L,t_0)
&=
\frac{1}{Z(\beta)}
\sum_{i}
\left\langle E_i\right|
e^{-(\beta-t_0)\hat{H}}
\hat{F}_\mu
e^{-t_0\hat{H}}
\hat{F}^\dagger_\mu
\left|E_i\right\rangle 
\nonumber\\
&=
\frac{1}{Z(\beta)}
\sum_{i}
e^{-\beta E_i}
\left\langle E_i\right|
\hat{F}_\mu
e^{-t_0(\hat{H}-E_i)}
\hat{F}^\dagger_\mu
\left|E_i\right\rangle.  
\end{align}
Note that the contribution from the mass of the probes is subtracted.

Let us write $F^\dagger_\mu\left|E_i\right\rangle$ as a linear combination of the energy eigenstates,
\begin{align}
\hat{F}^\dagger_\mu\left|E_i\right\rangle
=
\sum_{j}c_{ij}\left|E_j\right\rangle. 
\end{align}
Then, 
\begin{align}
W_\mu(L,t_0)
=
\frac{1}{Z(\beta)}
\sum_{i,j}
e^{-\beta E_i}
|c_{ij}|^2
e^{-t_0(E_j-E_i)}.  
\end{align}
Suppose the sum over all the states can be replaced by a typical energy eigenstate $|E_i\rangle$.  
Then,
\begin{align}
W_\mu(L,t_0)
\sim
\sum_j
|c_{ij}|^2
e^{-t_0(E_j-E_i)}.  
\end{align}
If the minimum excitation above $E_i$ increases linearly with $L$, we have $W_\mu(L,t_0)\sim e^{-c L}$, with some constant $c$.  
We expect this to happen in the confined sector, i.e., we expect $W_{\mu,{\rm con}}(L,t_0)\sim e^{-c L}$. 
If only the low-energy states contribute, we should have $c\propto t_0\sigma$, where $\sigma$ is called the string tension.  
In the Eguchi-Kawai model, by using the Polyakov line  
\begin{align}
{\cal P}
\equiv
{\rm diag}(e^{i\alpha_1},\cdots,e^{i\alpha_N}), 
\end{align}
we can write $W_{\mu}(L,\beta)$ as 
\begin{align}
W_{\mu}(L,\beta)
=
\left\langle
{\rm Tr}
\left(
{\cal P}(U_\mu(t))^{L}
{\cal P}^\dagger(U_\mu(t)^\dagger)^{L}
\right)
\right\rangle.   
\end{align} 
We can see the exponential decay of $W_{\mu}(L,\beta)$ in Fig.~\ref{fig:Completely_confined_T025} for the completely-confined case.

$\frac{1}{N}W_\mu(L,\beta)$ is equivalent to the two-point function of Polyakov loops, i.e., 
\begin{align}
\frac{1}{N}W_\mu(L,\beta)
=
\left\langle
{\rm Tr}{\cal P}(\vec{x})
\cdot
{\rm Tr}{\cal P}(\vec{x}+L\hat{\mu})
\right\rangle. 
\end{align}
Here $\hat{\mu}$ is the unit vector along the $\mu$-th dimension. 
One way to see this is to exchange the roles of temporal and spatial directions. 
Namely, we interpret the $\mu$-direction to be imaginary time. 
$W_\mu(L,t_0)$ is interpreted as the propagation along the imaginary time direction of an open string with length $t_0$ stretched along the spatial circle with circumference $\beta$. 
When $t_0=\beta$ and the color factors (Chan-Paton factors) at the endpoints of the open string are summed over, 
we get a closed string, or equivalently, the Polyakov loop. 

To see the properties of the confined and deconfined sectors separately, we define the Polyakov line in the deconfined sector,
\begin{align}
{\cal P}_{\rm dec}
\equiv
{\rm diag}(e^{i\alpha_1},\cdots,e^{i\alpha_M},0,\cdots,0),
\end{align}
and that in the confined sector,
\begin{align}
{\cal P}_{\rm con}
\equiv
{\rm diag}(0,\cdots,0,e^{i\alpha_{M+1}},\cdots,e^{i\alpha_N}).
\end{align}
By using these, we can define the counterparts of two-point functions in the lattice gauge theory
$\left\langle{\cal P}_{\rm dec}(\vec{x}){\cal P}_{\rm dec}(\vec{y})\right\rangle$, 
$\left\langle{\cal P}_{\rm con}(\vec{x}){\cal P}_{\rm con}(\vec{y})\right\rangle$
and
$\left\langle{\cal P}_{\rm con}(\vec{x}){\cal P}_{\rm dec}(\vec{y})\right\rangle$ as 
\begin{align}
W_{\mu,{\rm dec}}(L,\beta)
\equiv
\left\langle
{\rm Tr}
\left(
{\cal P}_{\rm dec}(U_\mu(t))^{L}
{\cal P}_{\rm dec}^\dagger(U_\mu(t)^\dagger)^{L}
\right) 
\right\rangle, 
\end{align}
\begin{align}
W_{\mu,{\rm con}}(L,\beta)
\equiv
\left\langle
{\rm Tr}
\left(
{\cal P}_{\rm con}(U_\mu(t))^{L}
{\cal P}_{\rm con}^\dagger(U_\mu(t)^\dagger)^{L}
\right) 
\right\rangle. 
\end{align}
and
\begin{align}
W_{\mu,{\rm mix}}(L,\beta)
\equiv
\left\langle
{\rm Tr}
\left(
{\cal P}_{\rm con}(U_\mu(t))^{L}
{\cal P}_{\rm dec}^\dagger(U_\mu(t)^\dagger)^{L}
\right) 
\right\rangle. 
\end{align}
We can also define $W_{\mu,{\rm dec}}(L,t_0)$, $W_{\mu,{\rm con}}(L,t_0)$ and $W_{\mu,{\rm mix}}(L,t_0)$ in a similar manner. 

Because $\frac{1}{N}W_\mu(L,\beta)$ is equivalent to the two-point function of Polyakov loops, and because the connected part of the two-point function is suppressed at long distance ($L\to\infty$), we will have 
\begin{align}
\lim_{L\to\infty}
\frac{1}{N}W_\mu(L,\beta)
=
|\langle P\rangle|^2. 
\end{align}
We expect $W_{\mu,{\rm con}}(L,\beta)$ and $W_{\mu,{\rm mix}}(L,\beta)$ to vanish at large $L$, and hence we expect 
\begin{align}
\lim_{L\to\infty}
\frac{1}{N}W_{\mu,{\rm dec}}(L,\beta)
=
|\langle P\rangle|^2. 
\end{align}

We expect that $W_{\mu,{\rm con}}(L,\beta)$ and $W_{\mu,{\rm mix}}(L,\beta)$ vanish in a very specific manner. 
We expect the exponential decay at long distance with the same string tension as the confined sector, i.e., we expect 
\begin{align}
\frac{1}{N}W_{\mu,{\rm con}}(L,\beta)
=
C_{\rm con}(N,M)\cdot\exp\left(
-\sigma L\beta
\right) 
\label{theory-confined}
\end{align}
and
\begin{align}
\frac{1}{N}W_{\mu,{\rm mix}}(L,\beta)
=
C_{\rm mix}(N,M)\cdot\exp\left(
-\sigma L\beta
\right)\ .    
\label{theory-mixed}
\end{align}
To obtain \eqref{theory-mixed}, we can interpret the $\mu$-direction to be the Euclidean time. Then, closed string in the confined sector is created by $\mathcal{P}_{\rm con}$ and propagate distance $L$. Then it is annihilated by $\mathcal{P}_{\rm dec}$. It is natural to expect that $\mathcal{P}_{\rm dec}$ contains a small but nonzero contribution from the lightest mode in $\mathcal{P}_{\rm con}$ because there is no reason that it is forbidden, and hence, we expect to see the propagation of the closed string in the confined sector that is associated with the decay factor $\exp\left(-\sigma L\beta\right)$.  

In the strong-coupling lattice gauge theory, with our normalization, 
\begin{align}
\sigma=\frac{1}{2}. 
\end{align}
Note that the large-$N$ limit should be taken first. 
Approximately, we expect 
$C_{\rm con}(N,M)\simeq\left(1-\frac{M}{N}\right)^2$ and
$C_{\rm mix}(N,M)\simeq\frac{M}{N}\left(1-\frac{M}{N}\right)$. 
To give a stronger constraint on $C_{\rm con}(N,M)$ and $C_{\rm mix}(N,M)$, let us assume that these factors do not depend on the size of the loops along the temporal direction, i.e., 
\begin{align}
\frac{1}{N}W_{\mu,{\rm con}}(L,t_0)
=
C_{\rm con}(N,M)\cdot\exp\left(
-\sigma Lt_0
\right)
\end{align}
and
\begin{align}
\frac{1}{N}W_{\mu,{\rm mix}}(L,t_0)
=
C_{\rm mix}(N,M)\cdot\exp\left(
-\sigma Lt_0
\right)\ . 
\end{align}
By definition, the sum of $W_{\mu,{\rm con}}(L,t_0=0)$ and $W_{\mu,{\rm mix}}(L,t_0=0)$ can be written as
\begin{align}
W_{\mu,{\rm con}}(L,t_0=0)
+
W_{\mu,{\rm mix}}(L,t_0=0)
&=
\left\langle
{\rm Tr}
\left(
\Pi_{\rm con}
(U_\mu(t))^{L}
\textbf{1}_N(U_\mu(t)^\dagger)^{L}
\right)
\right\rangle
\nonumber\\
&=
\left\langle
{\rm Tr}
\left(
\Pi_{\rm con}
\right)
\right\rangle
\nonumber\\
&=
N-M,
\end{align}
where $\Pi_{\rm con}=\textrm{diag}(\underbrace{0,\cdots,0}_{M},\underbrace{1,\cdots,1}_{N-M})$. Therefore, 
\begin{align}
C_{\rm con}(N,M)
+
C_{\rm mix}(N,M)
=
1-\frac{M}{N}\ ,  
\label{theory-con+mixed}
\end{align}
if $C_{\rm con}(N,M)$ and $C_{\rm mix}(N,M)$ do not depend on $t_0$. In Sec.~\ref{sec:flux-tube-simulation}, we will confirm this relation numerically.

\subsection{Simulation results}\label{sec:flux-tube-simulation}
In this subsection, we will show the simulation results. The expectation values of $W(L)$ used for the analyses are shown in Tables~\ref{table-W-T025}, \ref{table-W-T029-1}, \ref{table-W-T029-2}, and \ref{table-W-T029-3}. 
\subsubsection*{Completely-confined phase}
\begin{figure}[htbp]
  \begin{center}
  \includegraphics[width=0.95\textwidth]{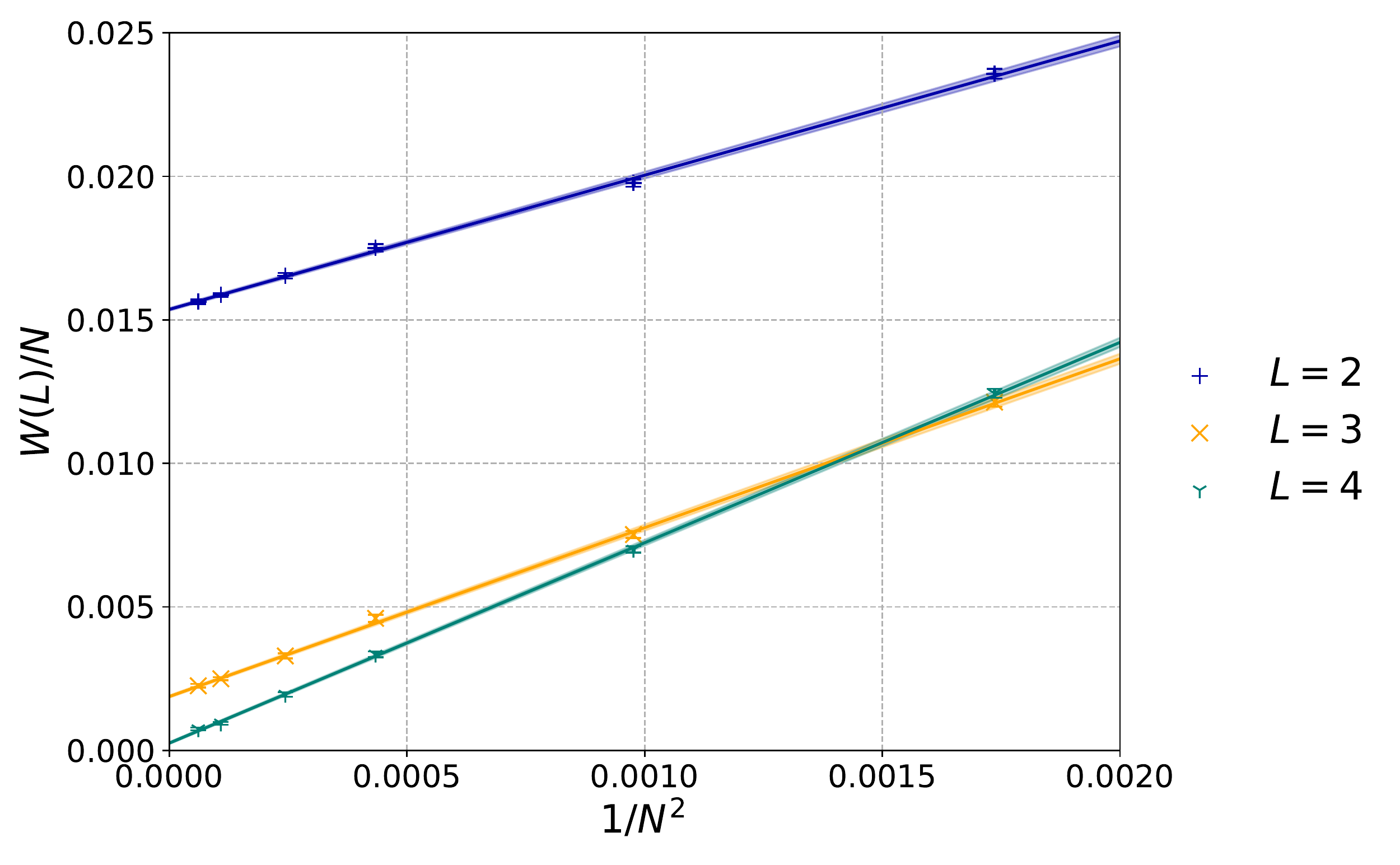}
  \includegraphics[width=0.95\textwidth]{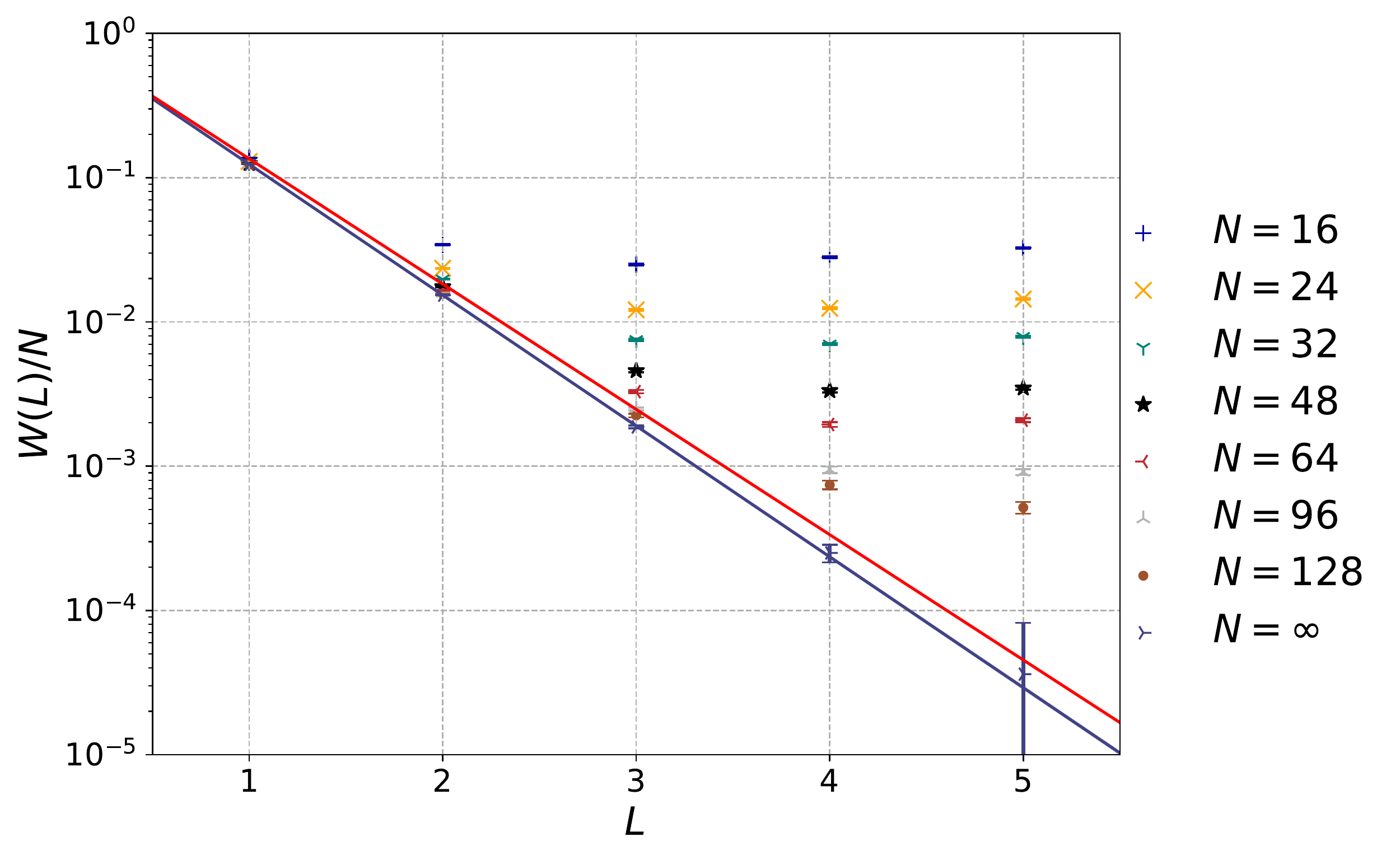}
  \end{center}
  \caption{$T=0.25$ and $n_t=24$, without constraint. Temperature is sufficiently low such that the completely-confined phase is realized automatically.
  For the extrapolation to $N=\infty$, we used data points at $N=24, 32, 48, 64, 96, 128$ for $L=1,2,3,4$ and $N=48, 64, 96, 128$ for $L=5$. 
  We used the fitting ansatz $\frac{1}{N}W(L)= a(L)+\frac{b(L)}{N^2}$.  
  By fitting the large-$N$ extrapolated results at $1\le L\le 4$ by the ansatz $a(L)=\exp(-cL+d)$, we obtained
$c=\cTZeroTwoFive$ and $d=\dTZeroTwoFive$. This fit is shown by the blue line. The red line is $\exp(-2L)$, which is the theoretical expectation for long distance (large $L$) and continuum limit ($n_t=\infty$). A small disagreement would be finite-$n_t$ effects. 
  }\label{fig:Completely_confined_T025}
\end{figure}

Let us start with the unconstrained simulation at $T=0.25$, where temperature is sufficiently low such that the completely-confined phase is obtained in the large-$N$ limit. We took the lattice size $n_t=24$, and studied $N=16, 24, 32, 48, 64, 96$ and $128$. We took the average over all spatial dimensions $\mu=1,2,3$ and studied $W=\frac{1}{3}\sum_{\mu=1}^3W_\mu$.  
As we can see in the first panel of Fig.~\ref{fig:Completely_confined_T025}, the $N$-dependence of $\frac{1}{N}W(L)$ can be fitted well to the fitting ansatz $\frac{1}{N}W(L)= a(L)+\frac{b(L)}{N^2}$. We used $N\ge 24$ to perform the extrapolation to $N=\infty$. We could achieve reliable extrapolations at $L=1,2,3$ and $4$. For $L\ge 5$, larger $N$ are needed for reliable extrapolations. 

Because the Eguchi-Kawai reduction is valid at $N=\infty$, we can compare the large-$N$ extrapolated values with the theoretical expectation \eqref{theory-confined} and $\sigma=\frac{1}{2}$, $\beta=\frac{1}{T}=4$, and $M=0$, i.e., $\frac{1}{N}W(L,\beta=4)=\exp(-2L)$. Note that $C_{\rm con}(N,M=0)=1$, because of \eqref{theory-con+mixed} and $C_{\rm mix}(N,M=0)=0$. In the second panel of Fig.~\ref{fig:Completely_confined_T025}, $\frac{1}{N}W(L)$ for each $N$ and the large-$N$ extrapolation are shown for $L=1,2,3,4,5$. By fitting the large-$N$ extrapolated results at $1\le L\le 4$ to the ansatz $a(L)=\exp(-cL+d)$, we obtained
$c=\cTZeroTwoFive$ and $d=\dTZeroTwoFive$. Small deviations from $\exp(-2L)$ would be finite-$n_t$ effects. 

\begin{figure}[htbp]
  \begin{center}
  \includegraphics[width=1.0\textwidth]{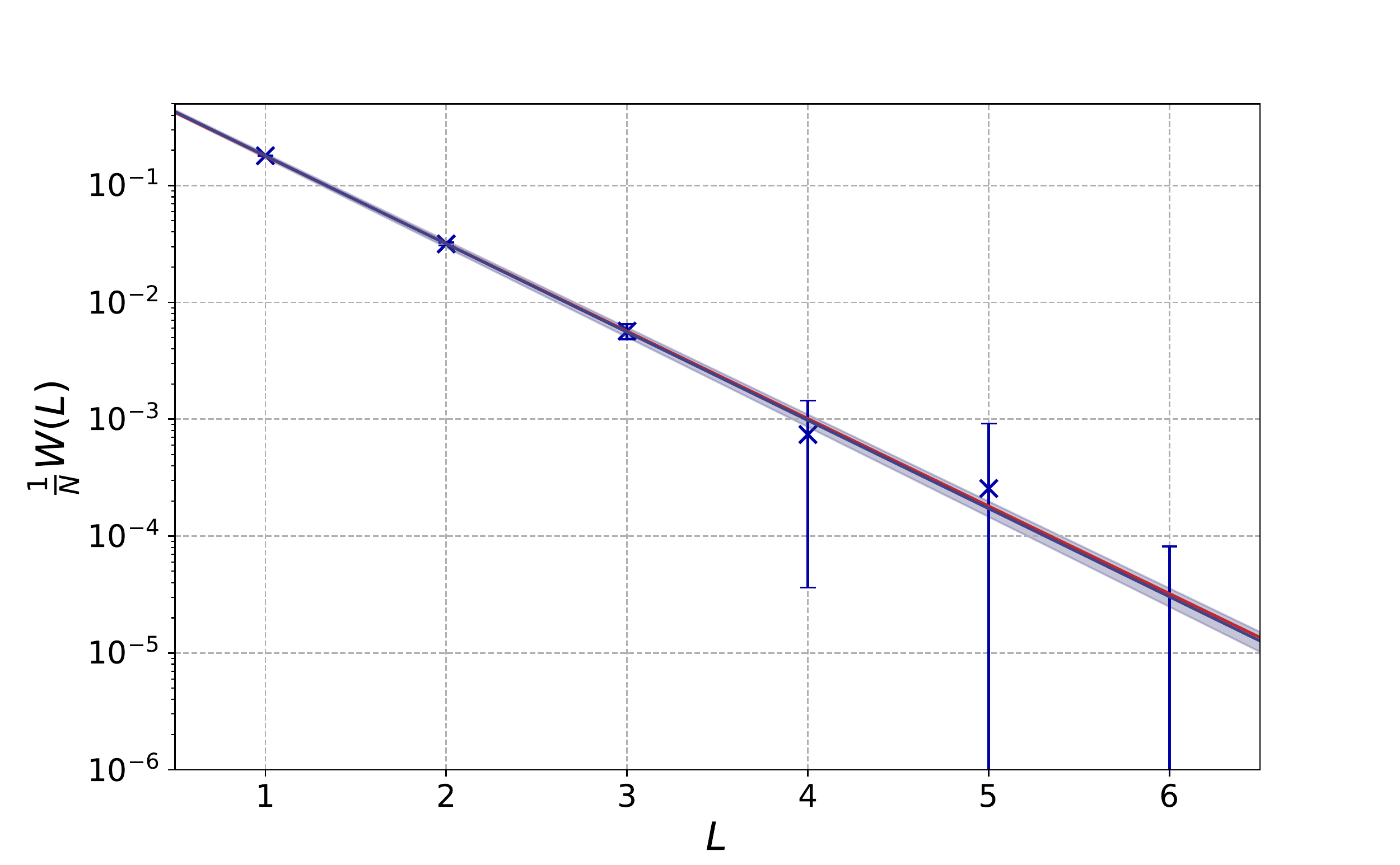}
  \end{center}
  \caption{$T=0.29$ with the constraint $P=0$, allowing us to probe the completely-confined phase near the transition. We have extrapolated to the large $N$ limit using $N=7, 10, 16, 24, 32$, and to the continuum limit using $n_t=16,24,32$ for each $N$, using the 2D interpolation $\frac{1}{N}W_{\rm con}(L) = a_{\rm con}(L) + \frac{b_{\rm 1,con}(L)}{N^2} + \frac{b_{\rm 2,con}(L)}{n_t} + \frac{b_{\rm 3,con}(L)}{n_t N^2}$. We fitted these interpolated values $a_{\rm con}(L)$ for $1\le L\le 4$ to the ansatz $a_{\rm con}(L)=\exp(-c_{\rm con}L+d_{\rm con})$, yielding $c_{\rm con}=\cCompletelyConfinedPZero$ and $d_{\rm con}=\dCompletelyConfinedPZero$. The fit is shown along with error bounds by the blue line. The theoretical prediction ($c_{\rm con}=1.724$, and $d_{\rm con}=0$ for $M=0$) is given by the red line, which is mostly obscured by the fit. 
   }\label{fig:Completely_confined_T029_P0}
\end{figure}

We can also probe the confined phase at $T=0.29$ by constraining $P$ to its confined value, $P=0$, as shown in Fig.~\ref{fig:Completely_confined_T029_P0}. This is more comparable to the constrained simulation, described below, that we perform at $T=0.29$ to probe the partial phase, but is here applied to examining the familiar completely-confined phase with its well-established theory and predictions.

For this, we took the large-$N$ and continuum limit at each $L$ by performing a weighted least-squares regression with the ansatz $\frac{1}{N}W_{\rm con}(L; n_t) = a_{\rm con}(L) + \frac{b_{\rm 1,con}(L)}{N^2} + \frac{b_{\rm 2,con}(L)}{n_t} + \frac{b_{\rm 3,con}(L)}{n_t N^2}$. The weights were derived from the error bars of the Monte-Carlo observables, where the integrated autocorrelation time measured by the Madras-Sokal algorithm~\cite{Sokal1996MonteCM} is taken into account. Then $a_{\rm con}(L)$ gives the extrapolated value in the $N, n_t \rightarrow \infty$ limit. Fitting the ansatz $a_{\rm con}(L) = \exp(-c_{\rm con}L + d_{\rm con})$, we obtain $c_{\rm con}=\cCompletelyConfinedPZero$ and $d_{\rm con}=\dCompletelyConfinedPZero$. This is consistent with the theoretical expectation $e^{\sigma\beta L}=e^{-\frac{1}{2\times 0.29}L}\simeq e^{-1.724L}$. 

\subsubsection*{Partially-deconfined phase}

\begin{figure}[htbp]
  \begin{center}
     \includegraphics[width=0.8\textwidth]{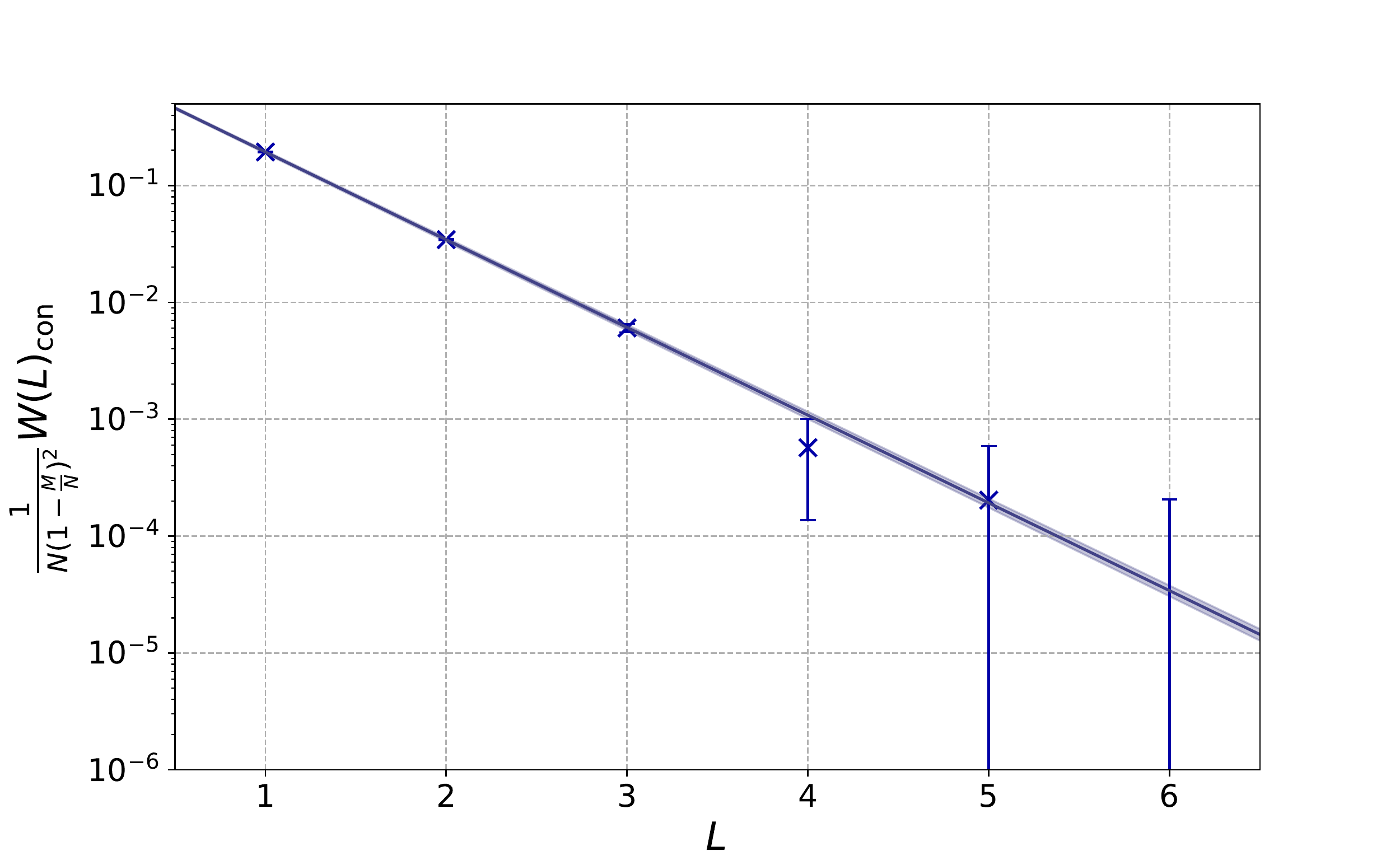}
     \includegraphics[width=0.8\textwidth]{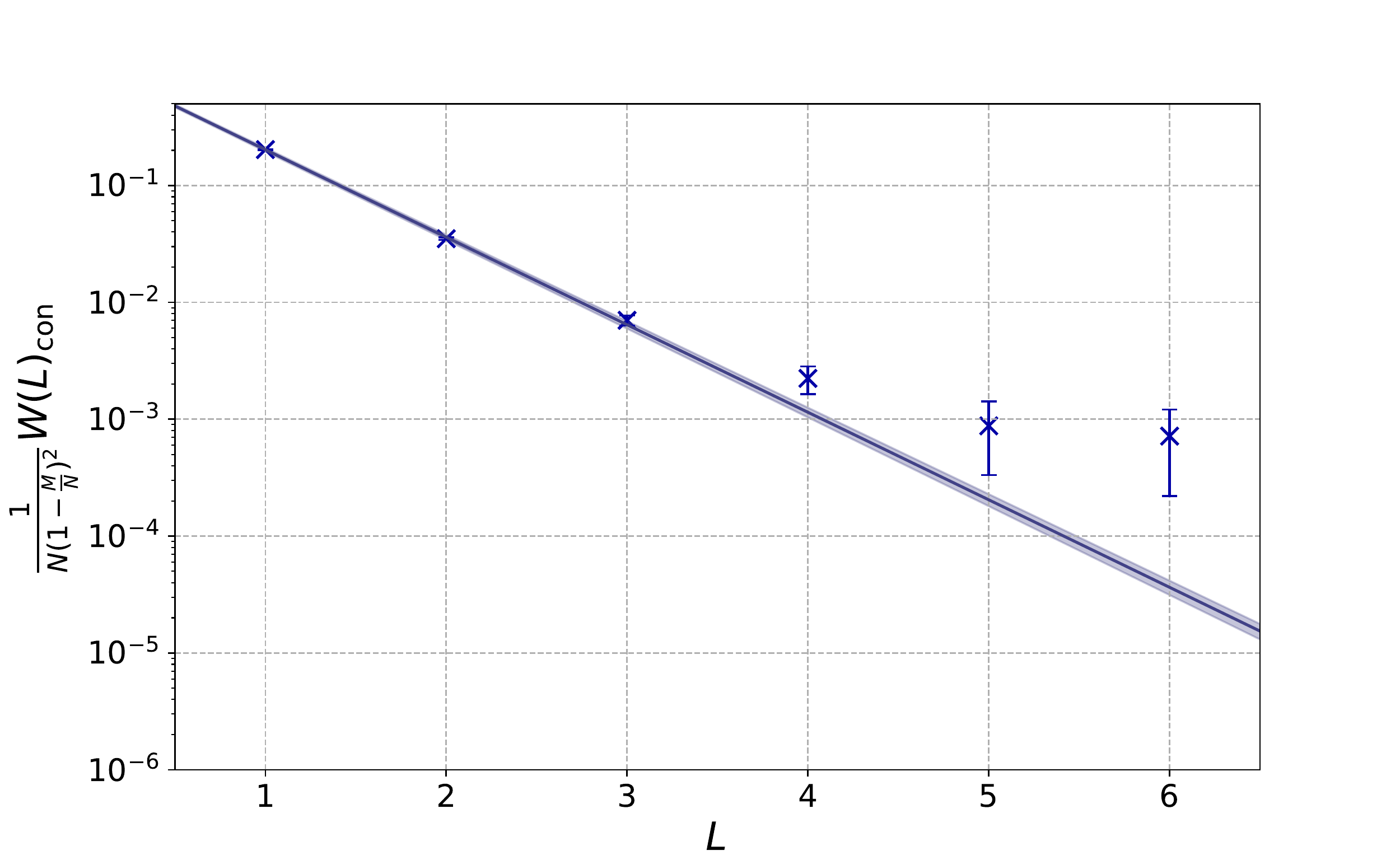}
  \end{center}
  \caption{
    Confined subsector Wilson loop, $W_{\rm con}$, at $T=0.29$, for $P=0.2$ and $P=0.25$, respectively. We have extrapolated to the large $N$ limit using $N=24, 32, 64$, and to the continuum limit using $n_t=16,24,32$ for each $N$, using the 2D interpolation $\frac{1}{N(1-\frac{M}{N})^2}W_{\rm con}(L) = \gamma_{\rm con}(L) + \frac{\beta_1}{N^2} + \frac{\beta_2}{n_t} + \frac{\beta_2}{n_t N^2}$. We fitted these interpolated values $\gamma_{\rm con}(L)$ for $1\le L\le 4$ to the ansatz $\gamma_{\rm con}(L)=\exp(-c_{\rm con}L+d_{\rm con})$ and obtained $c_{\rm con}=\cConPpointTwo$ and $c_{\rm con}=\cConPpointTwoFive$ for $P=0.2$ and $P=0.25$, respectively. The fit is shown along with error bounds by the blue line. }
    \label{fig:W-con-2D}
\end{figure}

In Fig.~\ref{fig:W-con-2D}, we show $W_{\rm con}$ at $T=0.29$, $P=0.2$ and $P=0.25$, which corresponds to $\frac{M}{N}\simeq 0.50$ and $\frac{M}{N}\simeq 0.61$. We performed a two-dimensional weighted least-squares regression to take the large $N$ and continuum ($n_t\to\infty$) limits simultaneously. Note that it is more difficult to take the large-$N$ limit at larger $L$, and/or when either $M$ or $N-M$ is small.   

Motivated by \eqref{theory-confined} with $C_{\rm con}(N,M)\simeq\left(1-\frac{M}{N}\right)^2$, we normalized the loop as $\frac{1}{N}W_{\rm con}(L)\times \left(1-\frac{M}{N}\right)^{-2}$. We will compare the loop normalized this way to $e^{\sigma\beta L}=e^{-\frac{1}{2\times 0.29}L}\simeq e^{-1.724L}$. 

The large-$N$ and continuum extrapolations were estimated at each $L$ by performing a weighted least-squares regression with the ansatz $\frac{1}{N}W_{\rm con}(L)\times \left(1-\frac{M}{N}\right)^{-2} = a_{\rm con}(L) + \frac{b_{\rm 1,con}(L)}{N(N-M)} + \frac{b_{\rm 2,con}(L)}{n_t} + \frac{b_{\rm 3,con}(L)}{n_t N(N-M)}$ \footnote{In the completely-confined phase, we used $N^2$ in the denominators of the regression. For the partial phase, we used $N(N-M)$ instead. We found that our data was much closer to being linear in $\frac{1}{N(N-M)}$, both for data points taken at the same $P$ but which differed slightly in $M/N$, and also when combining points at different $P$ and thus very different values of $M/N$.}. Then $a_{\rm con}(L)$ gives the extrapolated value of the Wilson loop in the $N, n_t \rightarrow \infty$ limit.

We plotted $a_{\rm con}(L)$ in Fig.~\ref{fig:W-con-2D}. We can perform one final linear regression with the ansatz $\log(a_{\rm con}(L)) = -c_{\rm con}L + d_{\rm con}$ to check the area law. Fitting to $L \leq 4$, we obtain $c_{\rm con}= \cConPpointTwo$ and $c_{\rm con}= \cConPpointTwoFive$ for $P=0.2$ and $P=0.25$, respectively, in good agreement with the theoretical expectation, $1.724$.
The values of $d$ were $d_{\rm con}=\dConPpointTwo$ and $d_{\rm con}=\dConPpointTwoFive$. That they are not zero is not a problem; what we expect instead is \eqref{theory-con+mixed}, which we will confirm shortly.

\begin{figure}[htbp]
  \begin{center}
     \includegraphics[width=0.8\textwidth]{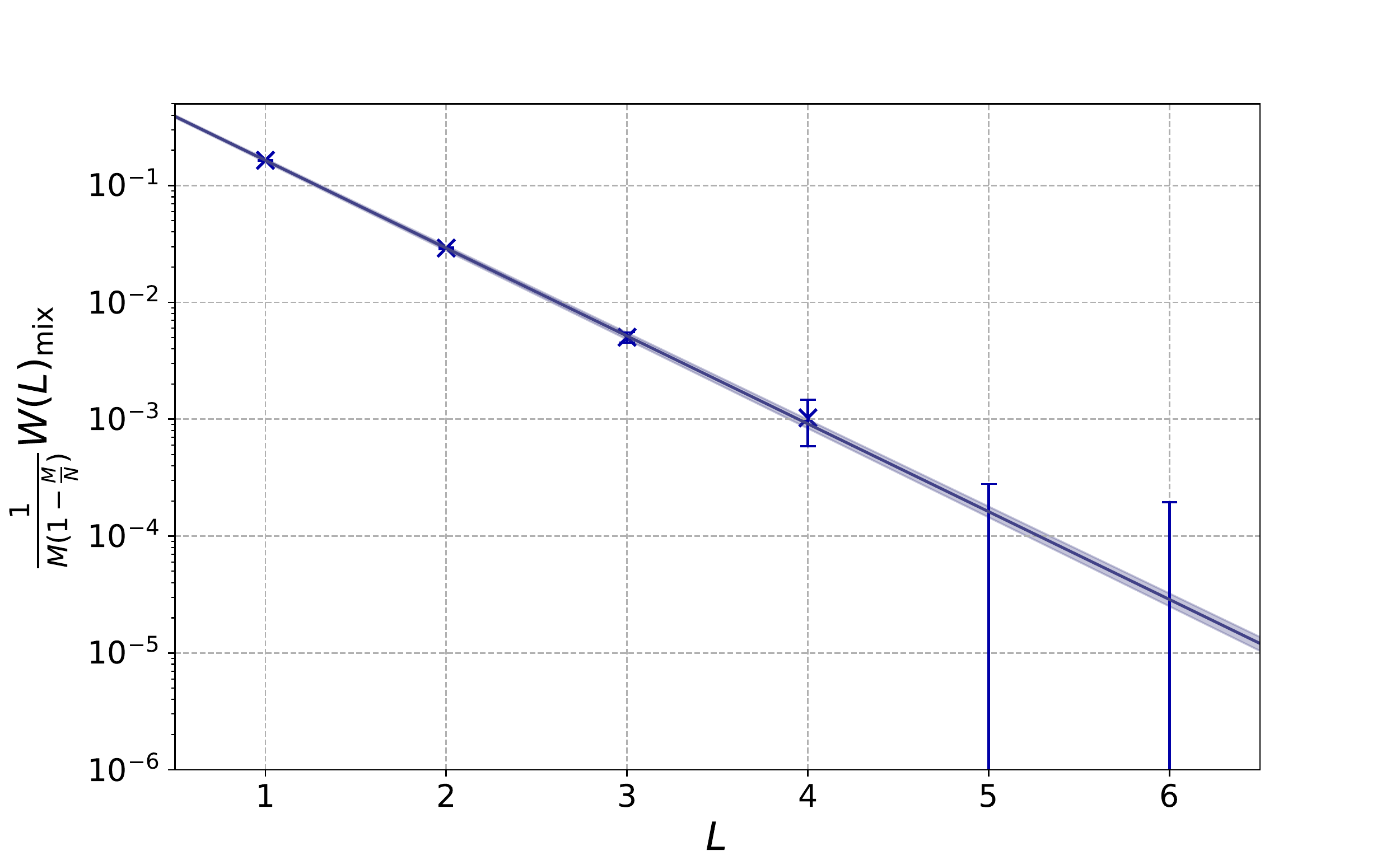}
     \includegraphics[width=0.8\textwidth]{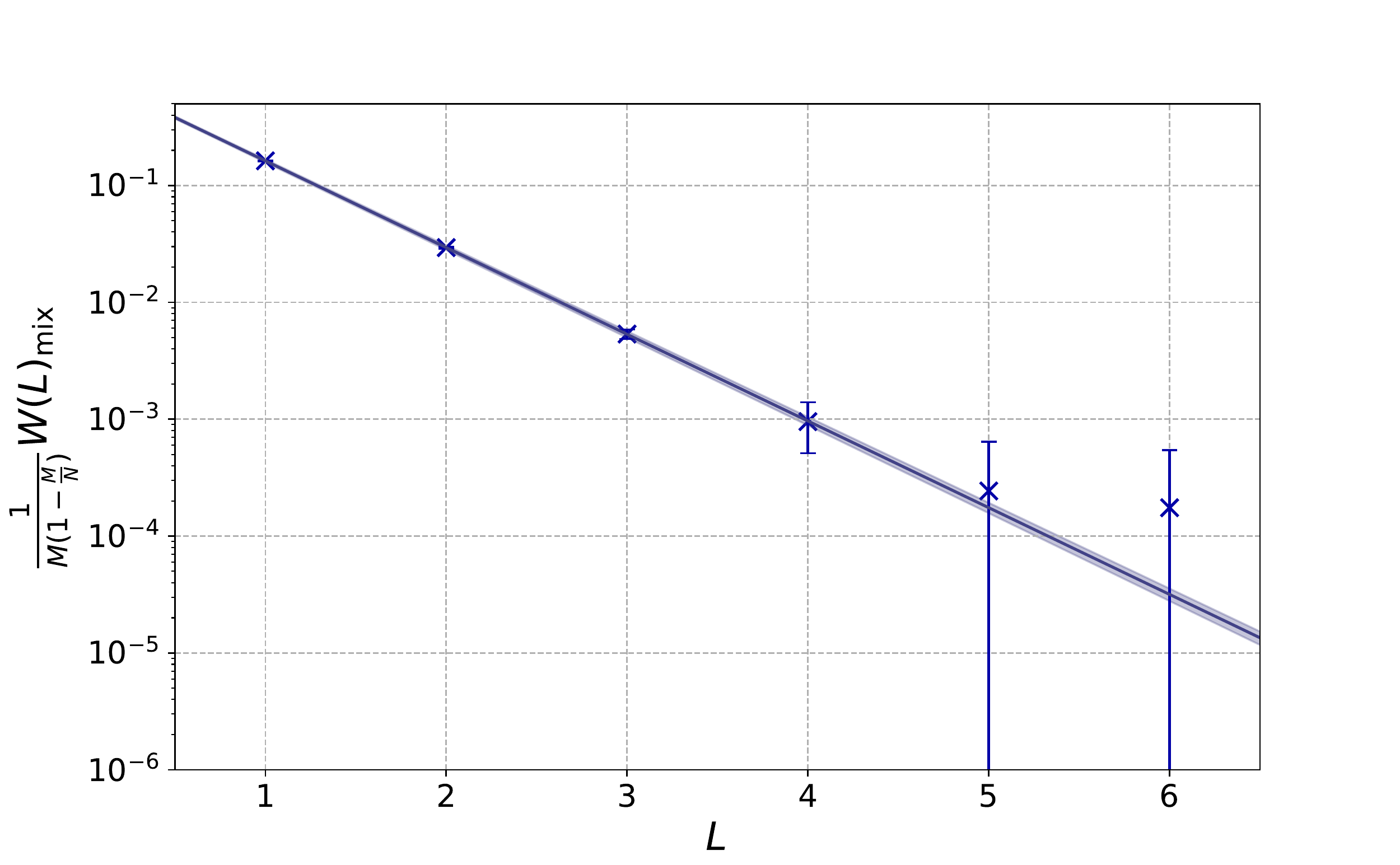}
  \end{center}
  \caption{
    Mixed subsector Wilson loop, $W_{\rm mix}$, at $T=0.29$, for $P=0.2$ and $P=0.25$, respectively. We have extrapolated to the large $N$ limit using $N=24, 32, 64$, and to the continuum limit using $n_t=16,24,32$ for each $N$, using the 2D interpolation $\frac{1}{M(1-\frac{M}{N})}W_{\rm mix}(L) = \gamma(L) + \frac{\beta_1}{N^2} + \frac{\beta_2}{n_t} + \frac{\beta_2}{n_t N^2}$. We fitted the interpolated values $\gamma(L)$ for $1\le L\le 4$ to the ansatz $\gamma(L)=\exp(-cL+d)$. We find $c_{\rm mix}=\cMixPpointTwo$ and $c_{\rm mix}=\cMixPpointTwoFive$ for $P=0.2$ and $P=0.25$, respectively. The fit is shown along with error bounds by the blue line.
  }\label{fig:W-mix-2D}
\end{figure}

The mixed-correlator $W_{\rm mix}(L)$ is shown in Fig.~\ref{fig:W-mix-2D}. 
Motivated by \eqref{theory-mixed} with $C_{\rm mix}(N,M)\simeq\frac{M}{N}\left(1-\frac{M}{N}\right)$, we normalized the loop as $\frac{1}{N}W_{\rm mix}(L)\times\left[\frac{M}{N}\left(1-\frac{M}{N}\right)\right]^{-1}$. We used a similar process as described for $W_{\rm con}$ above, but using this different normalisation. We find $c_{\rm mix}=\cMixPpointTwo$ and $c_{\rm mix}=\cMixPpointTwoFive$ for $P=0.2$ and $P=0.25$, respectively, in good agreement with the theoretical expectation, 1.724.
The values of $d_{\rm mix}$ were $d_{\rm mix}=\dMixPpointTwo$ and $d_{\rm mix}=\dMixPpointTwoFive$. Again, that they are not zero is not a problem; we will confirm \eqref{theory-con+mixed} next.

\begin{figure}[htbp]
  \begin{center}
     \includegraphics[width=0.8\textwidth]{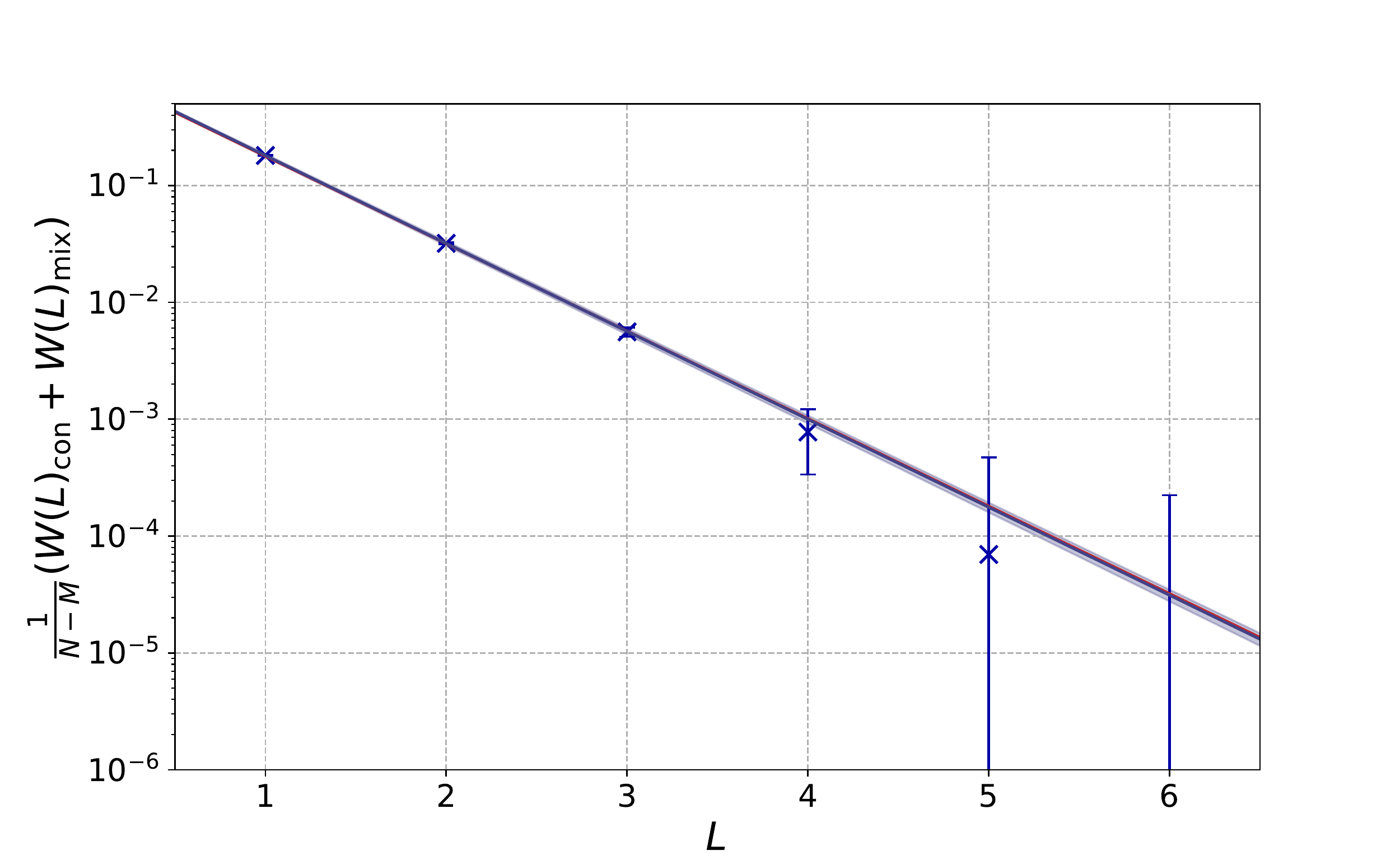}
     \includegraphics[width=0.8\textwidth]{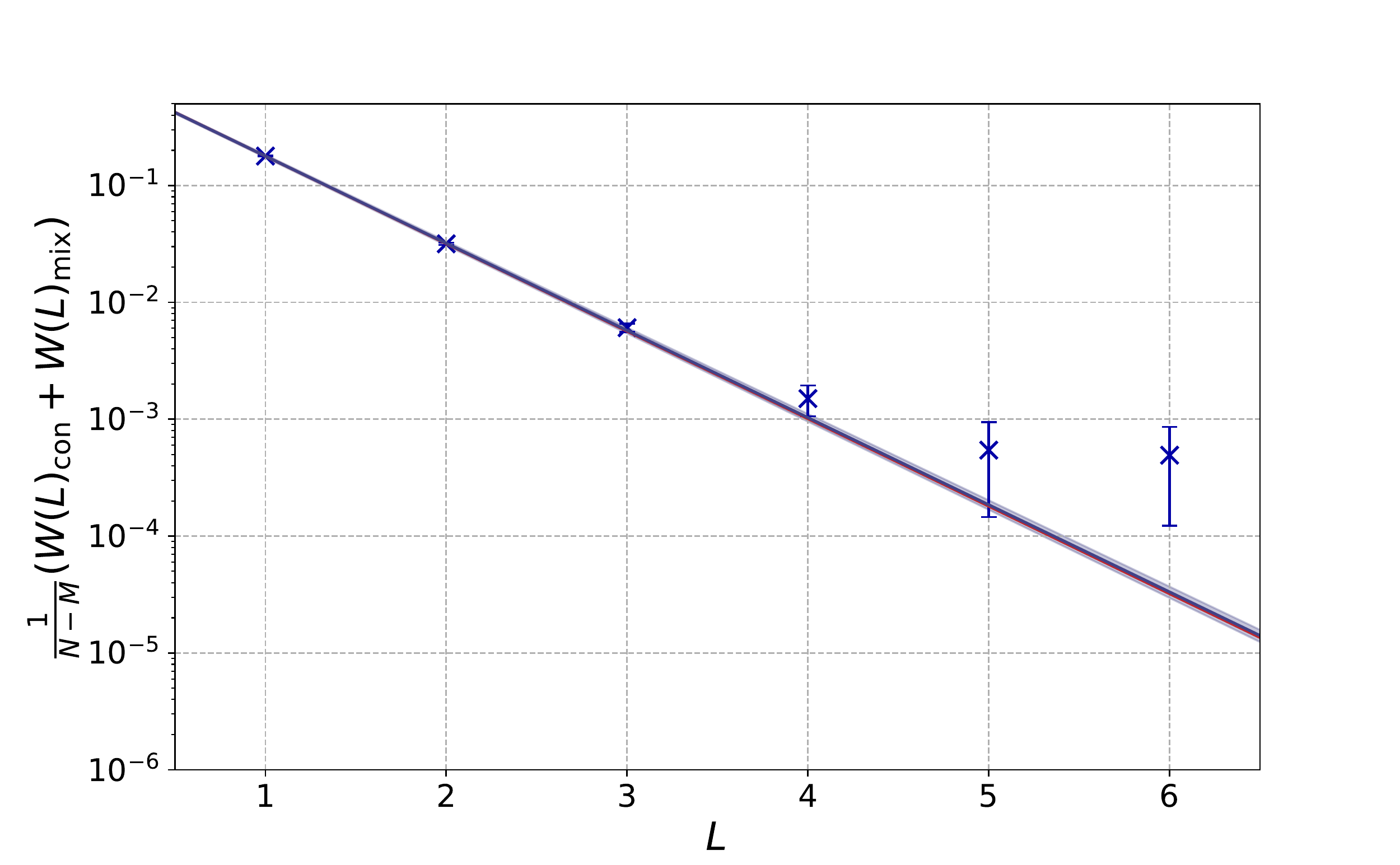}
  \end{center}
  \caption{
    Combined mixed and confined subsector Wilson loop, $W_{\rm con}+W_{\rm mix}$, at $T=0.29$, for $P=0.2$ and $P=0.25$, respectively. We have extrapolated to the large $N$ limit using $N=24, 32, 64$, and to the continuum limit using $n_t=16,24,32$ for each $N$, using the 2D interpolation $\frac{1}{M-N}(W_{\rm con} + W_{\rm mix})(L) = \gamma(L) + \frac{\beta_1}{N^2} + \frac{\beta_2}{n_t} + \frac{\beta_2}{n_t N^2}$. We fitted these interpolated values $\gamma(L)$ for $1\le L\le 4$ to the ansatz $\gamma(L)=\exp(-cL+d)$. We obtained $c=\cConPlusMixPpointTwo$ and $d=\dConPlusMixPpointTwo$ for $P=0.20$, and $c=\cConPlusMixPpointTwoFive$ and $d=\dConPlusMixPpointTwoFive$ for $P=0.25$. The fit is shown along with error bounds by the blue line. The theoretical prediction is shown by the red and is mostly obscured by the fit.
  }\label{fig:W-mix-con-2D}
\end{figure}

In Fig.~\ref{fig:W-mix-con-2D}, $\frac{W_{\rm con}+W_{\rm mix}}{N-M}$ is plotted. 
It is consistent with $e^{-1.724L}$ including the overall normalization factor. We obtained $c=\cConPlusMixPpointTwo$ and $d=\dConPlusMixPpointTwo$ for $P=0.20$, 
and $c=\cConPlusMixPpointTwoFive$ and $d=\dConPlusMixPpointTwoFive$ for $P=0.25$.
Now $d$ is consistent with zero.

\begin{figure}[htbp]
  \begin{center}
     \includegraphics[width=0.8\textwidth]{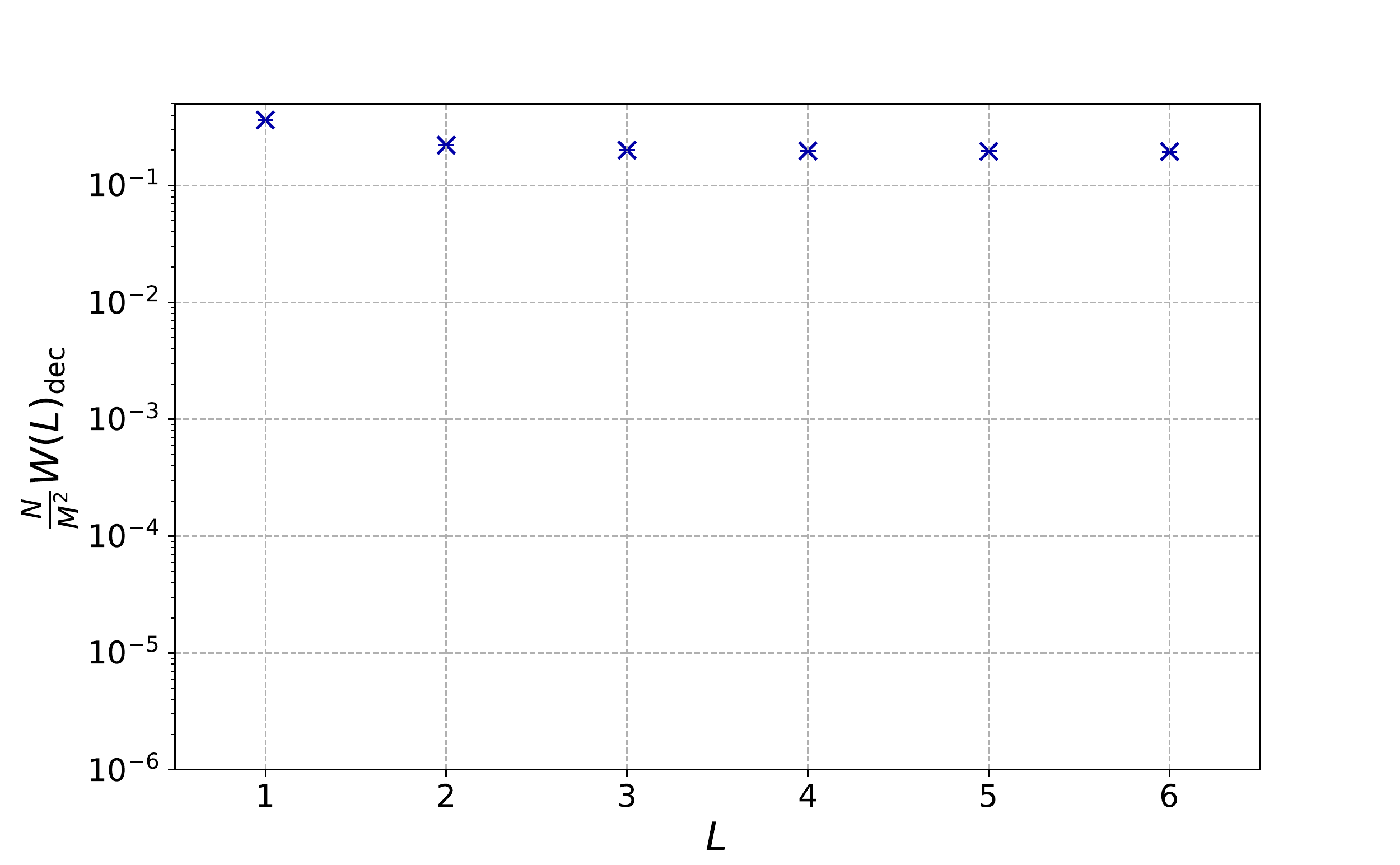}
     \includegraphics[width=0.8\textwidth]{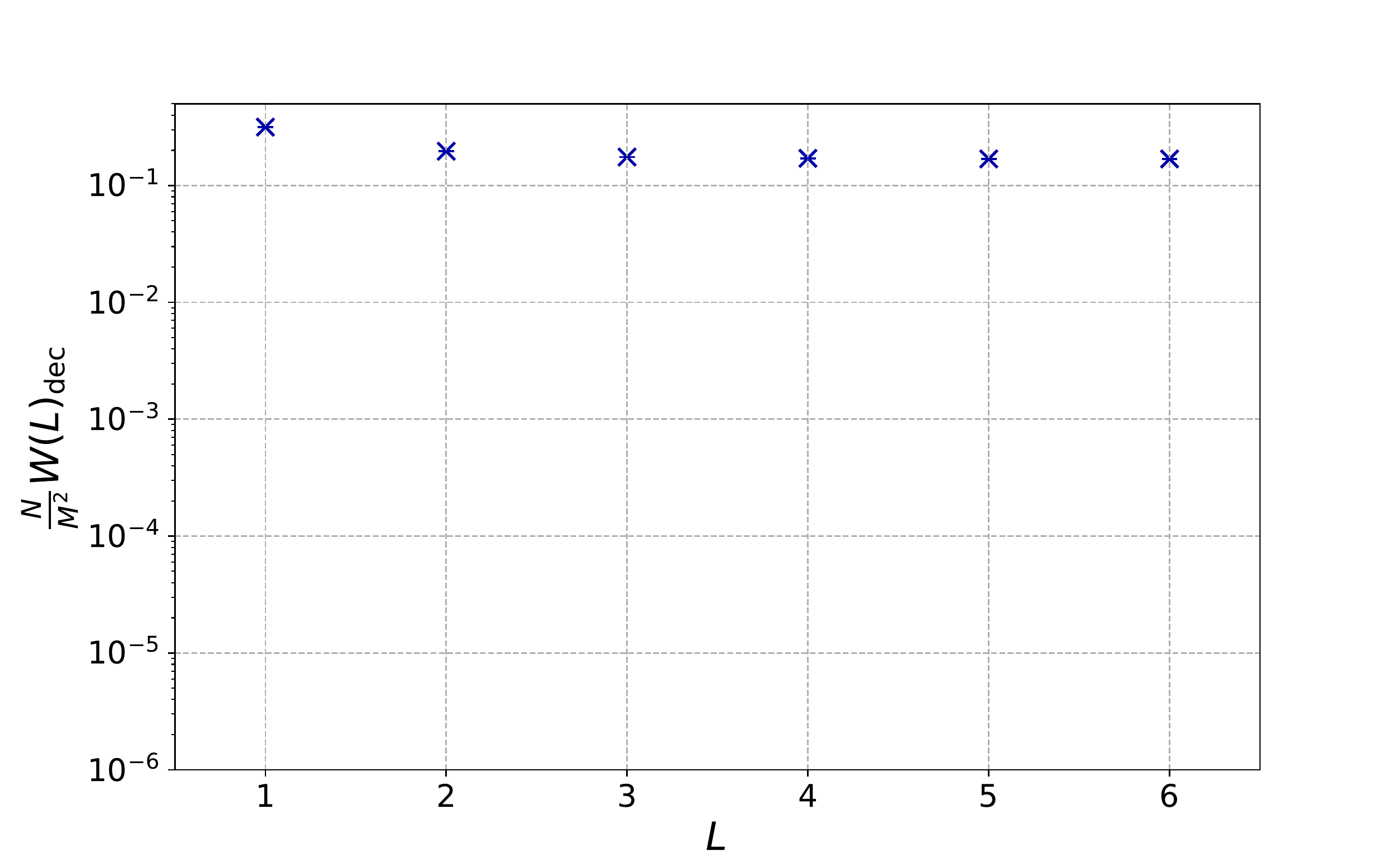}
  \end{center}
  \caption{
    Deconfined subsector Wilson loop, $W_{\rm dec}$, at $T=0.29$, for $P=0.2$ and $P=0.25$, respectively. We used the same scale as in Figs.~\ref{fig:W-con-2D}, \ref{fig:W-mix-2D}, and \ref{fig:W-mix-con-2D}. The potential is clearly nonconfining, and no fitting was attempted}
    \label{fig:W-dec-2D}
\end{figure}

To contrast with all of these, we also plotted the deconfined-correlator, $W_{\rm dec}$, in Fig.~\ref{fig:W-dec-2D}. Here, there is a total absence of any confining behaviour, in easy agreement with our conjecture.

In summary, we have observed that numerical data is consistent with nontrivial theoretical predictions made about partial confinement, i.e., \eqref{theory-confined}, \eqref{theory-mixed}, and \eqref{theory-con+mixed}. 
In particular, the agreement between $W_{\rm con}+W_{\rm mix}$ and the theoretical prediction could be confirmed without even performing a fit, as shown in Fig.~\ref{fig:W-mix-con-2D}. Although our data is not good enough to determine the values of Wilson loops at $L\ge 5$ at this moment, in principle we can study arbitrary large $L$ by taking $N$ larger and collecting sufficiently many statistics in Monte Carlo simulations.

\subsubsection*{Comments on temperature dependence}
In the model under consideration, the partially-deconfined saddle is the maximum of the free energy. The distribution of Polyakov line phases at the saddle changes with temperature. Therefore, strictly speaking, we need to study multiple values of $T$, choosing the value of $P$ exactly on top of the saddle. However, we explicitly confirmed that the expected behaviors \eqref{theory-confined}, \eqref{theory-mixed}, and \eqref{theory-con+mixed} at $T=0.29$ at $P=0$, $0.20$ and $0.25$, which suggests that this relation holds near the critical point regardless of the value of $P$, and hence also on the partially-deconfined saddle. It is straightforward to perform the more complete analyses if more computational resources are available. We will leave it for future work.   
See also comments at the end of Sec.~\ref{sec:determining-M}.
\section{Conclusions and discussions}
\hspace{0.51cm}
In this paper, we presented evidence for the formation of a flux tube and linear confinement potential in the confined sector of the partially-deconfined saddle of pure Yang-Mills theory by taking strongly-coupled lattice gauge theory as a concrete example. The linear confinement potential in the confined sector supports the argument given in Ref.~\cite{Hanada:2019kue} concerning the 't Hooft anomaly matching associated with the chiral symmetry breaking in the confined sector. Namely, quarks in the confined sector must form a bound state, and then the pion should be formed in the confined sector so that the anomaly-matching condition is satisfied. It can also explain the observation in Ref.~\cite{Hanada:2021ksu}, i.e., chiral symmetry breaks spontaneously at the complete deconfinement/partial deconfinement phase transition point, because the formation of chiral condensate in the confined sector is sufficient to break the chiral symmetry.

By studying large-$N$ QCD in the Veneziano limit with sufficiently light quarks, we should be able to see the stable partially-deconfined phase. It would be interesting to see whether chiral symmetry breaking and the complete deconfinement/partial deconfinement phase transition coincide. Another interesting direction would be to explore whether partial confinement is a sharp notion for SU(3) real-world QCD (see Ref.~\cite{Hanada:2021ksu}).

Our result means that, in the partially-deconfined phase, the probe quark-antiquark potential behaves differently in the confined and deconfined sectors. 
In the holographic calculation of the Wilson loop~\cite{Maldacena:1998im}, this would mean that different worldsheets are preferred in the confined and deconfined sectors. We have not understood how it affects the conventional computations on the gravity side.  

The original motivation~\cite{Hanada:2016pwv} for considering partial deconfinement was to understand the emergent geometry in gauge/gravity duality. Naturally, the deconfined sector is interpreted as a black hole. The findings in this paper are consistent with the expectation that short strings, which correspond to gravitons when there is a gravity dual, can live in the confined sector~\cite{Hanada:2021ipb,Hanada:2021swb,Gautam:2022akq}. When applied to the evaporating black hole, the Page curve~\cite{Page:1993wv} follows from this interpretation~\cite{Gautam:2022akq}. It would be interesting if we can deepen our understanding of the roles matrix degrees of freedom play in the emergent geometry.

\begin{center}
\textbf{Acknowledgement}
\end{center}
We thank Georg Bergner, Thomas Cohen, Robert Pisarski, and Hiromasa Watanabe for useful discussions and comments.
V.G. thanks STFC for Doctoral Training Programme funding (ST/W507854-2021 Maths DTP).
M.H. thanks the STFC Ernest Rutherford Grant ST/R003599/1.
J.H. thanks STFC for the support of a studentship.
E.R. is supported by Nippon Telegraph and Telephone Corporation (NTT) Research.
\begin{center}
\textbf{Data management}
\end{center}
No additional research data beyond the data presented and cited in this work are needed to validate the research findings in this work. Simulation data will be publicly available after publication on Zenodo.

\appendix

\section{Constrained simulations of the first kind and second kind}\label{appendix:constraint-sanity-check}

\begin{figure}[htbp]
  \begin{center}
     \includegraphics[width=0.8\textwidth]{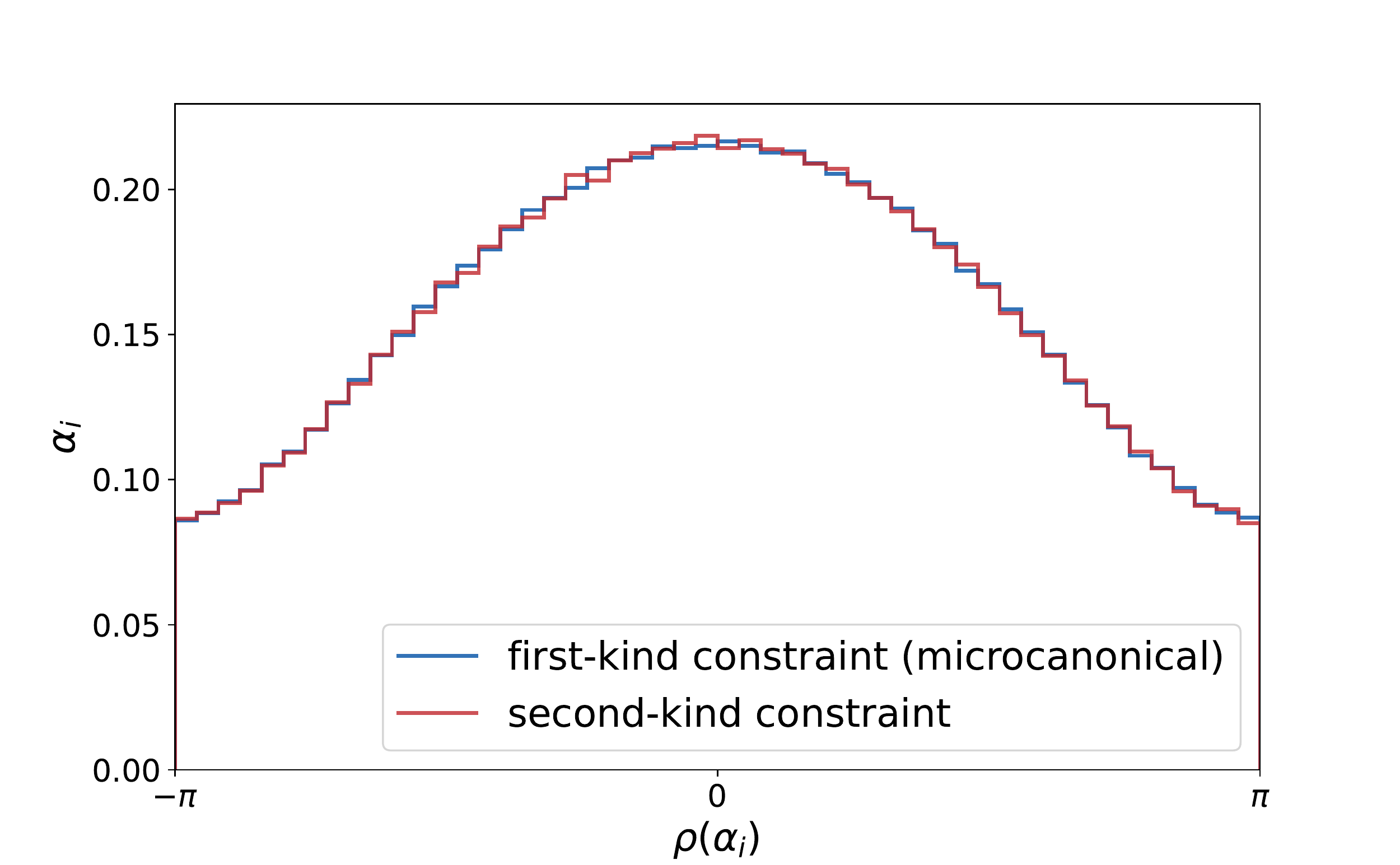}
  \end{center}
  \caption{
    Polyakov loop phase distributions for $N=64$, $n_t=32$, $P=0.2$, for constraints of the first and second kind.}
  \label{fig:polyakov-distributions-compared}
\end{figure}

\begin{figure}[htbp]
    \centering
     \begin{subfigure}[b]{0.3\textwidth}
         \centering
         \includegraphics[width=\textwidth]{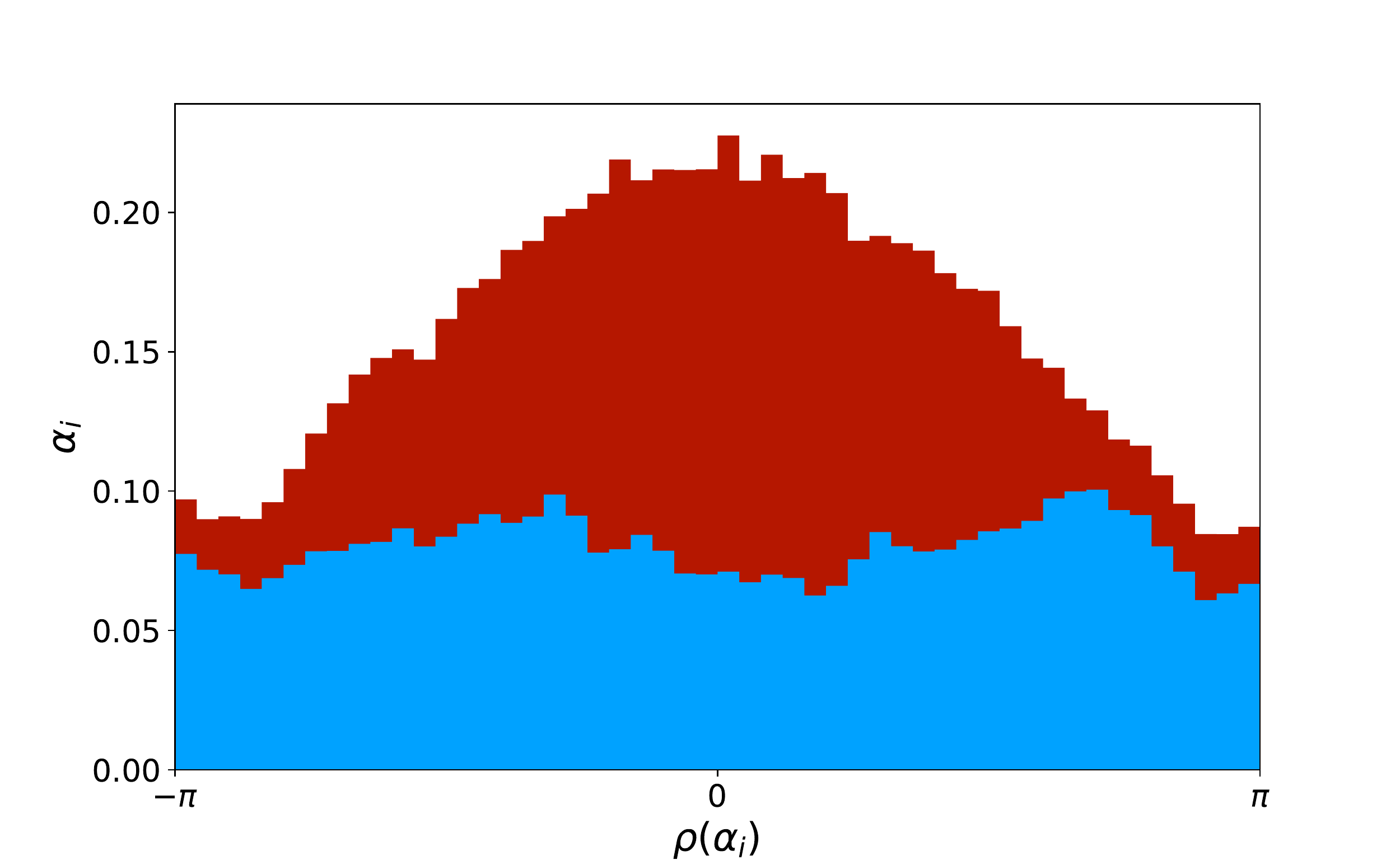}
         \caption{}
     \end{subfigure}
         \begin{subfigure}[b]{0.3\textwidth}
         \centering
         \includegraphics[width=\textwidth]{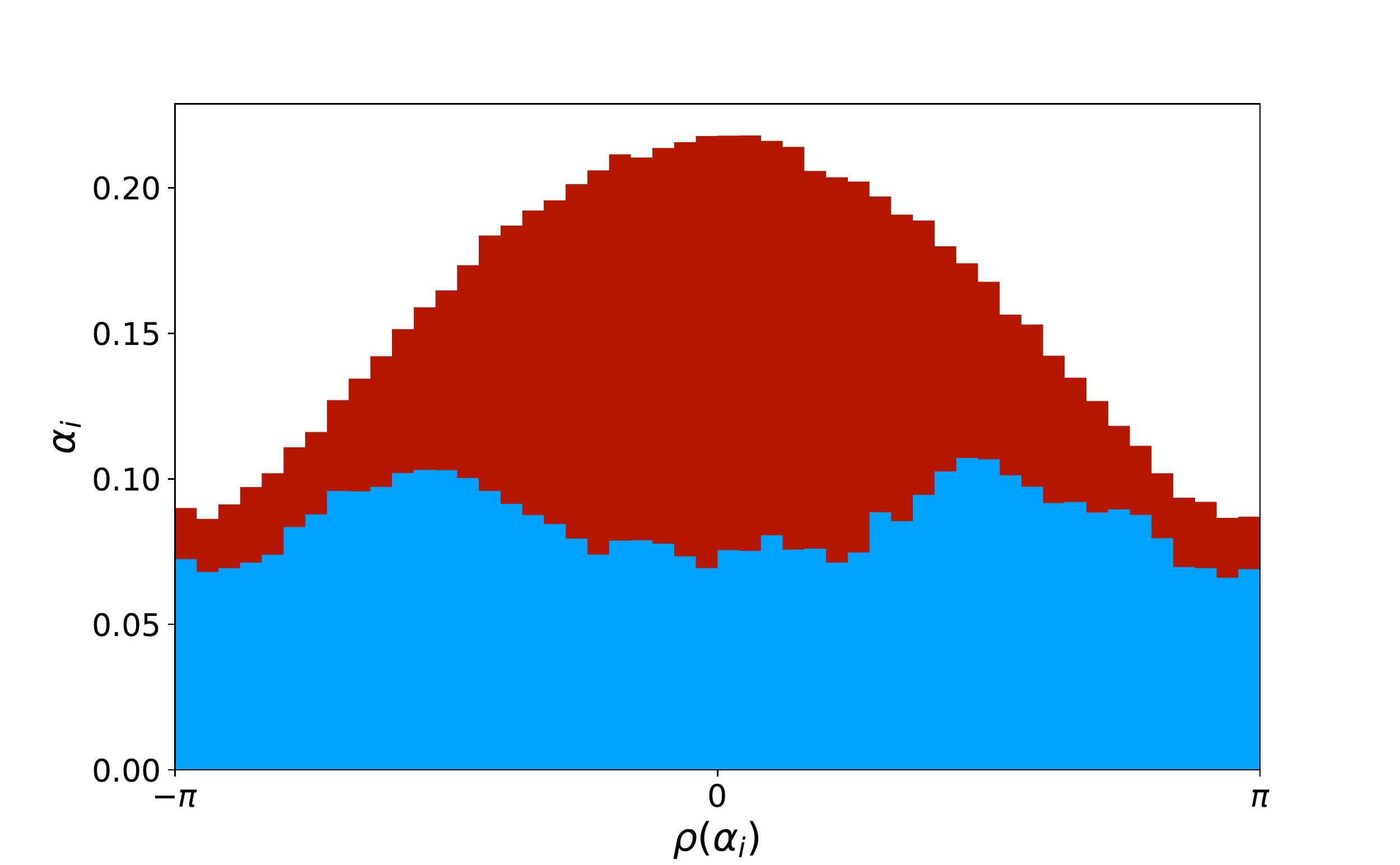}
         \caption{}
     \end{subfigure}
     \begin{subfigure}[b]{0.3\textwidth}
         \centering
         \includegraphics[width=\textwidth]{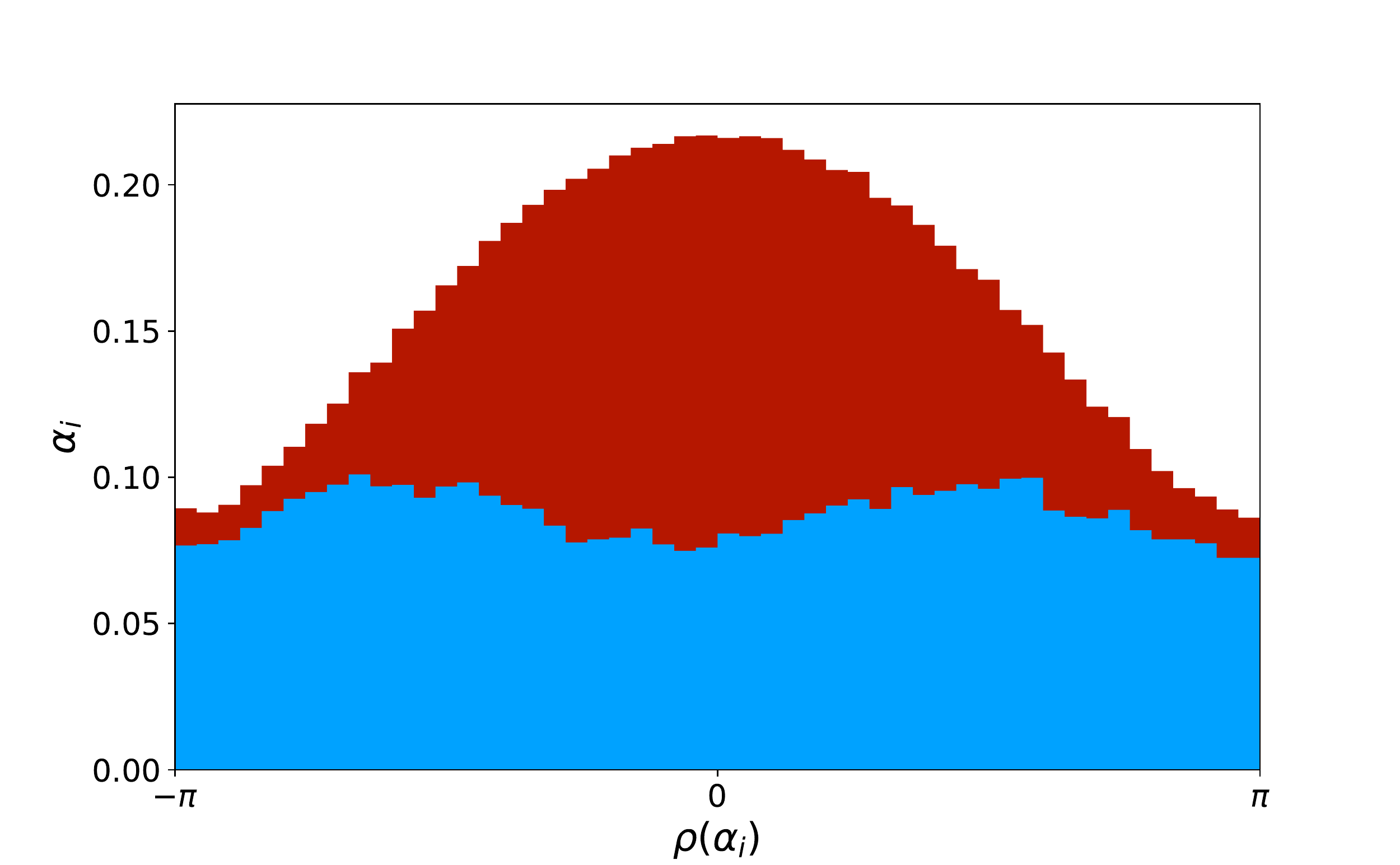}
         \caption{}
     \end{subfigure}

\begin{subfigure}[b]    {0.3\textwidth}
         \centering
         \includegraphics[width=\textwidth]{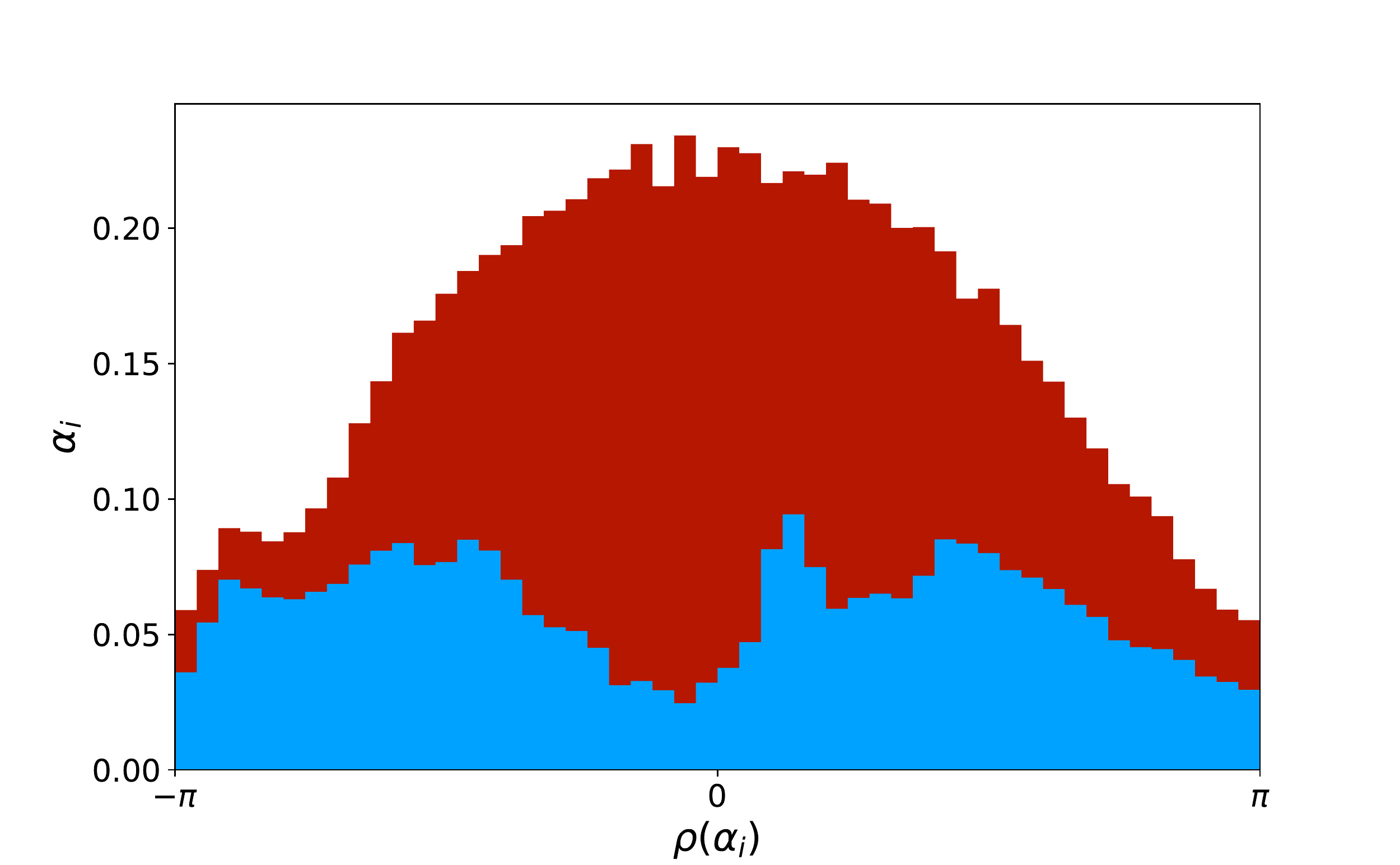}
         \caption{}
     \end{subfigure}
         \begin{subfigure}[b]{0.3\textwidth}
         \centering
         \includegraphics[width=\textwidth]{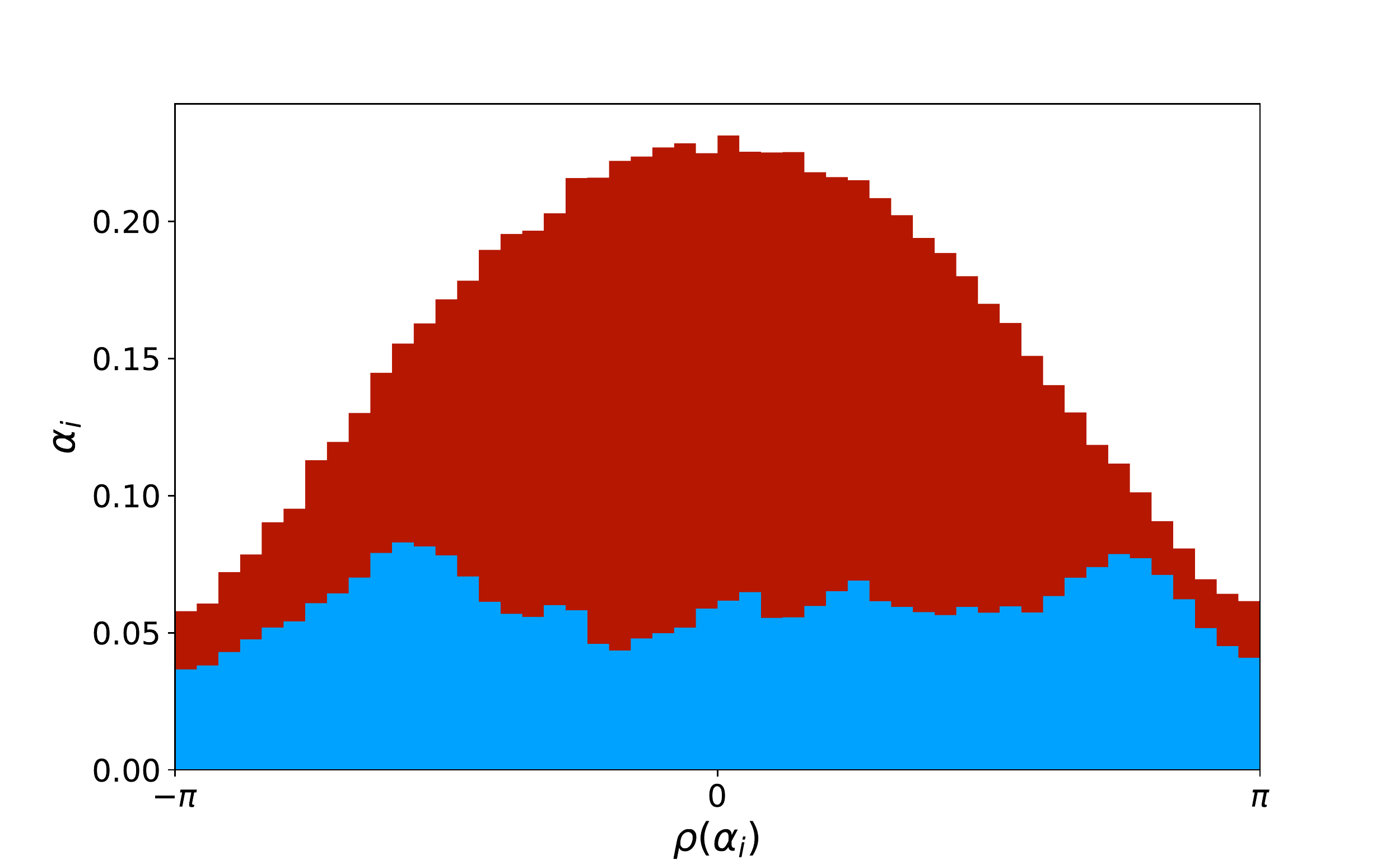}
         \caption{}
     \end{subfigure}
     \begin{subfigure}[b]{0.3\textwidth}
         \centering
         \includegraphics[width=\textwidth]{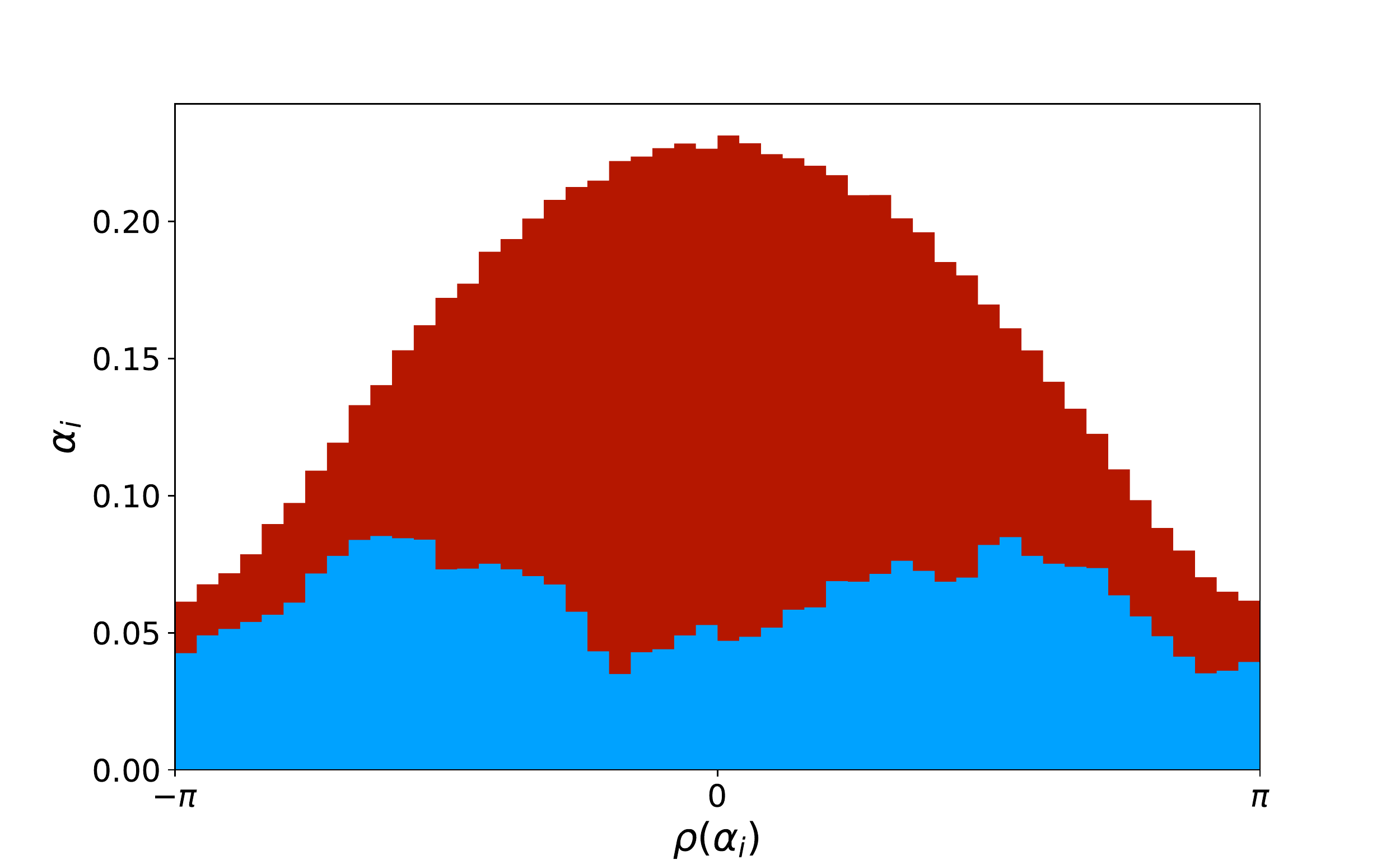}
         \caption{}
     \end{subfigure}
  \caption{
    Polyakov loop phase distributions for $N=32$ and $N=64$, $n_t=24$, $P=0.2$ (top) and $P=0.25$ (bottom), for constraints of the second kind. In addition to the distribution of all phases, the `deconfined part' ($\alpha_1,\cdots,\alpha_M$) and `confined part' ($\alpha_{M+1},\cdots,\alpha_N$) are shown. 
  }\label{fig:phase-splitting-2nd-constraint}
\end{figure}

A potential objection to our approach might be the apparent unnaturalness of the constraint simulation of the second kind.
Here, we justify the second constraint as being equivalent to fixing some of the gauge redundancy that remains after having taken static diagonal gauge. The second constraint imposes uniformity on some subset $N-M$ of the eigenvalues of the Polyakov loop. In a general configuration at finite $N$ outputted by the Monte-Carlo simulation under the constraint of the first kind, a subset of eigenvalues is not guaranteed to be exactly uniformly distributed in this way. However, at large $N$, the constraint of the second-kind and fixing of residual gauge symmetry should be equivalent, with fluctuations away from the mean value being suppressed. If the Polyakov eigenvalue distributions are identical between the microcanonical (first-kind constrained) and second-constrained simulation, we can use our method as an effective gauge fixing in the large-$N$ limit. As we can see from Fig.~\ref{fig:polyakov-distributions-compared}, constrained simulations with the 1st and 2nd constraints indeed give the same distribution of the Polyakov-loop phases. Furthermore, as we can see Fig.~\ref{fig:phase-splitting-2nd-constraint}, the distributions of $\alpha_1,\cdots,\alpha_M$ and $\alpha_{M+1},\cdots,\alpha_N$ exhibit the kind splitting depicted in Fig.~\ref{fig:phase-distribution}.

We can also check that the confined Polyakov loop correlator exhibits an area law in the microcanonical ensemble, without the imposition of uniformity on the confined subsector that is done in the constrained simulations of the second kind. We reasoned that the Wilson loop obtained in the constrained simulation of the first kind must be the sum of the contributions from $W_{\rm con}$, $W_{\rm dec}$ and $W_{\rm mix}$, and that the constrained simulation of second kind separates these pieces neatly. One may wonder whether this method of explicit separation does something more and spoils the partially-deconfined saddle. As a consistency check, we can subtract the deconfined-sector Wilson loop computed from the second-kind simulations from the first-kind Wilson loop to obtain a quantity with an area law, as expected from the confined and mixed sectors. More advantageously, we can calculate,
\begin{align} \label{eq:W_subtracted}
    W_{\rm subtracted} =   W_{\rm full, 1st} - W_{\rm dec, 2nd}- W_{\rm mix, 2nd}. 
\end{align}
This $W_{\rm subtracted}$ should be the sum of $W_{\rm con}$ and $W_{\rm mix}$, and hence, $W_{\rm subtracted}=(N-M)\exp(-\sigma L\beta)$ is expected. 

This demonstrates that the linear potential is not the result of an artifical imposition of uniformity on the eigenvalues, but is present even in the constrained simulations of the first kind. We also find by this analysis that the constraint of the second kind approximates the gauge-fixed microcanonical ensemble well even at $N=16$.

\begin{figure}[htbp]
  \begin{center} 
     \includegraphics[width=1.0\textwidth]{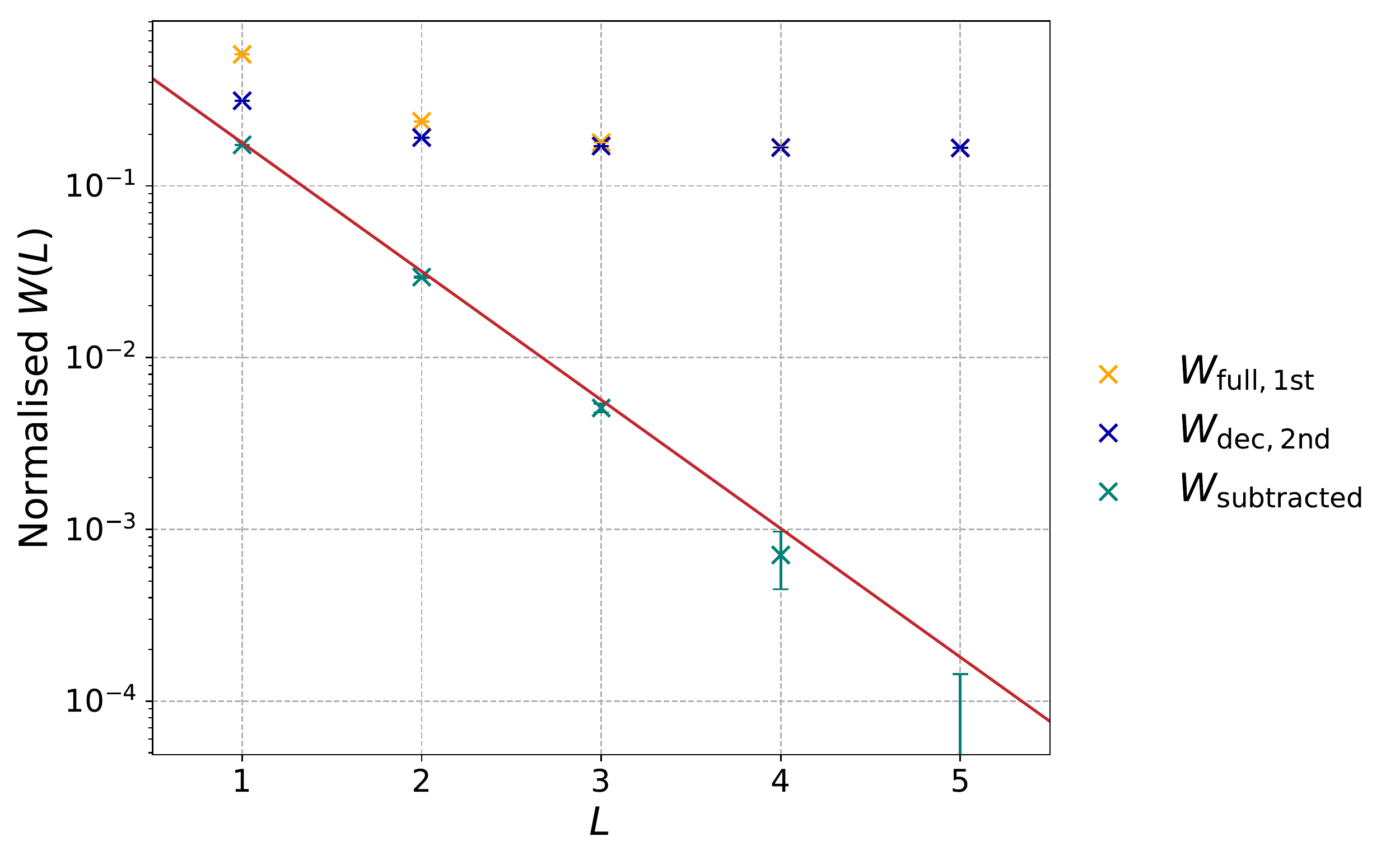}
  \end{center}
  \caption{
    The complete Wilson loop from first-kind simulations, the deconfined Wilson loop from second-kind simulations, and their difference, $n_t=32$ and $T=0.29$, extrapolated to large $N$. Note that the continuum limit has not been taken.
  }\label{fig:subtraction-method}
\end{figure}

In Fig.~\ref{fig:subtraction-method}, we plotted $W_{\rm subtracted}$. As expected, by subtracting the contribution from the deconfined sector, we find confining behaviour matching that of the mixed and confined subgroups from second-kind simulations.

\section{Stability of the extrapolations}\label{appendix:stability_extrapolations}
\hspace{0.51cm}

\subsubsection*{Small $N$ effects}
We collected data for $N=16,24,32,64$. For the large-$N$ extrapolation of $W_{\rm con}$ (respectively, $W_{\rm mix}$), we should use only the data points with sufficiently large $N-M$ (respectively, $M$ and $N-M$). However, for $P=0.25$, the $N=16$ data seemed to present a source of systematic error from finite $N$, $M$, and $N-M$ effects. In particular, for $N=16$ and $P=0.25$ we have $N-M=6$. Such errors would not be captured by the error bars. We can measure the contribution to the systematic error by removing data at different values of $N$ from the interpolation. The effect of this on the fitting parameter $c$ is shown in Tables~\ref{con+mix-with-nt=16},~\ref{con-with-nt=16}, and~\ref{mix-with-nt=16}.

\begin{figure}[htbp]
  \begin{center} 
     \includegraphics[width=0.49\textwidth]{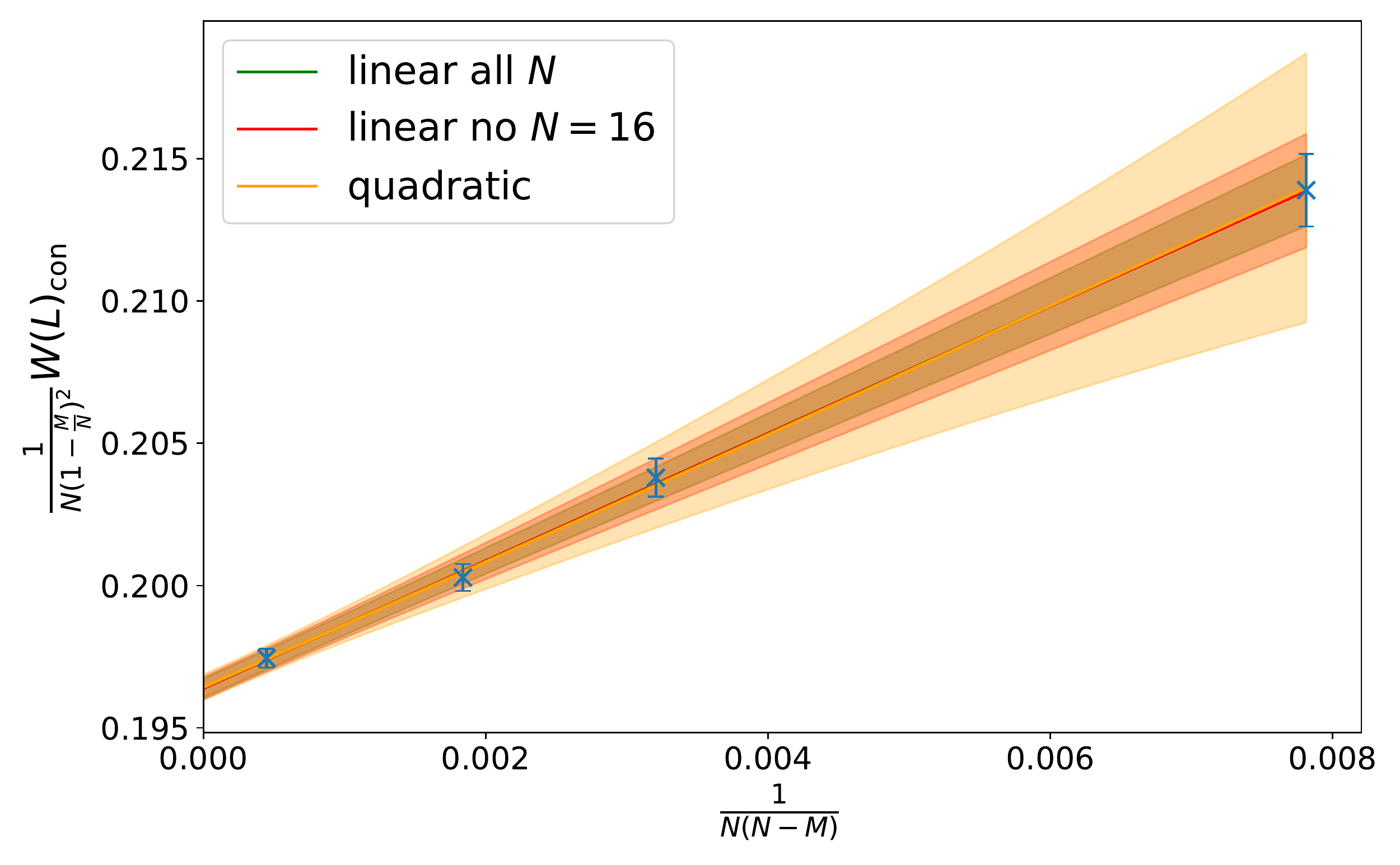}
     \includegraphics[width=0.49\textwidth]{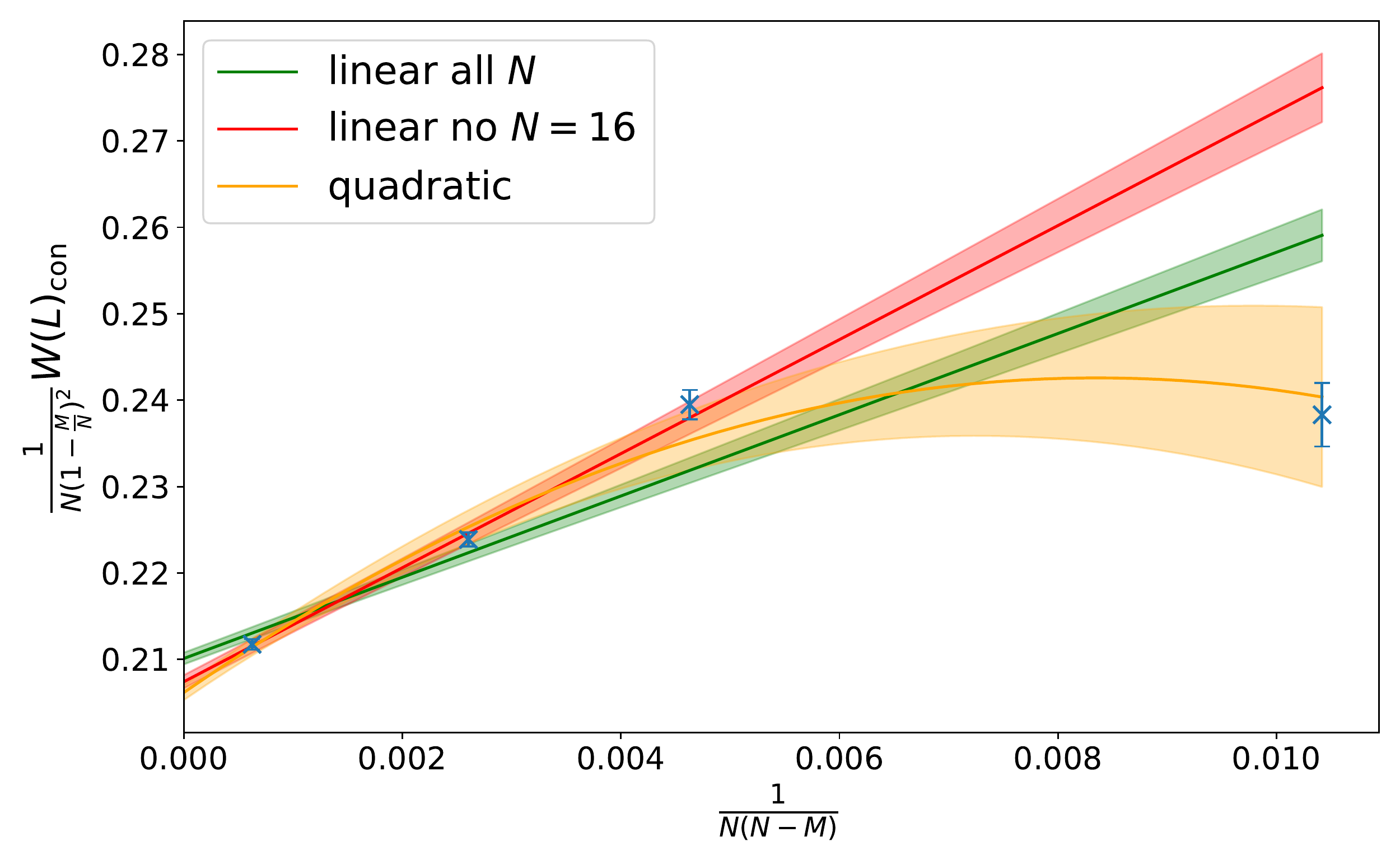}
  \end{center}
  \caption{
    Performing the large $N$ extrapolation for the confined subsector $P=0.2$ (left) and $P=0.25$ (right) for $L=1$. The continuum limit $n_t \rightarrow \infty$ has been taken. We try a linear fit of all points (green), a linear fit that ignores $N=16$ (red), and a quadratic fit of all points (orange). The destabilising effect of the $N=16$ (rightmost) point in the $P=0.25$ linear extrapolation is obvious and can be attributed to the small value $N-M=6$.
  }\label{fig:N16_removal_analysis}
\end{figure}

It is easy to understand how the $N=16$ data is distorting the results when we attempt a linear fit. We demonstrate this in Fig.~\ref{fig:N16_removal_analysis}. The small value of $N-M$ causes the $N=16$ point to be far separated from the others on the x axis. Consequently, it has a large effect on the linear fit. There is also clear non-linear behaviour. For $P=0.2$, $N-M$ is larger, and this explains why the distortion is not so severe.
We could have attempted a quadratic fit, but we see from Fig.~\ref{fig:N16_removal_analysis} that this does not significantly reduce the uncertainty, and potentially increases it.
Therefore, we performed the large-$N$ extrapolations in the main text without using $N=16$. 
\subsubsection*{Small $n_t$ effects}
We see a similar issue with respect to $n_t=16$. In Table~\ref{con+mix-with-nt=16} and Table~\ref{con-with-nt=16}, we see that $c$ for $P=0.25$ is smaller than the theoretical value $1.724$, whether $N=16$ is removed or not. We attribute this to the use of $n_t=16$. If we plot points at $n_t=16,24$ and $32$ with fixed $N$ and $L$ by taking $\frac{1}{n_t}$ as the horizontal axis, we can see that those three points are not always on a straight line and a linear fit is often very bad.
This happens for both $P=0.2$ and $P=0.25$. For the analyses in the main text, we did not include $n_t=16$ for either $P=0.2$ or $P=0.25$. The effect of removing $n_t=16$ can be seen by comparing Tables~\ref{con+mix-with-nt=16},~\ref{con-with-nt=16},~\ref{mix-with-nt=16} with Tables~\ref{con+mix-without-nt=16},~\ref{con-without-nt=16},~\ref{mix-without-nt=16}.

\begin{table}
    \centering
    \begin{tabular}{|c|c|c|c|c|c|}\hline%
    $P$ & all $N$ & rmv. $N=16$ & rmv. $N=24$ & rmv. $N=32$ & rmv. $N=64$ \\\hline
    \begin{filecontents*}{tables/c_con_plus_mix_removeN.csv}
P,all $N$,rmv. $N=16$,rmv. $N=24$,rmv. $N=32$,rmv. $N=64$
$0.2$,1.7248(59),1.7343(74),1.7271(59),1.7221(64),1.718(11)
$0.25$,1.7117(64),1.7215(82),1.7122(64),1.7133(72),1.692(12)
\end{filecontents*}
\csvreader[head to column names, late after line=\\]
    {tables/c_con_plus_mix_removeN.csv}{}{\csvcoli & \csvcolii & \csvcoliii & \csvcoliv & \csvcolv & \csvcolvi}%
    \hline
    \end{tabular}
    \caption{Values of $c$ for $W_{\rm con}+W_{\rm mix}$ with data for different $N$ removed. With $n_t=16$.}\label{con+mix-with-nt=16}
\end{table}

\begin{table}
    \centering
    \begin{tabular}{|c|c|c|c|c|c|}\hline%
    $P$ & all $N$ & rmv. $N=16$ & rmv. $N=24$ & rmv. $N=32$ & rmv. $N=64$ \\\hline
    \begin{filecontents*}{tables/c_con_only_removeN.csv}
P,all $N$,rmv. $N=16$,rmv. $N=24$,rmv. $N=32$,rmv. $N=64$
$0.2$,1.7281(55),1.7335(66),1.7288(55),1.7229(58),1.729(11)
$0.25$,1.7170(78),1.732(10),1.7193(79),1.7195(86),1.682(15)
\end{filecontents*}
\csvreader[head to column names, late after line=\\]
    {tables/c_con_only_removeN.csv}{}{\csvcoli & \csvcolii & \csvcoliii & \csvcoliv & \csvcolv & \csvcolvi}%
    \hline
    \end{tabular}
    \caption{Values of $c_{\rm con}$ for $W_{\rm con}$ with data for different $N$ removed. With $n_t=16$.}\label{con-with-nt=16}
\end{table}

\begin{table}
    \centering
    \begin{tabular}{|c|c|c|c|c|c|}\hline%
    $P$ & all $N$ & rmv. $N=16$ & rmv. $N=24$ & rmv. $N=32$ & rmv. $N=64$ \\\hline
    \begin{filecontents*}{tables/c_mix_removeN.csv}
P,all $N$,rmv. $N=16$,rmv. $N=24$,rmv. $N=32$,rmv. $N=64$
$0.2$,1.7340(67),1.7346(83),1.7341(68),1.7338(74),1.737(13)
$0.25$,1.7066(72),1.7116(91),1.7059(72),1.7062(80),1.692(14)
\end{filecontents*}
\csvreader[head to column names, late after line=\\]
    {tables/c_mix_removeN.csv}{}{\csvcoli & \csvcolii & \csvcoliii & \csvcoliv & \csvcolv & \csvcolvi}%
    \hline
    \end{tabular}
    \caption{Values of $c_{\rm mix}$ for $W_{\rm mix}$ with data for different $N$ removed. With $n_t=16$.}\label{mix-with-nt=16}
\end{table}

\begin{table}
    \centering
    \begin{tabular}{|c|c|c|c|c|c|}\hline%
    $P$ & all $N$ & rmv. $N=16$ & rmv. $N=24$ & rmv. $N=32$ & rmv. $N=64$ \\\hline
    \begin{filecontents*}{tables/c_con_plus_mix_removeN_removed_nt16.csv}
P,all $N$,rmv. $N=16$,rmv. $N=24$,rmv. $N=32$,rmv. $N=64$
$0.2$,1.730(15),1.733(19),1.730(15),1.729(17),1.729(28)
$0.25$,1.703(14),1.718(18),1.703(14),1.709(16),1.663(27)
\end{filecontents*}
\csvreader[head to column names, late after line=\\]
    {tables/c_con_plus_mix_removeN_removed_nt16.csv}{}{\csvcoli & \csvcolii & \csvcoliii & \csvcoliv & \csvcolv & \csvcolvi}%
    \hline
    \end{tabular}
    \caption{Values of $c$ for $W_{\rm con}+W_{\rm mix}$ with data for different $N$ removed. Without $n_t=16$.}\label{con+mix-without-nt=16}
\end{table}

\begin{table}
    \centering
    \begin{tabular}{|c|c|c|c|c|c|}\hline%
    $P$ & all $N$ & rmv. $N=16$ & rmv. $N=24$ & rmv. $N=32$ & rmv. $N=64$ \\\hline
    \begin{filecontents*}{tables/c_con_only_removeN_removed_nt16.csv}
P,all $N$,rmv. $N=16$,rmv. $N=24$,rmv. $N=32$,rmv. $N=64$
$0.2$,1.742(14),1.729(17),1.741(14),1.737(16),1.779(29)
$0.25$,1.701(18),1.724(23),1.702(18),1.707(19),1.596(35)
\end{filecontents*}
\csvreader[head to column names, late after line=\\]
    {tables/c_con_only_removeN_removed_nt16.csv}{}{\csvcoli & \csvcolii & \csvcoliii & \csvcoliv & \csvcolv & \csvcolvi}%
    \hline
    \end{tabular}
    \caption{Values of $c_{\rm con}$ for $W_{\rm con}$ with data for different $N$ removed. Without $n_t=16$.}\label{con-without-nt=16}
\end{table}

\begin{table}
    \centering
    \begin{tabular}{|c|c|c|c|c|c|}\hline%
    $P$ & all $N$ & rmv. $N=16$ & rmv. $N=24$ & rmv. $N=32$ & rmv. $N=64$ \\\hline
    \begin{filecontents*}{tables/c_mix_removeN_removed_nt16.csv}
P,all $N$,rmv. $N=16$,rmv. $N=24$,rmv. $N=32$,rmv. $N=64$
$0.2$,1.721(17),1.731(21),1.719(17),1.731(19),1.691(31)
$0.25$,1.702(16),1.708(20),1.702(16),1.703(17),1.703(31)
\end{filecontents*}
\csvreader[head to column names, late after line=\\]
    {tables/c_mix_removeN_removed_nt16.csv}{}{\csvcoli & \csvcolii & \csvcoliii & \csvcoliv & \csvcolv & \csvcolvi}%
    \hline
    \end{tabular}
    \caption{Values of $c_{\rm mix}$ for $W_{\rm mix}$ with data for different $N$ removed. Without $n_t=16$.}\label{mix-without-nt=16}
\end{table}

\section{Gauge fixing condition and separation of confined and deconfined sectors}\label{appendix:gauge_fixing}
\hspace{0.51cm}
As a starter, let us recall how the operator formalism and path-integral formalism are related (see e.g.,~\cite{Rinaldi:2021jbg} for the case of Hermitian variables). 
The thermal partition function of the Eguchi-Kawai model is 
\begin{align}
Z(T)
&=
\frac{1}{[{\rm volU}(N)]^K}
\int\left(\prod_{k=1}^Kdg_k\right)
{\rm Tr}_{{\cal H}_{\rm ext}}
\Bigl(
\hat{g}_{(K)}
e^{-\frac{H(\hat{E},\hat{U})}{TK}}
\hat{g}_{(K-1)}^{-1}\hat{g}_{(K-1)}
\nonumber\\
&
\quad
e^{-\frac{H(\hat{E},\hat{U})}{TK}}
\hat{g}_{(K-2)}^{-1}\hat{g}_{(K-2)}
\cdots
\hat{g}_{(1)}^{-1}\hat{g}_{(1)}
e^{-\frac{H(\hat{E},\hat{U})}{TK}}
\Bigl)
\nonumber\\
&=
\frac{1}{[{\rm volU}(N)]^K}
\int\left(\prod_{k=1}^Kdg_{(k)}\right)
\int\left(\prod_{k=1}^KdU_{(k)}\right)
\nonumber\\
&
\qquad
\langle U_{(K)}|
\hat{g}_{(K)}
e^{-\frac{H(\hat{E},\hat{U})}{TK}}
\hat{g}_{(K-1)}^{-1}
|U_{(K-1)}\rangle
\nonumber\\
&
\qquad
\times
\langle U_{(K-1)}|
\hat{g}_{(K-1)}
e^{-\frac{H(\hat{E},\hat{U})}{TK}}
\hat{g}_{(K-2)}^{-1}
|U_{(K-2)}\rangle
\nonumber\\
&
\qquad
\times
\cdots
\times
\langle U_{(1)}|
\hat{g}_{(1)}
e^{-\frac{H(\hat{E},\hat{U})}{TK}}
|U_{(K)}\rangle.
\end{align}
For $H(\hat{E},\hat{U})=\frac{1}{2}{\rm Tr}\hat{E}^2+V(\hat{U})$, we can rewrite each term in the product as follows.
\begin{align}
&
\langle U_{(k)}|
\hat{g}_{(k)}
e^{-\frac{H(\hat{E},\hat{U})}{TK}}
\hat{g}_{(k-1)}^{-1}
|U_{(k-1)}\rangle
\nonumber\\
&\quad=
\langle g^{-1}_{(k)}U_{(k)}g_{(k)}|
e^{-\frac{H(\hat{E},\hat{U})}{TK}}
|g^{-1}_{(k-1)}U_{(k-1)}g_{(k-1)}\rangle
\nonumber\\
&\quad=
e^{-\frac{1}{TK}V(g^{-1}_{(k-1)}U_{(k-1)}g_{(k-1)})}
\langle g^{-1}_{(k)}U_{(k)}g_{(k)}|
e^{-\frac{1}{2TK}{\rm Tr}\hat{E}^2}
|g^{-1}_{(k-1)}U_{(k-1)}g_{(k-1)}\rangle. 
\end{align}
By using
\begin{align}
i\epsilon
\equiv
\log\left(
g_{(k)}^{-1}U_{(k)}g_{(k)}g_{(k-1)}^{-1}U^{-1}_{(k-1)}g_{(k-1)}
\right), 
\end{align}
we can relate $|g^{-1}_{(k-1)}U_{(k-1)}g_{(k-1)}\rangle$ and $|g^{-1}_{(k)}U_{(k)}g_{(k)}\rangle$ as 
\begin{align}
    |g^{-1}_{(k)}U_{(k)}g_{(k)}\rangle
    =
    e^{i{\rm Tr}(\epsilon\hat{E})}
    |g^{-1}_{(k-1)}U_{(k-1)}g_{(k-1)}\rangle, 
\end{align}
and hence, 
\begin{align}
&
\langle g^{-1}_{(k)}U_{(k)}g_{(k)}|
e^{-\frac{1}{2TK}{\rm Tr}\hat{E}^2}
|g^{-1}_{(k-1)}U_{(k-1)}g_{(k-1)}\rangle
\nonumber\\
&=
\langle g^{-1}_{(k-1)}U_{(k-1)}g_{(k-1)}|
e^{-\frac{1}{2TK}{\rm Tr}\hat{E}^2-i{\rm Tr}(\epsilon\hat{E})}
|g^{-1}_{(k-1)}U_{(k-1)}g_{(k-1)}\rangle
\nonumber\\
&=
\langle U|
e^{-\frac{1}{2TK}{\rm Tr}\hat{E}^2-i{\rm Tr}(\epsilon\hat{E})}
|U\rangle,
\end{align}
where $U$ can be any element of the unitary group.\footnote{
Because ${\rm Tr}\hat{E}^2$ commutes with the shift on the group manifold, the first line could be $\langle Wg^{-1}_{(k)}U_{(k)}g_{(k)}|
e^{-\frac{1}{2TK}{\rm Tr}\hat{E}^2}
|Wg^{-1}_{(k-1)}U_{(k-1)}g_{(k-1)}\rangle$, where $W$ is an arbitrary element of U($N$). 
}
Up to a normalization constant, this can be written by using trace over the Hilbert space as 
\begin{align}
&
\textrm{Tr}
\exp\left(-\frac{1}{2KT}\sum_\alpha\hat{E}_\alpha^2 + i\sum_\alpha\epsilon^\alpha\hat{E}_\alpha\right)
\nonumber\\
&=
e^{-\frac{KT}{2}{\rm Tr}\epsilon^2}\times
{\rm Tr}
\exp\left(-\frac{1}{2KT}\sum_\alpha(\hat{E}_\alpha-iKT\epsilon_\alpha)^2\right). 
\end{align}
The second term (${\rm Tr}\cdots$) becomes constant in the limit of $KT\to\infty$ when $KT\epsilon^2$ is of order one (which is justified because of the first term $e^{-\frac{KT}{2}{\rm Tr}\epsilon^2}$), because the sum over momentum modes can be identified with usual integral in flat space. Omitting the second term and by approximating $\epsilon^2$ by $|g_{(k)}^{-1}U_{(k)}g_{(k)}-g_{(k-1)}^{-1}U_{(k-1)}g_{(k-1)}|^2$, we obtain
\begin{align}
&
e^{-\frac{KT}{2}{\rm Tr}[|g_{(k)}^{-1}U_{(k)}g_{(k)}-g_{(k-1)}^{-1}U_{(k-1)}g_{(k-1)}|^2]}
e^{-V(g_{(k)}^{-1}U_{(k)}g_{(k)})/(TK)}
\nonumber\\
&\quad\simeq
e^{-L[D_t(g_{(k)}^{-1}U_{(k)}g_{(k)}),(g_{(k)}^{-1}U_{(k)}g_{(k)})]/(TK)}
\nonumber\\
&\quad=
e^{-L[D_tU_{(k)},U_{(k)})]/(TK)}. 
\end{align}
Here we used
\begin{align}
g_{(k-1)}g_{(k)}^{-1}\equiv e^{-iA_{(k)}/(KT)}, 
\end{align}
and
\begin{align}
U_{(k)}-(g_{(k-1)}g_{(k)}^{-1})^{-1}U_{(k-1)}(g_{(k-1)}g_{(k)}^{-1})
\simeq
\frac{D_tU_{(k)}}{KT}. 
\end{align}
By taking $K\to\infty$ limit, we obtain
\begin{align}
Z(T)
&=
\int [dA] [dU]e^{-\int dt L[D_tU,U]}. 
\end{align}
Note that $g_{(K)}$ is the Polyakov loop. In the static diagonal gauge, 
\begin{align}
g_{(k)}={\rm diag}(e^{ik\theta_1/K},\cdots,e^{ik\theta_N/K}). 
\end{align}

In the confined sector, the distribution of the phases of the Polyakov loop is uniform~\cite{Hanada:2020uvt}. We fix the S$_N$ permutation symmetry such that the phase distribution does indeed satisfy this property. This guarantees the neat separation into confined and deconfined sectors at $t=\beta$ (equivalently, $t=0$).

Let us use the trace cyclicity in the extended Hilbert space and shift the initial time. In the static diagonal gauge, this does not change the Polyakov loop. Therefore, exactly the same separation into confined and deconfined sectors holds at any $t$. 

\section{Simulation algorithm}
\hspace{0.51cm}
Several aspects of the simulation algorithm used for this study are similar to the one used in Ref.~\cite{Berkowitz:2016jlq}, which was originally developed in unpublished work by one of the authors (MH), Takashi Kaneko, Jun Nishimura, and Asato Tsuchiya in 2013. 

The lattice action we consider is, before adding the constraint term for the Polyakov loop, 
\begin{align}
S
=
\frac{N}{2a}
\sum_{\mu=1}^{d}\sum_{t=1}^{n_t}
{\rm Tr}\left(
\textbf{1}_N
-
U_{\mu,t}VU_{\mu,t+1}^\dagger V^\dagger 
\right)
+
{\rm h.c.}
+
S_{F.P.}, 
\end{align}
where $V={\rm diag}(e^{i\alpha_1/n_t},\cdots,e^{i\alpha_N/n_t})$. 
Here, $a$ is the lattice spacing and $\beta=an_t$ is the inverse temperature, $\beta=T^{-1}$. 
We will focus on $d=3$.
The Polyakov line phases $\alpha_{1,2,\cdots,N}$ are constrained to be
\begin{align}
-\pi<\alpha_i\le\pi. 
\end{align} 
The Faddeev-Popov term $S_{F.P.}$ is given by  
\begin{align}
S_{F.P.}
= 
-
\sum_{i<j}2\log\left|\sin\left(\frac{\alpha_i-\alpha_j}{2}\right)\right|. 
\end{align}
If we just impose the constraint $-\pi < \alpha_i\le \pi$ as it is, then the simulation is not very efficient -- the `center of mass' $(\sum_i\alpha_i)/N$ randomly moves 
and hits $\pm\pi$. For this reason, we use the following trick. 

Firstly, let us recall that $\alpha_i$ and $\tilde{\alpha}_i\equiv\alpha_i-C$ give the same weight for any $C$.  
Here, $C$ must satisfy ${\rm min}(\tilde{\alpha}_i)+C > -\pi$ and ${\rm max}(\tilde{\alpha}_i)+C \le \pi$,  
because of the condition $-\pi < \alpha_i\le \pi$. Hence $C$ must sit in $(-\pi - {\rm min}(\tilde{\alpha}_i),\pi - {\rm max}(\tilde{\alpha}_i)]$, 
whose interval is $2\pi-\mu$, where $\mu\equiv{\rm max}(\tilde{\alpha}_i)-{\rm min}(\tilde{\alpha}_i)$. 
Therefore, we can replace the integral over $\alpha_i$ with that over $\tilde{\alpha}_i$, with an additional Boltzmann weight 
\begin{eqnarray}
w(\mu) = \left\{
\begin{array}{cc}
2\pi-\mu & (\mu<2\pi)\\
0 &  (\mu\ge 2\pi). 
\end{array}
\right.
\end{eqnarray}

For numerical calculations, this is not very nice because of the singularity at $\mu=2\pi$. 
Instead, in the molecular evolution, we use 
\begin{eqnarray}
\tilde{w}(\mu) = \left\{
\begin{array}{cc}
2\pi-\mu+\varepsilon & (\mu<2\pi)\\
\varepsilon e^{-g_\alpha(\mu-2\pi)} &  (\mu\ge 2\pi),  
\end{array}
\right.
\end{eqnarray}
with a large enough $g_\alpha$ (say $g_\alpha$=100) and $\varepsilon=g_\alpha^{-1}$.
For the Metropolis test, we use $w(\mu)$ instead of $\tilde{w}(\mu)$. 
The detailed balance condition is not violated and the correct path-integral weight is reproduced; see e.g., Ref.~\cite{Hanada-Matsuura}.

\subsection*{Permutation of Polyakov line phases in the constrained simulation of second kind}
\hspace{0.51cm}
When $\Delta S_{\rm dec}$ and $\Delta S_{\rm con}$ are added, the S$_N$ permutation symmetry is explicitly broken to S$_M\times$S$_{N-M}$. The ordering of the $\alpha$'s --- say $\alpha_{i_1}<\alpha_{i_2}<\alpha_{i_3}<\cdots$ --- does not change (or very rarely changes) in the HMC simulation, which can cause a problem when the permutation symmetry is explicitly broken. 
To avoid this problem, we choose $\alpha_i$ and $\alpha_j$ randomly from the deconfined and confined sectors and perform a permutation using the Metropolis algorithm. Note that the $i$-th and $j$-th rows and columns in $U_\mu$ have to be exchanged, too. We do this 10 times after each HMC steps. 
This method is the same as the one used in Ref.~\cite{Watanabe:2020ufk}.

\begin{table}
\small
\centering
\begin{tabular}{|c||l|l|l|l|l|}\hline%
 $N$ &  \ \ \quad $W$(1) & \ \ \quad $W$(2) & \ \ \quad $W$(3) & \ \ \quad $W$(4) & \ \ \quad $W$(5) \\\hline\hline
\begin{filecontents*}{tables/T025.csv}
N,W(1),W(2),W(3),W(4),W(5)
16,0.13755(41),0.03427(40),0.02506(39),0.02809(38),0.03269(38)
24,0.12955(19),0.02357(17),0.01213(16),0.01244(16),0.01442(15)
32,0.12674(17),0.01977(13),0.00752(13),0.00700(11),0.00791(11)
48,0.12548(17),0.01751(13),0.00460(12),0.00334(10),0.003483(96)
64,0.12505(13),0.016542(97),0.003288(88),0.001946(74),0.002086(68)
96,0.124261(79),0.015867(63),0.002493(52),0.000944(49),0.000910(42)
128,0.124144(98),0.015631(76),0.002248(63),0.000742(51),0.000517(48)
\end{filecontents*}
\csvreader[head to column names, late after line=\\]
{tables/T025.csv}{}{\csvcoli & \csvcolii & \csvcoliii & \csvcoliv & \csvcolv & \csvcolvi}%
\hline
\end{tabular}
\caption{Wilson loop for $L=1,2,3,4,5$, unconstrained, $n_t=24$, $T=0.25$. }\label{table-W-T025}
\end{table}

\begin{table}
\small
\centering
\begin{tabular}{|c|c|c||l|l|l|l|l|}\hline%
 $P$ & $N$ &  $n_t$ & \ \quad $W$(1) & \ \quad $W$(2) & \ \quad $W$(3) & \ \quad $W$(4) & \ \quad $W$(5) \\\hline\hline
\begin{filecontents*}{tables/con.csv}
P,N,S,W(1),W(2),W(3),W(4),W(5)
0.2,16,16,0.19720(56),0.04844(32),0.03111(27),0.03531(24),0.04399(23)
,,24,0.20151(44),0.05014(31),0.03150(27),0.03588(24),0.04466(23)
,,32,0.20688(71),0.05279(33),0.03250(27),0.03627(25),0.04480(23)
,24,16,0.18633(31),0.03648(18),0.01522(15),0.01498(13),0.01772(13)
,,24,0.19048(26),0.03914(18),0.01619(15),0.01541(13),0.01835(12)
,,32,0.19626(35),0.04076(19),0.01656(16),0.01549(14),0.01850(12)
,32,16,0.18234(22),0.03325(13),0.01070(11),0.00875(10),0.010341(91)
,,24,0.18740(25),0.03521(15),0.01139(12),0.00888(11),0.01047(10)
,,32,0.19162(22),0.03640(14),0.01173(11),0.009082(98),0.010560(92)
,64,16,0.17677(13),0.029452(90),0.006087(77),0.002606(65),0.002500(60)
,,24,0.18392(22),0.03170(12),0.006751(89),0.003008(82),0.002580(71)
,,32,0.18703(16),0.032779(98),0.006974(80),0.002905(67),0.002640(62)
0.25,16,16,0.2227(41),0.05909(47),0.04191(39),0.04820(34),0.05954(32)
,,24,0.2257(11),0.06278(44),0.04272(36),0.04798(33),0.05964(31)
,,32,0.22946(87),0.06356(45),0.04280(37),0.04818(34),0.05931(32)
,24,16,0.20105(64),0.04231(37),0.02091(31),0.02186(28),0.02626(26)
,,24,0.21621(80),0.04765(40),0.02205(33),0.02212(28),0.02672(26)
,,32,0.21863(93),0.04878(39),0.02327(31),0.02226(29),0.02645(26)
,32,16,0.19751(38),0.03760(20),0.01382(17),0.01228(14),0.01447(13)
,,24,0.20707(68),0.04061(24),0.01455(19),0.01256(16),0.01503(14)
,,32,0.21059(38),0.04224(21),0.01556(17),0.01279(15),0.01490(13)
,64,16,0.18902(31),0.03223(18),0.00721(14),0.00351(13),0.00380(12)
,,24,0.19792(33),0.03484(15),0.00780(12),0.003743(98),0.003560(89)
,,32,0.20000(29),0.03546(14),0.00828(11),0.004051(94),0.003723(87)
\end{filecontents*}
\csvreader[head to column names, late after line=\\]
{tables/con.csv}{}{\csvcoli & \csvcolii & \csvcoliii & \csvcoliv & \csvcolv & \csvcolvi & \csvcolvii & \csvcolviii}%
\hline
\end{tabular}
\caption{$W_{\rm con}$ for $L=1,2,3,4,5$, constrained, $T=0.29$.}\label{table-W-T029-1}
\end{table}

\begin{table}
\small
\centering
\begin{tabular}{|c|c|c||l|l|l|l|l|}\hline%
 $P$ & $N$ &  $n_t$ & \ \quad $W$(1) & \ \quad $W$(2) & \ \quad $W$(3) & \ \quad $W$(4) & \ \quad $W$(5) \\\hline\hline
\begin{filecontents*}{tables/mix.csv}
P,N,S,W(1),W(2),W(3),W(4),W(5)
0.2,16,16,0.13990(48),0.02331(27),0.00490(21),0.00277(19),0.00344(17)
,,24,0.14753(37),0.02590(26),0.00598(21),0.00303(19),0.00356(17)
,,32,0.14991(55),0.02700(26),0.00601(21),0.00315(19),0.00401(17)
,24,16,0.14023(36),0.02299(17),0.00422(14),0.00149(13),0.00146(11)
,,24,0.14921(25),0.02529(18),0.00494(14),0.00155(13),0.00181(11)
,,32,0.14864(33),0.02598(17),0.00490(15),0.00183(13),0.00155(12)
,32,16,0.14380(21),0.02357(13),0.00407(11),0.001061(97),0.000705(88)
,,24,0.15052(23),0.02533(15),0.00464(12),0.00098(11),0.000932(95)
,,32,0.15165(21),0.02624(13),0.00468(11),0.001279(93),0.000807(84)
,64,16,0.14602(14),0.023551(93),0.003613(78),0.000382(71),-0.000016(65)
,,24,0.15021(27),0.02512(12),0.004036(95),0.000452(80),0.000090(75)
,,32,0.15327(17),0.02607(10),0.004283(83),0.000645(72),0.000065(64)
0.25,16,16,0.13695(84),0.02376(27),0.00531(22),0.00275(19),0.00354(18)
,,24,0.14873(49),0.02639(27),0.00559(22),0.00311(20),0.00402(18)
,,32,0.15251(45),0.02780(28),0.00648(23),0.00347(20),0.00382(18)
,24,16,0.14656(39),0.02502(26),0.00480(21),0.00180(19),0.00157(17)
,,24,0.14711(50),0.02594(25),0.00530(21),0.00172(18),0.00172(16)
,,32,0.15323(41),0.02773(26),0.00584(21),0.00191(19),0.00167(16)
,32,16,0.14674(22),0.02455(13),0.00443(11),0.001313(97),0.000991(91)
,,24,0.15044(30),0.02625(15),0.00466(12),0.00118(11),0.001140(97)
,,32,0.15446(20),0.02732(14),0.00523(11),0.00139(10),0.000989(90)
,64,16,0.14876(21),0.02429(13),0.00388(11),0.000710(98),0.000189(90)
,,24,0.15305(19),0.02606(10),0.004208(88),0.000703(74),0.000110(66)
,,32,0.15572(18),0.027028(99),0.004563(80),0.000816(71),0.000190(65)
\end{filecontents*}
\csvreader[head to column names, late after line=\\]
{tables/mix.csv}{}{\csvcoli & \csvcolii & \csvcoliii & \csvcoliv & \csvcolv & \csvcolvi & \csvcolvii & \csvcolviii}%
\hline
\end{tabular}
\caption{$W_{\rm mix}$ for $L=1,2,3,4,5$, constrained, $T=0.29$.}\label{table-W-T029-2}
\end{table}

\begin{table}
\small
\centering
\begin{tabular}{|c|c|c||l|l|l|l|l|}\hline%
 $P$ & $N$ &  $n_t$ & \ \quad $W$(1) & \ \quad $W$(2) & \ \quad $W$(3) & \ \quad $W$(4) & \ \quad $W$(5) \\\hline\hline
\begin{filecontents*}{tables/dec.csv}
P,N,S,W(1),W(2),W(3),W(4),W(5)
0.2,16,16,0.31420(51),0.19824(28),0.18637(25),0.19100(22),0.19865(21)
,,24,0.31659(41),0.19991(30),0.18692(24),0.19110(22),0.19882(21)
,,32,0.32003(52),0.20061(29),0.18713(24),0.19126(22),0.19927(21)
,24,16,0.33554(37),0.21962(20),0.20452(17),0.20506(15),0.20854(14)
,,24,0.33592(29),0.22067(19),0.20486(17),0.20536(15),0.20895(13)
,,32,0.34119(37),0.22169(20),0.20528(17),0.20555(15),0.20885(14)
,32,16,0.32265(26),0.20719(14),0.19100(12),0.19007(11),0.19192(10)
,,24,0.32574(25),0.20918(16),0.19166(13),0.19041(12),0.19206(11)
,,32,0.32924(24),0.20992(14),0.19185(12),0.19072(11),0.192287(95)
,64,16,0.32819(16),0.21602(11),0.198914(93),0.196599(80),0.196673(80)
,,24,0.33471(53),0.21786(13),0.19951(12),0.19681(10),0.196764(92)
,,32,0.33630(20),0.21869(11),0.199764(96),0.196940(84),0.196834(77)
0.25,16,16,0.30670(34),0.19424(23),0.18135(19),0.18456(17),0.19108(16)
,,24,0.30748(33),0.19576(23),0.18229(19),0.18534(17),0.19154(17)
,,32,0.30880(33),0.19676(23),0.18264(19),0.18571(18),0.19191(16)
,24,16,0.29786(28),0.18699(20),0.17150(17),0.17095(15),0.17347(14)
,,24,0.30475(37),0.18914(21),0.17190(17),0.17158(15),0.17376(14)
,,32,0.30477(29),0.18926(22),0.17214(17),0.17185(15),0.17418(14)
,32,16,0.29571(16),0.18425(11),0.167607(92),0.166404(79),0.167424(75)
,,24,0.30153(20),0.18604(12),0.16829(10),0.166629(91),0.167586(88)
,,32,0.30259(16),0.18686(11),0.168547(90),0.166706(79),0.167646(75)
,64,16,0.29883(16),0.18903(11),0.171897(94),0.169310(87),0.169291(81)
,,24,0.30404(15),0.190567(90),0.172290(74),0.169675(74),0.169527(63)
,,32,0.30702(16),0.191836(92),0.172973(74),0.169992(67),0.169785(68)
\end{filecontents*}
\csvreader[head to column names, late after line=\\]
{tables/dec.csv}{}{\csvcoli & \csvcolii & \csvcoliii & \csvcoliv & \csvcolv & \csvcolvi & \csvcolvii & \csvcolviii}%
\hline
\end{tabular}
\caption{$W_{\rm dec}$ for $L=1,2,3,4,5$, constrained, $T=0.29$.}\label{table-W-T029-3}
\end{table}

\bibliographystyle{jhep}
\bibliography{EK-model}

\providecommand{\href}[2]{#2}\begingroup\raggedright\begin{thebibliography}{10}

\bibitem{Witten:1998zw}
E.~Witten, \emph{{Anti-de Sitter space, thermal phase transition, and
  confinement in gauge theories}},
  \href{https://doi.org/10.4310/ATMP.1998.v2.n3.a3}{\emph{Adv. Theor. Math.
  Phys.} {\bfseries 2} (1998) 505}
  [\href{https://arxiv.org/abs/hep-th/9803131}{{\ttfamily hep-th/9803131}}].

\bibitem{Sundborg:1999ue}
B.~Sundborg, \emph{{The Hagedorn transition, deconfinement and N=4 SYM
  theory}}, \href{https://doi.org/10.1016/S0550-3213(00)00044-4}{\emph{Nucl.
  Phys. B} {\bfseries 573} (2000) 349}
  [\href{https://arxiv.org/abs/hep-th/9908001}{{\ttfamily hep-th/9908001}}].

\bibitem{Aharony:2003sx}
O.~Aharony, J.~Marsano, S.~Minwalla, K.~Papadodimas and M.~Van~Raamsdonk,
  \emph{{The Hagedorn - deconfinement phase transition in weakly coupled large
  N gauge theories}},
  \href{https://doi.org/10.4310/ATMP.2004.v8.n4.a1}{\emph{Adv. Theor. Math.
  Phys.} {\bfseries 8} (2004) 603}
  [\href{https://arxiv.org/abs/hep-th/0310285}{{\ttfamily hep-th/0310285}}].

\bibitem{Polyakov:1978vu}
A.M.~Polyakov, \emph{{Thermal Properties of Gauge Fields and Quark
  Liberation}}, \href{https://doi.org/10.1016/0370-2693(78)90737-2}{\emph{Phys.
  Lett. B} {\bfseries 72} (1978) 477}.

\bibitem{Susskind:1979up}
L.~Susskind, \emph{{Lattice Models of Quark Confinement at High Temperature}},
  \href{https://doi.org/10.1103/PhysRevD.20.2610}{\emph{Phys. Rev. D}
  {\bfseries 20} (1979) 2610}.

\bibitem{Hanada:2016pwv}
M.~Hanada and J.~Maltz, \emph{{A proposal of the gauge theory description of
  the small Schwarzschild black hole in AdS$_5\times$S$^5$}},
  \href{https://doi.org/10.1007/JHEP02(2017)012}{\emph{JHEP} {\bfseries 02}
  (2017) 012} [\href{https://arxiv.org/abs/1608.03276}{{\ttfamily
  1608.03276}}].

\bibitem{Berenstein:2018lrm}
D.~Berenstein, \emph{{Submatrix deconfinement and small black holes in AdS}},
  \href{https://doi.org/10.1007/JHEP09(2018)054}{\emph{JHEP} {\bfseries 09}
  (2018) 054} [\href{https://arxiv.org/abs/1806.05729}{{\ttfamily
  1806.05729}}].

\bibitem{Hanada:2018zxn}
M.~Hanada, G.~Ishiki and H.~Watanabe, \emph{{Partial Deconfinement}},
  \href{https://doi.org/10.1007/JHEP03(2019)145}{\emph{JHEP} {\bfseries 03}
  (2019) 145} [\href{https://arxiv.org/abs/1812.05494}{{\ttfamily
  1812.05494}}].

\bibitem{Hanada:2019czd}
M.~Hanada, A.~Jevicki, C.~Peng and N.~Wintergerst, \emph{{Anatomy of
  Deconfinement}}, \href{https://doi.org/10.1007/JHEP12(2019)167}{\emph{JHEP}
  {\bfseries 12} (2019) 167}
  [\href{https://arxiv.org/abs/1909.09118}{{\ttfamily 1909.09118}}].

\bibitem{Hanada:2020uvt}
M.~Hanada, H.~Shimada and N.~Wintergerst, \emph{{Color confinement and
  Bose-Einstein condensation}},
  \href{https://doi.org/10.1007/JHEP08(2021)039}{\emph{JHEP} {\bfseries 08}
  (2021) 039} [\href{https://arxiv.org/abs/2001.10459}{{\ttfamily
  2001.10459}}].

\bibitem{Aoki:2006we}
Y.~Aoki, G.~Endrodi, Z.~Fodor, S.D.~Katz and K.K.~Szabo, \emph{{The Order of
  the quantum chromodynamics transition predicted by the standard model of
  particle physics}}, \href{https://doi.org/10.1038/nature05120}{\emph{Nature}
  {\bfseries 443} (2006) 675}
  [\href{https://arxiv.org/abs/hep-lat/0611014}{{\ttfamily hep-lat/0611014}}].

\bibitem{Hanada:2019kue}
M.~Hanada and B.~Robinson, \emph{{Partial-Symmetry-Breaking Phase
  Transitions}}, \href{https://doi.org/10.1103/PhysRevD.102.096013}{\emph{Phys.
  Rev. D} {\bfseries 102} (2020) 096013}
  [\href{https://arxiv.org/abs/1911.06223}{{\ttfamily 1911.06223}}].

\bibitem{Kogut:1974ag}
J.B.~Kogut and L.~Susskind, \emph{{Hamiltonian Formulation of Wilson's Lattice
  Gauge Theories}}, \href{https://doi.org/10.1103/PhysRevD.11.395}{\emph{Phys.
  Rev. D} {\bfseries 11} (1975) 395}.

\bibitem{Hanada:2021ipb}
M.~Hanada, \emph{{Bulk geometry in gauge/gravity duality and color degrees of
  freedom}}, \href{https://doi.org/10.1103/PhysRevD.103.106007}{\emph{Phys.
  Rev. D} {\bfseries 103} (2021) 106007}
  [\href{https://arxiv.org/abs/2102.08982}{{\ttfamily 2102.08982}}].

\bibitem{Hanada:2021swb}
M.~Hanada, \emph{{Large-$N$ limit as a second quantization}},
  \href{https://arxiv.org/abs/2103.15873}{{\ttfamily 2103.15873}}.

\bibitem{Gautam:2022akq}
V.~Gautam, M.~Hanada, A.~Jevicki and C.~Peng, \emph{{Matrix Entanglement}},
  \href{https://arxiv.org/abs/2204.06472}{{\ttfamily 2204.06472}}.

\bibitem{Langfeld:2012ah}
K.~Langfeld, B.~Lucini and A.~Rago, \emph{{The density of states in gauge
  theories}}, \href{https://doi.org/10.1103/PhysRevLett.109.111601}{\emph{Phys.
  Rev. Lett.} {\bfseries 109} (2012) 111601}
  [\href{https://arxiv.org/abs/1204.3243}{{\ttfamily 1204.3243}}].

\bibitem{Aarts:2023vsf}
G.~Aarts et~al., \emph{{Phase Transitions in Particle Physics -- Results and
  Perspectives from Lattice Quantum Chromo-Dynamics}},  in \emph{{Progress in
  Particle and Nuclear Physics}}, 1, 2023
  [\href{https://arxiv.org/abs/2301.04382}{{\ttfamily 2301.04382}}].

\bibitem{Hanada:2014noa}
M.~Hanada, J.~Maltz and L.~Susskind, \emph{{Deconfinement transition as black
  hole formation by the condensation of QCD strings}},
  \href{https://doi.org/10.1103/PhysRevD.90.105019}{\emph{Phys. Rev. D}
  {\bfseries 90} (2014) 105019}
  [\href{https://arxiv.org/abs/1405.1732}{{\ttfamily 1405.1732}}].

\bibitem{Eguchi:1982nm}
T.~Eguchi and H.~Kawai, \emph{{Reduction of Dynamical Degrees of Freedom in the
  Large N Gauge Theory}},
  \href{https://doi.org/10.1103/PhysRevLett.48.1063}{\emph{Phys. Rev. Lett.}
  {\bfseries 48} (1982) 1063}.

\bibitem{Watanabe:2020ufk}
H.~Watanabe, G.~Bergner, N.~Bodendorfer, S.~Shiba~Funai, M.~Hanada, E.~Rinaldi
  et~al., \emph{{Partial deconfinement at strong coupling on the lattice}},
  \href{https://doi.org/10.1007/JHEP02(2021)004}{\emph{JHEP} {\bfseries 02}
  (2021) 004} [\href{https://arxiv.org/abs/2005.04103}{{\ttfamily
  2005.04103}}].

\bibitem{Mason:2022aka}
D.~Mason, B.~Lucini, M.~Piai, E.~Rinaldi and D.~Vadacchino, \emph{{The density
  of state method for first-order phase transitions in Yang-Mills theories}},
  \href{https://doi.org/10.22323/1.430.0216}{\emph{PoS} {\bfseries LATTICE2022}
  (2023) 216} [\href{https://arxiv.org/abs/2212.01074}{{\ttfamily
  2212.01074}}].

\bibitem{skilling2004nested}
J.~Skilling, \emph{Nested sampling},  in \emph{Aip conference proceedings},
  vol.~735, pp.~395--405, American Institute of Physics, 2004.

\bibitem{Buchner2021nestedsampling}
J.~{Buchner}, \emph{{Nested Sampling Methods}}, {\emph{arXiv e-prints} (2021)
  arXiv:2101.09675} [\href{https://arxiv.org/abs/2101.09675}{{\ttfamily
  2101.09675}}].

\bibitem{Buchner2021ultranest}
J.~{Buchner}, \emph{{UltraNest - a robust, general purpose Bayesian inference
  engine}}, \href{https://doi.org/10.21105/joss.03001}{\emph{The Journal of
  Open Source Software} {\bfseries 6} (2021) 3001}
  [\href{https://arxiv.org/abs/2101.09604}{{\ttfamily 2101.09604}}].

\bibitem{Nair:2005iw}
V.P.~Nair, \emph{{Quantum field theory: A modern perspective}}, {Springer New
  York, NY} (2005).

\bibitem{Sokal1996MonteCM}
A.~Sokal, \emph{Monte carlo methods in statistical mechanics: Foundations and
  new algorithms},  in \emph{Functional Integration: Basics and Applications},
  C.~DeWitt-Morette, P.~Cartier and A.~Folacci, eds., (Boston, MA),
  pp.~131--192, Springer US (1997),
  \href{https://doi.org/10.1007/978-1-4899-0319-8_6}{DOI}.

\bibitem{Hanada:2021ksu}
M.~Hanada, J.~Holden, M.~Knaggs and A.~O'Bannon, \emph{{Global symmetries and
  partial confinement}},
  \href{https://doi.org/10.1007/JHEP03(2022)118}{\emph{JHEP} {\bfseries 03}
  (2022) 118} [\href{https://arxiv.org/abs/2112.11398}{{\ttfamily
  2112.11398}}].

\bibitem{Maldacena:1998im}
J.M.~Maldacena, \emph{{Wilson loops in large N field theories}},
  \href{https://doi.org/10.1103/PhysRevLett.80.4859}{\emph{Phys. Rev. Lett.}
  {\bfseries 80} (1998) 4859}
  [\href{https://arxiv.org/abs/hep-th/9803002}{{\ttfamily hep-th/9803002}}].

\bibitem{Page:1993wv}
D.N.~Page, \emph{{Information in black hole radiation}},
  \href{https://doi.org/10.1103/PhysRevLett.71.3743}{\emph{Phys. Rev. Lett.}
  {\bfseries 71} (1993) 3743}
  [\href{https://arxiv.org/abs/hep-th/9306083}{{\ttfamily hep-th/9306083}}].

\bibitem{Rinaldi:2021jbg}
E.~Rinaldi, X.~Han, M.~Hassan, Y.~Feng, F.~Nori, M.~McGuigan et~al.,
  \emph{{Matrix-Model Simulations Using Quantum Computing, Deep Learning, and
  Lattice Monte Carlo}},
  \href{https://doi.org/10.1103/PRXQuantum.3.010324}{\emph{PRX Quantum}
  {\bfseries 3} (2022) 010324}
  [\href{https://arxiv.org/abs/2108.02942}{{\ttfamily 2108.02942}}].

\bibitem{Berkowitz:2016jlq}
E.~Berkowitz, E.~Rinaldi, M.~Hanada, G.~Ishiki, S.~Shimasaki and P.~Vranas,
  \emph{{Precision lattice test of the gauge/gravity duality at large-$N$}},
  \href{https://doi.org/10.1103/PhysRevD.94.094501}{\emph{Phys. Rev. D}
  {\bfseries 94} (2016) 094501}
  [\href{https://arxiv.org/abs/1606.04951}{{\ttfamily 1606.04951}}].

\bibitem{Hanada-Matsuura}
M.~Hanada and S.~Matsuura, \emph{{MCMC from scratch --- a practical
  introduction to Markov Chain Monte Carlo method}}, Springer (2022).

\end{thebibliography}\endgroup

\end{document}